\documentclass[a4paper, 11pt]{article}
\usepackage[T1]{fontenc}
\usepackage[utf8]{inputenc}
\usepackage[english]{babel}
\usepackage{amsmath}
\usepackage{amssymb}
\usepackage{braket}
\usepackage{mathtools}
\usepackage{dsfont}
\usepackage{subcaption}
\usepackage{array}
\usepackage{booktabs}
\usepackage{yfonts}
\usepackage{setspace}
\usepackage{verbatim}

\usepackage[usenames,dvipsnames]{color}

\numberwithin{equation}{section}

\def\be{\begin{equation}}
\def\ee{\end{equation}}
\def\rme{{\rm e}}
\newcommand{\nn}{\nonumber}
\newcommand{\diff}{\mathrm{d}}
\def\constA{\xi}
\def\Oa{\mathcal{O}(\alpha)}

\def\aa{\mathtt{a}}
\def\cc{\mathtt{c}}

\usepackage{jheppub}                   % jheppub.sty

\makeatletter
\gdef\@fpheader{\ }                    % hack the jhep header
\makeatother

\title{Corrections to AdS$_5$ Black Hole Thermodynamics from Higher-Derivative Supergravity}

\author[a]{Davide Cassani,}
\author[a,b]{Alejandro Ruip\'erez,}
\author[b]{Enrico Turetta,}
\emailAdd{davide.cassani, alejandro.ruiperez @pd.infn.it, enrico.turetta@studenti.unipd.it}

\affiliation[a]{INFN, Sezione di Padova, Via Marzolo 8, 35131 Padova, Italy}
\affiliation[b]{Dipartimento di Fisica e Astronomia ``Galileo Galilei'',\\ Universit\`a di Padova, Via Marzolo 8, 35131 Padova, Italy}

\abstract{
We study four-derivative corrections to five-dimensional minimal gauged supergravity. We evaluate the on-shell action of the AdS$_5$ black hole solution with two independent angular momenta and one electric charge at linear order in the corrections. After imposing supersymmetry, we are able to recast the action in terms of the supersymmetric chemical potentials and match the result obtained from the dual superconformal index on the second sheet. To achieve this, we exploit the freedom to implement field redefinitions to recast the action in a much simpler form, as well as the fact that the two-derivative solution is enough. We use the on-shell action to determine the corrections to the black hole thermodynamics, including those to the entropy and the charges. We then specialize to the supersymmetric and extremal case and find a simple expression for the microcanonical entropy. In particular, for the case with one independent angular momentum the corrections are entirely encoded in the dual superconformal anomaly coefficients. We corroborate this result for the entropy by constructing the corrected near-horizon solution and applying Wald's formula.
}

\begin{document}

\maketitle

%%%%%%%%%%%%%%%%%%%%%%%%%%%%%%%%%%%%%%%%%%%%%%%%%%%%%%%%%%%%%%%%%%%%%%%%%%%%%%%%%%%%%%
\section{Introduction}\label{sec:intro}

The fundamental theory of quantum gravity is expected to manifest itself at low energies via a series of higher-derivative curvature corrections to the universal two-derivative theory. Investigating the structure of such corrections and determining how they affect the physical observables is key for gaining insight into the UV complete theory. For instance, studying the consistency of the low-energy effective theory led to bounds on the corrections and was a motivation for the weak-gravity conjecture \cite{Adams:2006sv,Arkani-Hamed:2006emk}. 

One avenue to make progress in this arduous endeavour is to concentrate on setups allowing for enhanced control. For instance, imposing supersymmetry the higher-derivative terms are greatly constrained, and can in principle be constructed systematically whenever an off-shell formulation of supergravity is available. Further, when considering supergravity in asymptotically AdS spaces one can combine the principles of holography with exact quantum field theory results to obtain a valuable guidance in exploring the gravitational higher-derivative effective theory. 

The present paper builds on this approach and focuses on the higher-derivative corrections to five-dimensional minimal gauged supergravity and the thermodynamics of its asymptotically AdS black hole solutions. 
 Some steps in this direction have previously been taken in \cite{Baggio:2014hua,Bobev:2021qxx,Liu:2022sew}, see also \cite{Melo:2020amq} for a study starting from $\alpha'{}^{\,3}$ corrections to type IIB supergravity on $S^5$.\footnote{A similar program for asymptotically locally AdS$_4$ solutions has been carried out in \cite{Bobev:2020egg,Bobev:2021oku,Genolini:2021urf}.}

Both supersymmetric and non-supersymmetric asymptotically AdS black hole solutions to five-dimensional minimal gauged supergravity are known~\cite{Gutowski:2004ez,Chong:2005hr}.
Via holography, the microstates accounting for the entropy of such black holes are expected to correspond to states of four-dimensional $\mathcal{N}=1$ superconformal field theories (SCFT's) at large $N$. Since minimal gauged supergravity arises as a consistent truncation around any supersymmetric AdS$_5$ solution to ten- or eleven-dimensional supergravity, one can see the black hole as a universal deformation of such higher-dimensional solutions, obtained by turning on suitable chemical potentials for angular momentum and graviphoton charge.
The corresponding field theory statement is that the dual microstates should exist in any holographic $\mathcal{N}=1$ SCFT, and furthermore that they are counted by a generating function depending on the same chemical potentials as the black hole.

In the last few years, there has been a lot of progress in defining and studying the SCFT partition function counting the microstates for supersymmetric black holes in AdS spaces, see e.g.~\cite{Zaffaroni:2019dhb} for a review. Starting with~\cite{Cabo-Bizet:2018ehj,Choi:2018hmj,Benini:2018ywd}, different methods have been devised to address  AdS$_5$ black holes and the dual four-dimensional SCFT's. 
The relevant SCFT generating function turns out to be the superconformal index~\cite{Romelsberger:2005eg,Kinney:2005ej} in a regime where the chemical potentials are complex. This is also computed by the path integral of the theory on $S^1\times S^3$ \cite{Closset:2013vra,Assel:2014paa}, with sources chosen so as to match the black hole asymptotics \cite{Cabo-Bizet:2018ehj}.  Working at leading order in the large-$N$ limit, it has been possible to isolate the saddles in the SCFT index corresponding to the contribution of supersymmetric black holes to the gravitational partition function, and successfully match the corresponding Bekenstein-Hawking entropy.

However, the index is given at finite $N$ and should thus compute the full quantum gravity partition function with the assigned boundary conditions. It should thus be possible to extract the higher-derivative and quantum corrections to the two-derivative theory by studying the large-$N$ expansion of the index order by order.
This is a hard problem as it has been found that there exist many competing large-$N$ contributions  \cite{Benini:2018ywd,Cabo-Bizet:2019eaf,ArabiArdehali:2019orz,Cabo-Bizet:2020nkr}, which may dominate in different regimes of the chemical potentials and are expected to correspond to different solutions contributing to the gravitational partition function \cite{Aharony:2021zkr}. 

One way to pick up the black hole saddle we are interested in is to perform a specific Cardy-like limit of small chemical potentials before taking the large-$N$ limit. The index of a four-dimensional $\mathcal{N}=1$ SCFT is a branched function of two complex chemical potentials
$\omega_1,\omega_2$, sourcing the combinations of angular momenta and R-charge that commute with the supercharge used to define it. One can move from one sheet to the other of this branched function by making the transformation $\omega_1\to\omega_1-2\pi i$. On the ``second sheet'', the index $\mathcal{I}$ may be written as 
\cite{Cabo-Bizet:2018ehj,Choi:2018hmj,Kim:2019yrz,Cabo-Bizet:2019osg} 
\be\label{unusual_index}
\mathcal{I} \,=\, {\rm Tr}\, \rme^{-\pi i Q_{R}}  \,\rme^{ 
  \,\omega_1\left(J_1 + \frac{1}{2}Q_{R} \right)+ \omega_2 \left(J_2 + \frac{1}{2}Q_{R} \right) }\,,
\ee
where the trace is taken over states in the Hilbert space of the theory that are annihilated by a chosen supercharge, % in radial quantization, 
$J_1,J_2$ are the angular momenta in the Cartan of the SO(4) symmetry of $S^3$, and $Q_{R}$ is the U(1) R-charge. Note that  as a consequence of shifting $\omega_1$,
the $(-1)^F$ familiar from the Witten index is replaced by $\rme^{-\pi i Q_{R}}$, that is why \eqref{unusual_index} was also dubbed the ``R-charge index'' in~\cite{Cassani:2021fyv}. 
Using a three-dimensional effective field theory approach, it was shown in~\cite{Cassani:2021fyv} (see also \cite{ArabiArdehali:2021nsx,GonzalezLezcano:2020yeb,Amariti:2021ubd,Ohmori:2021sqg} for related results) that under mild assumptions the small chemical potential regime $\omega_1,\omega_2\to 0$ of \eqref{unusual_index} can be expressed as\footnote{The different signs appearing in the first line of this expression compared to the analog expression (1.8) of~\cite{Cassani:2021fyv} are due to the redefinition $\omega_{1,2}^{\rm there}=-\omega_{1,2}^{\rm here}$ that we made here so as to match the gravity conventions used below.}
\be
\begin{aligned}\label{cardyresult}
\log{\mathcal{I}} \,&=\, -{\rm Tr}\mathcal{R}^3\, \frac{  (\omega_1+\omega_2- 2\pi i)^3}{48\,\omega_1\omega_2}  +  {\rm Tr}\mathcal{R}\,\frac{  (\omega_1+\omega_2 - 2\pi i )  (\omega_1^2 + \omega_2^2-4\pi^2 )}{48\,\omega_1\omega_2}  \\[1mm]
\,&\,\quad + \log |\mathcal{G}|+ \text{exp-terms}\,,
\end{aligned}
\ee
where ``exp-terms'' denotes exponentially suppressed terms, that have not been computed so far. 
 The expression only depends on the SCFT through the R-symmetry anomaly coefficients ${\rm Tr}\mathcal{R}^3$, ${\rm Tr}\mathcal{R}$, and through the rank $|\mathcal{G}|$ of the discrete one-form symmetry group $\mathcal{G}$ that the SCFT may have. The dependence of these quantities on the gauge group rank-parameter $N$ is sensitive to the details of the SCFT. However, it is always true  that for holographic theories  ${\rm Tr}\mathcal{R}$ is subleading in the large-$N$ expansion, hence one only keeps the ${\rm Tr}\mathcal{R}^3$ term when working at leading order. It was shown in \cite{Hosseini:2017mds} that the Legendre transform of this term indeed gives the Bekenstein-Hawking entropy of the supersymmetric black hole solutions %with two angular momenta and one electric charge
 of~\cite{Gutowski:2004ez,Chong:2005hr}. The same term was also matched with the supersymmetric  two-derivative on-shell action in~\cite{Cabo-Bizet:2018ehj}. A main point of~\cite{Cabo-Bizet:2018ehj} is that the on-shell action can be evaluated by imposing supersymmetry without also taking the extremal limit if one allows for a complexification of the black hole solution; this directly matches the dual superconformal index at leading order in the large-$N$ expansion.

However, since the expression \eqref{cardyresult} is valid at finite $N$, its large-$N$ expansion  provides a prediction for the higher-derivative and quantum corrections to the semiclassical  black hole contribution to the gravitational partition function.

In the present work we develop the holographic counterpart of this story by matching the full first line in \eqref{cardyresult} via a gravitational computation. The needed ingredients are contained in five-dimensional minimal gauged supergravity and its supersymmetric higher-derivative corrections. In particular, the chiral R-symmetry anomaly controlled by ${\rm Tr}\,\mathcal{R}^3$ is matched  by the two-derivative Chern-Simons term $\epsilon^{\mu\nu\rho\sigma\lambda} F_{\mu\nu}  F_{\rho\sigma}A_\lambda$ constructed out of the graviphoton $A_\mu$ (that is, the U(1) gauge field entering in the minimal supergravity multiplet), while the mixed gauge-gravitational anomaly controlled by ${\rm Tr}\,\mathcal{R}$ is reproduced by varying the four-derivative term $\epsilon^{\mu\nu\rho\sigma\lambda} {\rm} R_{\mu\nu}{}^{\alpha\beta} R_{\rho\sigma\alpha\beta}A_\lambda$. One can thus suspect that the  supersymmetrization of these terms is all what is needed in order to match the first line of~\eqref{cardyresult}. As we will see, this is indeed the case.

Given the intrinsic complication of dealing with higher-derivative supersymmetry invariants, we devote the first part of our work to recast the four-derivative supergravity action in the simplest possible form. In order to do so, we start from off-shell supergravity in the standard Weyl formulation and eliminate the auxiliary fields. Then we implement a series of field redefinitions so as to simplify the action. The resulting expression, that will be the basis of our study, is given in Eq.~\eqref{eq:4daction2} below. Here, the corrections are proportional to a parameter $\alpha$ with dimensions of length$^2$. Their precise contribution to the action depends on two dimensionless parameters, $\lambda_1,\lambda_2$,  reflecting the fact that we have started our analysis from a generic linear combination of two four-derivative off-shell invariants. The parameters $\lambda_1,\lambda_2$ are not fixed in the effective theory and should in principle be determined from the UV completion.
Although a third off-shell supersymmetry invariant exists, we will provide a physical argument showing that our action is in fact the most general one, at least for the purposes of the present paper. 

Then we evaluate the action at linear order in $\alpha$ for the non-supersymmetric black hole of~\cite{Chong:2005hr}. This is possible although we are lacking the corrected black hole solution. Indeed, adapting to the theory of interest an argument of~\cite{Reall:2019sah}, we show that it is sufficient to evaluate the action on the uncorrected solution. An important aspect of the proof is that it requires fixed boundary conditions. (This is in line with the fact that we are working in the grand-canonical ensemble and thus keep the thermodynamic potentials fixed.)
Because of this requirement, when implementing field redefinitions to simplify our four-derivative action, we have been careful in restricting to redefinitions that preserve the asymptotic structure of the metric and gauge field.

In order to remove the divergences from the action integral we use holographic renormalization. Although the general set of higher-derivative holographic counterterms for gravitational theories comprising the metric and a gauge field is not known, we argue that the ones used e.g.\ in~\cite{Cremonini:2009ih} are sufficient for our purposes. This is because the only purely gravitational term appearing in our action~\eqref{eq:4daction2} is the Gauss-Bonnet term, for which both the Gibbons-Hawking boundary term and the holographic counterterms are known. Moreover, the gauge field is sufficiently suppressed asymptotically \hbox{so as not to affect the boundary terms.}

Then, using the approach of~\cite{Cabo-Bizet:2018ehj}, we impose just the supersymmetry condition and, after translating the gravitational couplings into SCFT anomaly coefficients, show that the resulting on-shell action precisely matches the first line of \eqref{cardyresult}.

We then use the on-shell action to study the thermodynamics. In  Euclidean quantum gravity, the renormalized on-shell action with Dirichlet boundary conditions is interpreted as the logarithm of the thermodynamical partition function in the grand-canonical ensemble~\cite{Gibbons:1976ue}. Thus the corrected action provides the corrections to the black hole thermodynamics, both with and without supersymmetry.  Assuming that the first law and the quantum statistical relation continue to hold at linear order in $\alpha$, we determine the corrected charges and entropy. Then we obtain their supersymmetric and extremal (BPS) expressions; these can be found in Subsection~\ref{aeqb_sect} for the case of one independent angular momentum, and in Subsection~\ref{aneqb_sect} for the case of general angular momenta. 

In the BPS limit, the angular momentum and electric charges satisfy a non-linear relation that corrects to $\Oa$ the relation already known at the two-derivative level. We also find that the entropy can be nicely expressed in terms of the charges, thus obtaining its microcanonical form. 
When converted to field theory units, the BPS entropy reads
\be\label{eq:microcanonicalentropyCFT_intro}
\mathcal S \,=\,\pi\sqrt{3Q_R^2 -8 \mathtt a \left(J_1 + J_2\right) - 16\, \mathtt a \left(\mathtt{a}-\mathtt c\right) \frac{(J_1-J_2)^2}{Q_R^2-2\,\mathtt a\left(J_1 + J_2\right)}}\,\,,
\ee
while the non-linear relation between the charges can be written as
\be\label{corrected_nonlinear_rel_BPScharges_intro}
\begin{aligned}
& \left[3 Q_R  + 4\left(2\,\aa-\cc\right) \right]\left[ 3 Q_R^2 -  8\cc\, (J_1+J_2)\right]   \\[1mm]
&= Q_R^3 + 16 \left(3\cc-2\aa\right)J_1J_2 +\,64\aa\, (\aa-\cc)\frac{(Q_R+\aa)(J_1-J_2)^2}{Q_R^2-2\aa (J_1+J_2)}\,,
\end{aligned}
\ee
where $\aa,\cc$ are the SCFT central charges, related to the R-symmetry anomaly coefficients appearing in \eqref{cardyresult} by the well-known map $\aa = \frac{3}{32}(3\,{\rm Tr}\mathcal{R}^3 - {\rm Tr}\mathcal{R})\,,$ $\cc = \frac{1}{32}(9\,{\rm Tr}\mathcal{R}^3 - 5\, {\rm Tr}\mathcal{R})$. The previously known leading-order version of \eqref{eq:microcanonicalentropyCFT_intro}, \eqref{corrected_nonlinear_rel_BPScharges_intro} is recovered by setting $\aa=\cc$, and the formulae above hold at linear order in $\aa-\cc$. Besides resolving the degeneracy between these two anomaly coefficients, our result shows that the corrections to the microcanonical entropy at linear order in the four-derivative corrections is encoded in the corrections to $\aa$ and $\cc$, as well as in a new term proportional to $(J_1-J_2)^2$, which thus vanishes for equal angular momenta.

In order to verify our result for the BPS entropy we also construct the corrected near-horizon solution for the black hole of \cite{Gutowski:2004ez}. 
 Evaluating the Wald entropy formula on this solution we find perfect agreement with the expression obtained from the on-shell action.

We further validate our results by directly performing the Legendre transform of the supersymmetric on-shell action, or equivalently of the first line of~\eqref{cardyresult}, at linear order in ${\rm Tr}\,\mathcal{R}$. Besides representing a nice consistency check, this computation provides a very direct way to obtain the BPS entropy, extending to higher-order the method already proven useful in~\cite{Cabo-Bizet:2018ehj,Cassani:2019mms}.

It would be very interesting to investigate  how the expressions \eqref{eq:microcanonicalentropyCFT_intro}, \eqref{corrected_nonlinear_rel_BPScharges_intro} are modified beyond the first subleading order in the corrections. To test this, one should study higher orders in $\alpha$ and incorporate in the gravitational effective action supersymmetric terms comprising more than four derivatives, as well as quantum corrections. While this is of course a very hard task, a simpler way to assess it would be to  determine the exact Laplace transform of \eqref{cardyresult}. The fact that the four-derivative corrections already match the full functional form of the first line of~\eqref{cardyresult} may indicate some intriguing simplifications, perhaps based on non-renormalization theorems.

The remainder of the paper is organized as follows. In Section~\ref{TwoDerReview} we review the solution of~\cite{Chong:2005hr} together with its supersymmetric thermodynamics. In Section~\ref{FourDerAction} we work out our four-derivative effective action starting from off-shell Poincar\'e supergravity and exploiting field redefinitions.
In Section~\ref{sec:onshellact} we evaluate the four-derivative action at linear order in $\alpha$ on the black hole solution and match the field theory result \eqref{cardyresult} after imposing supersymmetry. In Section~\ref{sec:BPSlimit} we use the black hole thermodynamical relations to obtain the charges and the entropy, focusing on their BPS values. This leads us to the microcanonical form of the BPS entropy. In Section~\ref{sec:near_horizon} we construct the corrected near-horizon geometry for the black hole of \cite{Gutowski:2004ez} and check our result for the BPS entropy via Wald's formula. In Section~\ref{constrained_transform} we evaluate the Legendre transform of the supersymmetric on-shell action. We conclude in Section~\ref{sec:discussion}. In Appendix~\ref{sec:holographicdictionary}
we discuss the holographic dictionary between the gravitational couplings and the SCFT anomaly coefficients, in Appendix~\ref{app:non_susy_charges} we give the non-supersymmetric charges and entropy, while in Appendix~\ref{app:eoms} we provide the higher-derivative equations of motion.

\bigskip

{\bf Note added in v1.} While the present paper was under completion we became aware of \cite{Bobev:2022bjm}, which has considerable overlap. While our results for the corrected on-shell action agree, our expression for the BPS entropy disagrees with the one presented there. The reason may be that in \cite{Bobev:2022bjm} the formula for the entropy is evaluated on the two-derivative solution.

\medskip

{\bf Note added in v2.} The results for independent angular momenta, given in Subsection~\ref{aneqb_sect}, as well as the Legendre transform of the supersymmetric action, discussed in Section~\ref{constrained_transform}, have been derived in the revised version of the present paper.

%%%%%%%%%%%%%%%%%%%%%%%%%%%%%%%%%%%%%%%%%%%%%%%%%%%%%%%%%%%%%%%%%%%%%%%%%%%%%%%%%%%%%%
\section{The two-derivative solution and supersymmetric thermodynamics}\label{TwoDerReview}

In this section we briefly review the AdS$_5$ black hole solution to minimal five-dimensional gauged supergravity given in~\cite{Chong:2005hr}, as well as its supersymmetric thermodynamics presented in~\cite{Cabo-Bizet:2018ehj}. We will only present the features that are important for our discussion, referring to the original references for further details.

The two-derivative bosonic action is given by\footnote{We are using the same conventions as~\cite{Chong:2005hr}. The Levi-Civita tensor is defined by $\epsilon^{01234}=-e^{-1}$, where $e$ denotes the determinant of the f\"unfbein $e^{a}{}_{\mu}$.}
\begin{equation}\label{eq:2daction}
S \,=\, \frac{1}{16\pi G}\int \diff^5x \,e \left[R+12g^2-\frac{1}{4}F^2-\frac{1}{12\sqrt{3}}\epsilon^{\mu\nu\rho\sigma\lambda}F_{\mu\nu}F_{\rho\sigma} A_{\lambda}\right]\, ,
\end{equation}
with equations of motion
\begin{equation}\label{EoM2der}
\begin{aligned}
{\cal E}_{\mu\nu}\,&\equiv\, R_{\mu\nu}+4g^2 g_{\mu\nu} -\frac{1}{2}F_{\mu\rho}F_{\nu}{}^{\rho}+\frac{1}{12}g_{\mu\nu}F^2=0\,,\\
{\cal E}^{\mu}\,&\equiv\,\nabla_{\nu}F^{\nu \mu}-\frac{1}{4\sqrt{3}}\epsilon^{\mu\nu\rho\sigma\lambda}F_{\nu\rho}F_{\sigma\lambda}=0\,.
\end{aligned}
\end{equation}
Here $A_\mu$ is an Abelian gauge field, $F_{\mu\nu}=2\partial_{[\mu}A_{\nu]}$, and $F^2=F_{\mu\nu}F^{\mu\nu}$. The parameter $g$ controlling the cosmological constant is normalized so that the AdS solution has radius $1/g$.

This theory arises as a universal consistent truncation of ten- or eleven-dimensional supergravity on any internal geometry $M$ allowing for a supersymmetric AdS$_5\times M$ solution (where the product $\times$ may be warped) \cite{Gauntlett:2007ma,Cassani:2019vcl}. In particular, it arises as a consistent truncation of type IIB supergravity on any Sasaki-Einstein five-dimensional manifold, including~$S^5$.

The most general known asymptotically AdS$_5$ black hole solution to this theory was given in  \cite{Chong:2005hr} and reads in the coordinates $t,\theta,\phi,\psi,r$:
%%%%%
\begin{eqnarray}
\diff s^2 &=& -\frac{\Delta_\theta\, [(1+g^2 r^2)\rho^2 \diff t + 2q \nu]
\, \diff t}{\Xi_a\, \Xi_b \, \rho^2} + \frac{2q\, \nu\omega}{\rho^2}
+ \frac{f}{\rho^4}\Big(\frac{\Delta_\theta \, \diff t}{\Xi_a\Xi_b} -
\omega\Big)^2 + \frac{\rho^2 \diff r^2}{\Delta_r} +
\frac{\rho^2 \diff \theta^2}{\Delta_\theta}\nn\\
&& + \frac{r^2+a^2}{\Xi_a}\sin^2\theta\, \diff \phi^2 +
      \frac{r^2+b^2}{\Xi_b} \cos^2\theta\, \diff \psi^2\ ,\label{5met}\\[2mm]
A &=& \frac{\sqrt 3 q}{\rho^2}\,
         \Big(\frac{\Delta_\theta\, \diff t}{\Xi_a\, \Xi_b}
       - \omega\Big) + \constA \, \diff t\ ,\label{gaugepot}
\end{eqnarray}
%%%%%
where
%%%%%
\begin{align}\label{CCLPfunctions}
\nu &= b\sin^2\theta\, \diff \phi + a\cos^2\theta\, \diff \psi\,,\qquad\quad
\omega = a\sin^2\theta\, \frac{\diff \phi}{\Xi_a} +
              b\cos^2\theta\, \frac{\diff \psi}{\Xi_b}\ ,\nn\\
\Delta_r &= \frac{(r^2+a^2)(r^2+b^2)(1+g^2 r^2) + q^2 +2ab q}{r^2} - 2m
\ ,\nn\\
\Delta_\theta &= 1 - a^2 g^2 \cos^2\theta -
b^2 g^2 \sin^2\theta\,,\qquad
\rho^2 = r^2 + a^2 \cos^2\theta + b^2 \sin^2\theta\,,\nn\\
\Xi_a &=1-a^2 g^2\,,\quad \Xi_b = 1-b^2 g^2\ ,\ \ \qquad
f= 2 m \rho^2 - q^2 + 2 a b q g^2 \rho^2\ .
\end{align}
%%%%%
The angular coordinates $\phi,\psi$ are $2\pi$-periodic, while $\theta \in [0,\pi/2]$. %, so that together these parameterize a three-sphere $S^3$. 
 The constant $\constA$ appearing in $A$ parameterizes a gauge choice that needs to be made so that the Euclidean section of the analytically continued solution is globally well-defined (more later).

The solution depends on the four parameters $a, b,m, q$, with  $a^2g^2<1, b^2g^2<1$. These control  four independent conserved charges: the energy $E$, the angular momenta $J_1, J_2\,$ associated with rotations in the $\phi$ and $\psi$ directions, respectively, and the electric charge $Q$. Their expressions are:
\begin{align}\label{CCLPcharges}
E &= \frac{m\pi (2\Xi_a +2\Xi_b - \Xi_a\,\Xi_b) +2\pi qabg^2(\Xi_a+\Xi_b)}{4G\Xi_a^2\,\Xi_b^2}\ ,\qquad Q = \frac{\sqrt 3\pi q}{4G \Xi_a\, \Xi_b}\ ,\nn\\[2mm]
&\ \  J_1 = \frac{\pi[2am + qb(1+a^2 g^2) ]}{4G \Xi_a^2\, \Xi_b}\ ,\qquad
J_2 = \frac{\pi[2bm + qa(1+b^2 g^2) ]}{4G \Xi_b^2\, \Xi_a}\ .
\end{align}

The position $r=r_+$ of the outer event horizon is the largest positive root of the equation $\Delta_r(r)=0$.\footnote{In practice, since solving for $r_+$ as a function of the parameters $a,b,m,q$ is very cumbersome, we rather solve $\Delta_r=0$ for $m$ in terms of $r_+$ and the other parameters. This gives
\be
m = \frac{(r_+^2+a^2)(r_+^2+b^2)(1+ g^2r_+^2) + q^2 +2ab q}{2r_+^2}\nn \ .
\ee
 Hence the solution will be regarded as controlled by $a,b,r_+,q$.\label{foot:solm}}
This directly enters in the expressions for the thermodynamic potentials, namely the inverse Hawking temperature $\beta$, the angular velocities $\Omega_1,\Omega_2$, and the electrostatic potential $\Phi$,
\be\label{temperature_CCLP}
T \equiv\beta^{-1} = \frac{r_+^4[(1+ g^2(2r_+^2 + a^2+b^2)] -(ab + q)^2}{2\pi\,
         r_+\, [(r_+^2+a^2)(r_+^2+b^2) + abq]}\ ,
\ee
\be\label{angular_velocities_CCLP}
\Omega_1 = \frac{a(r_+^2+ b^2)(1+g^2 r_+^2) + b q}{
               (r_+^2+a^2)(r_+^2+b^2)  + ab q}\ ,\qquad
\Omega_2 = \frac{b(r_+^2+ a^2)(1+g^2 r_+^2) + a q}{
               (r_+^2+a^2)(r_+^2+b^2)  + ab q}\ ,
\ee
\be\label{electrostatic_pot_CCLP}
\Phi = \frac{\sqrt{3}\, q \,r_+^2}{(r_+^2 + a^2)(r_+^2 + b^2)+abq}\, .
\ee
Of course, $r_+$ also appears in the expression for the area of the horizon and thus in the Bekenstein-Hawking entropy,
\be\label{entropyCCLP}
{\cal S}= \frac{\rm Area}{4G} =\frac{\pi^2 [(r_+^2 +a^2)(r_+^2 + b^2) +a b q]}{2G\Xi_a \Xi_b r_+}
\ .
\ee

These quantities satisfy the first law of thermodynamics,
\be\label{firstlaw}
\diff E = T \diff {\cal S} + \Omega_1\, \diff J_1+ \Omega_2 \,\diff J_2 + \Phi\, \diff Q\ ,
\ee
as well as the quantum statistical relation
\be\label{QSR_CCLP}
I = \beta E - {\cal S} - \beta \Omega_1J_1 - \beta\Omega_2J_2 - \beta\Phi Q\ ,
\ee
where $I$ is the Euclidean on-shell action~ \cite{Gibbons:1976ue}. The divergences contained in the latter can be renormalized either via the background subtraction method~\cite{Chen:2005zj}, or using holographic renormalization \cite{deHaro:2000vlm,Bianchi:2001kw} and subtracting the on-shell action of the AdS$_5$ vacuum; the two approaches yield the same expression~\cite{Cassani:2019mms}. This reads
\be\label{OnShAction_CCLP}
I = \frac{\pi\beta}{4G\Xi_a\Xi_b}
\Big[m - g^2 (r_+^2 + a^2)(r_+^2 + b^2) -
\frac{q^2 r_+^2}{(r_+^2 + a^2)(r_+^2 + b^2)+abq}\Big]\ .
\ee
In evaluating the action, one should fix the gauge so that the following regularity condition at the horizon is satisfied,
\begin{equation}
V^{\mu}A_{\mu}|_{r=r_{+}}=0\,,
\end{equation}
where $V = \partial_t +\Omega_1\partial_\phi+\Omega_2\partial_\psi$ is the Killing vector generating the horizon.
The condition is satisfied by fixing the parameter $\xi$ in \eqref{gaugepot} as
\be\label{regularxi}
\xi = - \Phi \,,
\ee
implying that the electrostatic potential can be read from the gauge field at the conformal boundary of the solution, $\Phi = -V^\mu A_\mu|_{r\to\infty}$.

Combining \eqref{firstlaw} and \eqref{QSR_CCLP}, one finds that variation of the on-shell action with respect to the chemical potentials gives the charges,
\be\label{charges_from_I}
E = \frac{\partial I}{\partial\beta}\ ,\qquad J_1 = -\frac{1}{\beta}\frac{\partial I}{\partial\Omega_1}\ ,\qquad J_2 = -\frac{1}{\beta}\frac{\partial I}{\partial \Omega_2} \ ,\qquad Q = -\frac{1}{\beta}\frac{\partial I}{\partial\Phi}\ .
\ee
Hence the on-shell action can be regarded as a saddle of the grand-canonical partition function, $I = -\log Z_{\rm grand}$, and is a function of the chemical potentials $\beta,\Omega_1, \Omega_2,\Phi$.
This interpretation is in harmony with the one that regards the action of an asymptotically (locally) AdS solution as a function of the boundary values of the bulk fields (for a Dirichlet variational problem), see e.g.~\cite{Papadimitriou:2005ii}. Indeed after imposing regularity of the Euclidean solution, the quantities $\beta,\Omega_1, \Omega_2,\Phi$ appear in the boundary data (that is, the boundary metric and gauge field together with the global identification of the coordinates), and the action is a function of such boundary values, so that $I=I(\beta,\Omega_1,\Omega_2,\Phi)$. For instance, $\beta$ is given by the length of the Euclidean time circle, which can be read at the boundary; see e.g.~\cite{Cabo-Bizet:2018ehj} for a more detailed discussion. This point of view will be important for us later, when we will need to decide which variables are held fixed while including the higher-derivative corrections.

The solution above is supersymmetric if
\be\label{susyCCLP}
\begin{aligned}
q &= \frac{m}{1+ag+bg}\\
 &= -(a- i r_+)(b- i r_+)(1- i gr_+)\,,
\end{aligned}
\ee
where in the second line we have used the expression of $m$ in terms of $r_+$ and chosen one of the two roots of the resulting quadratic equation for $q$ (choosing the other root would just give expressions where $i$ is replaced by $-i$).
This relation does not automatically imply extremality, namely vanishing of the Hawking temperature. 
 Extremality only follows after further imposing that the Lorentzian solution is well-behaved~\cite{Chong:2005hr}, and is reached by taking 
 \be\label{BPSlimit}
 r_+\to r_*\,,
 \ee 
 with the BPS horizon radius
being\footnote{Following the notation of \cite{Cabo-Bizet:2018ehj}, we call ``BPS'' and denote by a $*$ the quantities arising by imposing both the supersymmetry and the extremality conditions.
}
\be\label{r_BPS}
r_* =  \sqrt{\frac{1}{g}(a+b+abg)} \ .
\ee
In this limit, the chemical potentials take the values:
\be\label{leadingterms_chempot}
\beta \to \infty\ ,\qquad \Omega_1\to \Omega^*_1 = g\ ,\qquad \Omega_2\to \Omega^*_2 = g\ ,\qquad \Phi \to \Phi^* = \sqrt{3}\ .
\ee
It  turns out that $\Omega_1^*,\Omega_2^*,\Phi^*$ are precisely the coefficients appearing in the superalgebra,\footnote{This is related to the fact that in the BPS limit the Killing vector $V$ generating the horizon coincides with the one constructed as a bilinear of the Killing spinor of the solution, see~\cite{Cabo-Bizet:2018ehj} for a comparison.}
\be\label{superalgebra}
\{\mathcal{Q},\overline{\mathcal{Q}}\} \,\propto\, E - g J_1 - g J_2 - \sqrt{3}\, Q\,,
\ee
Comparing the gravitational and SCFT superalgebras, one can see that the electric charge $Q$ is related to the canonically normalized R-charge $Q_R$  appearing in the index~\eqref{unusual_index}, under which the supercharge has eigenvalue 1, via
\be\label{QandRcharge}
Q = \frac{\sqrt{3} g}{2}\,Q_R\,.
\ee

It follows from \eqref{superalgebra} that in supersymmetric solutions the conserved charges satisfy the linear relation
\be\label{susyrel_charges}
E - g J_1 - g J_2 - \sqrt{3}\, Q = 0\,.
\ee
Using this to eliminate $E$, the quantum statistical relation \eqref{QSR_CCLP} and the first law \eqref{firstlaw} restricted to the supersymmetric ensemble become 
\be
I =  - S - \omega_1J_1 - \omega_2J_2 - \frac{2}{\sqrt 3 \,g}\varphi\, Q\ ,
\ee
\be
\diff S + \omega_1\, \diff J_1+ \omega_2\, \diff J_2 +  \frac{2}{\sqrt 3 \,g}\varphi\, \diff Q=0\,,
\ee
where we have introduced the supersymmetric chemical potentials
\begin{equation}\label{susychemicalpotentials}
\begin{aligned}
\omega_1 \,&=\, \beta(\Omega_1-\Omega_1^*) \,=\, \frac{2\pi(ag-1)(b-ir_+)}{2(1+ag+bg)r_+ + ig (r_*^2-3r_+^2)}\,,\\[2mm]
\omega_2 \,&=\, \beta(\Omega_2-\Omega_2^*) \,=\,  \frac{2\pi(bg-1)(a-ir_+)}{2(1+ag+bg)r_+ + ig (r_*^2-3r_+^2)}\ ,\\[2mm]
\varphi \,&=\, \frac{\sqrt{3}g}{2} \beta(\Phi-\Phi^*) \,=\, \frac{3\pi(a-i r_+)(b-ir_+)}{2(1+ag+bg)r_+ + ig (r_*^2-3r_+^2)}\ ,
\end{aligned}
\ee
which measure the departure from the extremal values.
These satisfy the relation
\be\label{linear_relation}
\omega_1 +\omega_2 -2 \varphi =  2\pi i \ .
\ee
One can show that this relation is in fact required to ensure regularity of the Killing spinor in the topology of the Euclidean solution~\cite{Cabo-Bizet:2018ehj}; it also ensures that the combination of the charges that are defined by varying $I$ with respect to $\omega_1,\omega_2$, that is
\be
-\frac{\partial I}{\partial\omega_{1,2}} = J_{1,2} + \frac{1}{\sqrt3\, g}Q\,,
\ee
 is supersymmetric (namely, it commutes with the supercharges when promoted to operator). 
 It further follows that $I$ must be just a function of $\omega_1,\omega_2$. Indeed, using the supersymmetry condition \eqref{susyCCLP}, one can check that the on-shell action \eqref{OnShAction_CCLP} can be expressed simply as~\cite{Cabo-Bizet:2018ehj}
\be\label{Isusy}
I = \frac{2\pi}{27G g^3}\,\frac{\varphi^3}{\omega_1\omega_2}\ ,
\ee
where $\varphi$ can be eliminated via \eqref{linear_relation}.

A feature of assuming \eqref{susyCCLP} without also taking the extremal limit is that both the five-dimensional metric and gauge field are complexified. However, after establishing the supersymmetric thermodynamics above one can take the BPS limit by sending $r_+\to r_*$; in these variables the limit is perfectly smooth and gives back a real solution (when Wick-rotating the Euclidean time back to Lorentzian signature), while the chemical potentials $\omega_1,\omega_2$ remain complex.
The BPS entropy reads
\begin{align}
\mathcal{S}^* &= \frac{\pi^2 (a+b)r_*}{2Gg(1-ag)(1-bg)}
\nn\\[2mm]
&= \frac{\pi}{g} \sqrt{4{ Q^*}^2 -\frac{\pi}{G g} \big(J^*_1+J^*_2\big)}\label{S_BPS}\ ,
\end{align}
where in the second line we have given its expression in terms of the BPS charges \cite{Kim:2006he}.
%
%The BPS values of the charges are:
%\begin{align}\label{BPScharges}
%J^*_1 &=  \frac{\pi(a+b)(2a+b+abg)}{4G g(1-ag)^2(1-bg)}    \ , \qquad
%J^*_2 = \frac{\pi(a+b)(a+2b+abg)}{4G g(1-ag)(1-bg)^2} \ , \nn\\[1mm]
%Q^* & =  \frac{\sqrt{3}\pi (a+b)}{4G g(1-ag)(1-bg)}\ .
%\end{align} 
These charges satisfy the non-linear relation
\be\label{nonlinear_rel_BPScharges}
 \left(\frac{2\sqrt3}{g} Q^*  + \frac{\pi}{2Gg^3} \right)\left( \frac{4}{g^2} {Q^*}^2 -  \frac{\pi}{Gg^3}(J^*_1+J^*_2)\right) \,=\, \left(\frac{2}{\sqrt 3 \,g} Q^*\right)^3 + \frac{2\pi}{Gg^3}J^*_1 J^*_2\ .
\ee

The supersymmetric version of the thermodynamics summarized above is key for matching the index of dual SCFT's as explained in Section~\ref{sec:intro}. The Legendre transform of \eqref{Isusy}, supplemented by a reality condition, yields both \eqref{S_BPS} and \eqref{nonlinear_rel_BPScharges} \cite{Hosseini:2017mds}, \cite{Cabo-Bizet:2018ehj}.

%%%%%%%%%%%%%%%%%%%%%%%%%%%%%%%%%%%%%%%%%%%%%%%%%%%%%%%%%%%%%%%%%%%%%%%%%%%%%%%%%%%%%%%%%%%%
\section{The four-derivative effective action}\label{FourDerAction}

In this section we give the most general four-derivative correction to minimal gauged supergravity which is compatible with its local symmetries, namely diffeomorphism, gauge invariance and supersymmetry. 

We start by supplementing the two-derivative action (\ref{eq:2daction}) with four-derivative corrections, 
\begin{equation}\label{eq:4daction}
S=\frac{1}{16\pi G}\int \diff^5x \, e \left[d_{0}R+12g^2d_{1}-\frac{d_2}{4}F^2-\frac{d_3}{12\sqrt{3}}\epsilon^{\mu\nu\rho\sigma\lambda}F_{\mu\nu}F_{\rho\sigma} A_{\lambda}+\alpha \,{\mathcal L}_{4\partial}\right]\, ,
\end{equation}
where ${\mathcal L}_{4\partial}$ contains all the possible four-derivative terms constructed out of the low-energy degrees of freedom which are compatible with the local symmetries of the two-derivative theory. The parameter $\alpha$ has dimensions of length$^2$ and we assume that $\alpha {\cal R}\ll 1$, where $\cal R$ here denotes  the curvature scale of the solution. This guarantees us that terms with six or more derivatives can be safely neglected as they will be subleading with respect to the four-derivative terms that we have included. The most general four-derivative correction compatible with supersymmetry that we can write down is a linear combination of three independent supersymmetric invariants $\{{\cal L}^{(i)}_{4\partial}\}_{i=1,2,3}$\,,
\begin{equation}\label{eq:L4dSUSY}
{\cal L}_{4\partial}=\lambda_{1} \,{\cal L}^{(1)}_{4\partial}+\lambda_{2} \;{\cal L}^{(2)}_{4\partial}+ \lambda_{3} \,{\cal L}^{(3)}_{4\partial}\, ,
\end{equation}
where $\lambda_{1}$, $\lambda_{2}$ and $\lambda_{3}$ are dimensionless couplings. In addition to the four-derivative corrections, since $\alpha g^2$ is a dimensionless quantity, we can expect possible corrections to the two-derivative terms controlled by this parameter (hence, we are also assuming $\alpha g^2<<1$). This kind of corrections appear through the dimensionless couplings $d_i$, which are of the form
\begin{equation}
d_{i}=1+\alpha g^2 \delta d_{i}\,,
\end{equation}
being $\delta d_{i}$ a linear combination of $\lambda_1$, $\lambda_2$ and $\lambda_3$, as we will see later.

The canonical approach to construct a basis of supersymmetric invariants $\{{\cal L}^{(i)}_{4\partial}\}_{i=1,2,3}$ would be to start with the off-shell formulation of ${\cal N}=2$ gauged supergravity, include the four-derivative invariants \cite{Ozkan:2013nwa} and finally integrate out the auxiliary degrees of freedom.\footnote{We will see that it is precisely the integration of the auxiliary fields what generates corrections to the two-derivative terms.} After this step has been completed, one is free to perform field redefinitions of the form
\begin{equation}\label{eq:fieldredefinition}
g_{\mu\nu}\rightarrow g_{\mu\nu}+\alpha\, \Delta_{\mu\nu}\, , \hspace{1cm} A_{\mu}\rightarrow A_{\mu}+\alpha\, \Delta_{\mu}\, ,
\end{equation}
which allow us to eliminate most of the four-derivative terms, drastically simplifying the form of the action. Advancing results, we will show that the action (\ref{eq:4daction}) can be brought to the following form using field redefinitions,
\begin{equation}\label{eq:4daction1}
\begin{aligned}
&S=\frac{1}{16\pi G}\int \diff^5x \,e \left\{c_{0}R+12c_1g^2-\frac{c_2}{4}F^2-\frac{c_3}{12\sqrt{3}}\epsilon^{\mu\nu\rho\sigma\lambda}F_{\mu\nu}F_{\rho\sigma} A_{\lambda}\right.\\
&\left.+\,\lambda_1\alpha\left[R_{\mu\nu\rho\sigma}R^{\mu\nu\rho\sigma}-\frac{1}{2}R_{\mu\nu\rho\sigma}F^{\mu\nu}F^{\rho\sigma}+\frac{5}{36}\left(F^2\right)^2-\frac{13}{24}F^4-\frac{1}{2\sqrt{3}}\epsilon^{\mu\nu\rho\sigma\lambda}R_{\mu\nu\alpha\beta}R_{\rho\sigma}{}^{\alpha\beta} A_{\lambda}\right]\right\} ,
\end{aligned}
\end{equation}
where $F^4= F_{\mu\nu}F^{\nu\rho}F_{\rho\sigma}F^{\sigma\mu}$. The coefficients in front of the two-derivative terms are given by $c_i=1+\alpha g^2{\delta c}_{i}$, where
\begin{equation}
{\delta c}_{0}=4\lambda_2\, , \hspace{.5cm}\delta c_{1}=-\frac{10\lambda_1}{3}+4\lambda_2 \,, \hspace{.5cm} {\delta c}_{2}=\frac{32\lambda_1}{3}+4\lambda_2 \,, \hspace{.5cm}{\delta c}_{3}=-12\lambda_1+4\lambda_2 \,.
\end{equation}
Note that of the three coefficients in~\eqref{eq:L4dSUSY}, only $\lambda_1,\lambda_2$ appear in~\eqref{eq:4daction1}. In fact, below we will give an argument proving that it is possible to choose the basis of invariants such that the one controlled by $\lambda_3$ (whose explicit form has in fact not been worked out in components in the standard Weyl formulation) yields a vanishing contribution after implementing appropriate field redefinitions.  Alternatively, considering the same basis of four-derivative invariants as in Ref.~\cite{Liu:2022sew}, we obtain
\begin{equation}\label{eq:4daction2}
\begin{aligned}
S\,=&\,\frac{1}{16\pi G}\int \diff^5x \,e \left\{{\tilde c}_0 R+12{\tilde c}_1g^2-\frac{{\tilde c}_2}{4}F^2-\frac{{\tilde c}_3}{12\sqrt{3}}\epsilon^{\mu\nu\rho\sigma\lambda}F_{\mu\nu}F_{\rho\sigma} A_{\lambda}\right.\\[2mm]
&\left.\,+\,\lambda_1 \alpha \left[{\cal X}_{\text{GB}}-\frac{1}{2}C_{\mu\nu\rho\sigma}F^{\mu\nu}F^{\rho\sigma}+\frac{1}{8}F^4-\frac{1}{2\sqrt{3}}\epsilon^{\mu\nu\rho\sigma\lambda}R_{\mu\nu\alpha\beta}R_{\rho\sigma}{}^{\alpha\beta} A_{\lambda}\right]\right\}\, ,
\end{aligned}
\end{equation}
where ${\cal X}_{\text{GB}}=R_{\mu\nu\rho\sigma}R^{\mu\nu\rho\sigma}-4 R_{\mu\nu}R^{\mu\nu}+R^2$ is the Gauss-Bonnet invariant,  $C_{\mu\nu\rho\sigma}=R_{\mu\nu\rho\sigma}-\frac{2}{3}\left(R_{\mu[\rho}g_{\sigma]\nu}+R_{\nu[\sigma}g_{\rho]\mu}\right)+\frac{1}{6}Rg_{\mu[\rho}g_{\sigma]\nu}$ is the Weyl tensor and ${\tilde c}_{i}=1+\alpha g^2 {\delta \tilde c}_{i}$, with
\begin{equation}\label{ctildes}
{\delta \tilde c}_{0}=4\lambda_2\, , \hspace{.5cm}{\delta\tilde c}_{1}=-10\lambda_1+4\lambda_2 \,, \hspace{.5cm} {\delta \tilde c}_{2}=4\lambda_1+4\lambda_2 \,, \hspace{.5cm}{\delta 
\tilde c}_{3}=-12\lambda_1+4\lambda_2 \,.
\end{equation}
We observe that the corrections controlled by $\lambda_2$ are proportional to the two-derivative Lagrangian in both \eqref{eq:4daction1} and \eqref{eq:4daction2}. Thus, they can be simply interpreted as corrections to the Newton's constant $G$, which do not modify the solutions of the two-derivative theory.

We further note that if one ignores boundary conditions, the action \eqref{eq:4daction2} is equivalent (up to an ambiguous sign in the mixed Chern-Simons term) to the action presented in Ref.~\cite{Liu:2022sew}. Indeed we can always rescale the metric and gauge field in a way such that the two-derivative terms can be recast as in the original two-derivative action \eqref{eq:2daction}, up to a redefinition of gauge coupling $g$ and the Newton's constant $G$. However, these field redefinitions are dangerous for our present purposes here as they change the asymptotic structure of the gauge field. A main consequence of this is that the value of the thermodynamical quantities is not left invariant by these field redefinitions.\footnote{For the particular case of the Wald entropy, this aspect was discussed in \cite{Jacobson:1993vj}.} This would explain why the authors of \cite{Liu:2022sew} obtained a non-vanishing Gibbs free energy for the BPS black hole of~\cite{Gutowski:2004ez}. As we will show, this problem does not arise when using either one of \eqref{eq:4daction1}, \eqref{eq:4daction2}. Furthermore, it is  not difficult to check that the field redefinitions one needs to bring the action to these two equivalent forms do not modify the value of the on-shell action, therefore all the thermodynamical quantities remain invariant. This is a strong consistency check of our results and of our effective actions \eqref{eq:4daction1}, \eqref{eq:4daction2}.

In what follows, we give the details of the derivation of our main results of this section. The reader not interested in them can safely ignore the rest of this section. The plan is the following. In Section~\ref{sec:fieldred} we discuss the general effect of field redefinitions, \textit{i.e.}\ without imposing supersymmetry. In Section~\ref{sec:susyinvariants} we consider off-shell ${\cal N}=2$ Poincar\'e supergravity in the standard Weyl formulation including four-derivative invariants and integrate out the auxiliary fields in order to obtain the supersymmetric action at linear order in $\alpha$. Then, we exploit the power of field redefinitions to show that this action can be mapped to either \eqref{eq:4daction1} or \eqref{eq:4daction2}. 
Some related computations have previously appeared in~\cite{Cremonini:2008tw,Baggio:2014hua,Bobev:2021qxx,Liu:2022sew}.

\subsection{Field redefinitions}\label{sec:fieldred}

Field redefinitions of the form~(\ref{eq:fieldredefinition}) induce the following terms into the action
\begin{equation}
S\rightarrow S-\frac{\alpha}{16\pi G}\int \diff^5x\,e\left[\left({\cal E}_{\mu\nu}-\frac{1}{2}g_{\mu\nu}\,{\cal E}\right)\Delta^{\mu\nu}-{\cal E}^{\mu}\Delta_{\mu}\right]+ {\cal O}(\alpha^2)\,,
\end{equation}
where ${\cal E}_{\mu\nu},\, {\cal E}^\mu$ are the two-derivative equations of motion \eqref{EoM2der}, and ${\cal E}=g^{\mu\nu}{\cal E}_{\mu\nu}$. The induced terms can be used to eliminate all four-derivative terms of the form ${\cal E}_{\mu\nu}K^{\mu\nu}$ and ${\cal E}_{\mu}L^{\mu}$ by choosing $\Delta_{\mu\nu}$ and $\Delta_{\mu}$ appropriately, which amounts to take
\begin{equation}
\Delta_{\mu\nu}=K_{\mu\nu}-\frac{1}{3}\, g_{\mu\nu} K\,, \hspace{1cm}\Delta_{\mu}=-L_{\mu}\,,
\end{equation}
where $K=g^{\mu\nu}K_{\mu\nu}$.
\noindent
In practice, this means that we can use the two-derivative equations of motion in the piece of the action of order $\alpha$ since all the terms that vanish on-shell can be eliminated with a field redefinition. Let us consider as an instance a term of the form $R_{\mu\nu}K^{\mu\nu}$. This term can be replaced in the action by 
\begin{equation}\label{eq:masterrule1}
R_{\mu\nu}K^{\mu\nu}\rightarrow \frac{1}{2}F_{\mu\rho}F_{\nu}{}^{\rho}K^{\mu\nu}-\left(4g^2+\frac{1}{12}F^2\right) K \, ,
\end{equation}
since the difference between the left and right-hand sides is ${\cal E}_{\mu\nu}K^{\mu\nu}$. For the same reason, a term $\nabla_{\rho}F^{\rho \mu}L_{\mu}$ in the action can be replaced by 
\begin{equation}\label{eq:masterrule2}
\nabla_{\rho}F^{\rho \mu}L_{\mu}\rightarrow \frac{1}{4\sqrt{3}}\epsilon^{\mu\nu\rho\sigma\lambda}F_{\nu\rho}F_{\sigma\lambda}L_{\mu}\, .
\end{equation}
In order to implement field redefinitions in a systematic fashion, it is convenient to derive first a set of replacement rules for all the terms that can appear in (\ref{eq:4daction}) modulo total derivatives and the use of Bianchi and Ricci identities. Using (\ref{eq:masterrule1}) and (\ref{eq:masterrule2}), one can derive the following rules for particular choices of the tensors $K_{\mu\nu}$ and $L_{\mu}$: 
\begin{equation}
\begin{aligned}
R^{\mu\nu}R_{\mu\nu}&\ \rightarrow\  \frac{1}{4}F^4-\frac{7}{144}\left(F^2\right)^2-4g^2 R-\frac{1}{3}g^2 F^2\, ,\\
R^2&\ \rightarrow\  \frac{1}{144}\left(F^2\right)^2-20g^2 R-\frac{5}{3}g^2 F^2\, ,\\
RF^2&\ \rightarrow\  \frac{1}{2}\left(F^2\right)^2-20g^2 F^2\, ,\\
R^{\mu\nu}F_{\mu\rho}F_{\nu}{}^{\rho}&\ \rightarrow\  \frac{1}{2}F^4-\frac{1}{12}\left(F^2\right)^2-4g^2 F^2\, ,\\
\nabla_{\rho}F^{\rho\mu}\nabla_{\sigma}F^{\sigma}{}_{\mu} &\ \rightarrow\  \frac{1}{3}F^4-\frac{1}{6}\left(F^2\right)^2\, ,\\
\epsilon^{\mu\nu\rho\sigma\lambda}F_{\nu\rho}F_{\sigma\lambda}\nabla_{\delta}F^{\delta}{}_{\mu} &\ \rightarrow\  \frac{4}{\sqrt{3}}F^4-\frac{2}{\sqrt{3}}\left(F^2\right)^2\, .
\end{aligned}
\end{equation}
Using Bianchi identities, integrations by parts and Ricci identities\footnote{Our conventions for the Riemann tensor are such that $[\nabla_{\mu}, \nabla_{\nu}]\xi^{\sigma}=-R_{\mu\nu\alpha}{}^{\sigma}\xi^{\alpha}$.}, one can further derive the following rules 
\begin{equation}
\begin{aligned}
F^{\nu \rho}[\nabla_{\mu}, \nabla_{\nu}]F^{\mu}{}_{\rho}&\ \rightarrow\ -\frac{1}{2}R_{\mu\nu\rho\sigma}F^{\mu\nu}F^{\rho\sigma}+\frac{1}{2}F^4-\frac{1}{12}\left(F^2\right)^2-4g^2 F^2\,,\\
\nabla_{\mu}F_{\nu\rho}\nabla^{\mu}F^{\nu\rho}&\ \rightarrow\  R_{\mu\nu\rho\sigma}F^{\mu\nu}F^{\rho\sigma}-\frac{1}{3}F^4-\frac{1}{6}\left(F^2\right)^2+8g^2 F^2\,,\\
\epsilon^{\mu\nu\rho\sigma\lambda}F_{\mu\nu}F_{\rho}{}^{\alpha}\nabla_{\sigma}F_{\lambda\alpha} &\ \rightarrow\  -\frac{1}{\sqrt{3}}F^4+\frac{1}{2\sqrt{3}}\left(F^2\right)^2\,.
\end{aligned}
\end{equation}
This set of rules is enough since the four-derivative part of the action \eqref{eq:4daction} can always be written (up to total derivatives) as a linear combination of the following terms
\begin{equation}\label{eq:genexpL4d}
\begin{aligned}
{\mathcal L}_{4\partial}=\,&{a}_{1}\,R_{\mu\nu\rho\sigma}R^{\mu\nu\rho\sigma}+{a}_{2}\,R_{\mu\nu\rho\sigma}F^{\mu\nu}F^{\rho\sigma}+a_3 \,\left(F^2\right)^2+a_4\,F^4+a_5\, \epsilon^{\mu\nu\rho\sigma\lambda}R_{\mu\nu\alpha\beta}R_{\rho\sigma}{}^{\alpha\beta} A_{\lambda}\\
&+b_1\,R_{\mu\nu}R^{\mu\nu}+b_2\,R^2+b_3 \,RF^2+b_4\, R^{\mu\nu}F_{\mu\rho}F_{\nu}{}^{\rho}+b_5\,\nabla_{\mu}F_{\nu\rho}\nabla^{\mu}F^{\nu\rho}+b_6\,\nabla_{\rho}F^{\rho\mu}\nabla_{\sigma}F^{\sigma}{}_{\mu}\\
&+b_7\, F^{\nu \rho}[\nabla_{\mu}, \nabla_{\nu}]F^{\mu}{}_{\rho}+b_8 \,\epsilon^{\mu\nu\rho\sigma\lambda}F_{\nu\rho}F_{\sigma\lambda}\nabla_{\delta}F^{\delta}{}_{\mu}+b_9\,\epsilon^{\mu\nu\rho\sigma\lambda}F_{\mu\nu}F_{\rho}{}^{\alpha}\nabla_{\sigma}F_{\lambda\alpha}\, .
\end{aligned}
\end{equation}
Making use of the replacement rules that we have just derived, one can show that there is a field redefinition such that the action (\ref{eq:4daction}) ---with ${\mathcal L}_{4\partial}$ given by (\ref{eq:genexpL4d})--- in terms of the new fields reads
\begin{equation}\label{eq:new4daction}
S=\frac{1}{16\pi G}\int \diff^5x \, e \left[c_{0}R+12c_{1} g^2-\frac{c_{2}}{4}F^2-\frac{c_{3}}{12\sqrt{3}}\epsilon^{\mu\nu\rho\sigma\lambda}F_{\mu\nu}F_{\rho\sigma} A_{\lambda}+\alpha \,{\mathcal L}'_{4\partial}\right]\, ,
\end{equation}
where 
\begin{equation}\label{eq:L4dprime}
{\mathcal L}'_{4\partial}=\,{a}'_{1}\,R_{\mu\nu\rho\sigma}R^{\mu\nu\rho\sigma}+{a}'_{2}\,R_{\mu\nu\rho\sigma}F^{\mu\nu}F^{\rho\sigma}+a'_3 \,\left(F^2\right)^2+a'_4\,F^4+a'_5\, \epsilon^{\mu\nu\rho\sigma\lambda}R_{\mu\nu\alpha\beta}R_{\rho\sigma}{}^{\alpha\beta} A_{\lambda}\, ,
\end{equation}
and 
\begin{equation}\label{eq:rulesa'}
\begin{aligned}
a'_1=\,&a_1\,,\\
a'_2=\,&a_2+b_5-\frac{b_7}{2}\,,\\
a'_3=\,&a_3-\frac{7b_1}{144}+\frac{b_2}{144}+\frac{b_3}{12}-\frac{b_4}{12}-\frac{b_5}{6}-\frac{b_6}{6}-\frac{b_7}{12}-\frac{2b_8}{\sqrt{3}}+\frac{b_9}{2\sqrt{3}}\,,\\
a'_4=\,&a_4+\frac{b_1}{4}+\frac{b_4}{2}-\frac{b_5}{3}+\frac{b_6}{3}+\frac{b_7}{2}+\frac{4b_8}{\sqrt{3}}-\frac{b_9}{\sqrt{3}}\,,\\
a'_5=\,&a_5\, .
\end{aligned}
\end{equation}
Finally, the $c_{i}=1+\alpha g^2 \delta c_{i}$ coefficients are given by
\begin{equation}\label{eq:rulesc}
\begin{aligned}
{\delta c}_{1}=\,&{\delta d}_{1}+\frac{5}{3} \left(4b_{1}+20 b_2+{\delta c}_{0}-\delta d_0\right)\,,\\
{\delta c}_{2}=\,&{\delta d}_{2}+\frac{8b_1}{3}+\frac{40b_2}{3}+80b_3+16b_4-32b_5+16b_7+\frac{\delta c_0}{3}-\frac{\delta d_0}{3}\,,\\
{\delta c}_{3}=\,&{\delta d}_{3}\,.
\end{aligned}
\end{equation}
When using the basis of four-derivative invariants in \eqref{eq:4daction2}, the resulting action is 
\begin{equation}\label{eq:new4daction2}
S=\frac{1}{16\pi G}\int \diff^5x \, e \left[{\tilde c}_{0}R+12{\tilde c}_{1} g^2-\frac{{\tilde c}_{2}}{4}F^2-\frac{{\tilde c}_{3}}{12\sqrt{3}}\epsilon^{\mu\nu\rho\sigma\lambda}F_{\mu\nu}F_{\rho\sigma} A_{\lambda}+\alpha \,{\mathcal L}''_{4\partial}\right]\, ,
\end{equation}
where 
\begin{equation}\label{eq:L4dprime}
{\mathcal L}''_{4\partial}=\,{a}''_{1}\,{\cal X}_{\rm {GB}}+{a}''_{2}\,C_{\mu\nu\rho\sigma}F^{\mu\nu}F^{\rho\sigma}+a''_3 \,\left(F^2\right)^2+a''_4\,F^4+a''_5\, \epsilon^{\mu\nu\rho\sigma\lambda}R_{\mu\nu\alpha\beta}R_{\rho\sigma}{}^{\alpha\beta} A_{\lambda}\, ,
\end{equation}
with 
\begin{equation}\label{eq:rulesa''}
\begin{aligned}
a''_1=\,&a_1\,,\\
a''_2=\,&a_2+b_5-\frac{b_7}{2}\,,\\
a''_3=&\frac{1}{144} \!\left(-29 a_1-18 a_2+144 a_3-7 b_1+b_2+12 b_3-12 b_4-42 b_5-24 b_6-3 b_7-96 \sqrt{3} b_8+24 \sqrt{3} b_9\right),\\
a''_4=\,&\frac{1}{12} \left(12 a_1+8 a_2+12 a_4+3 b_1+6 b_4+4 b_5+4 b_6+2 b_7+16 \sqrt{3} b_8-4 \sqrt{3} b_9\right)\,,\\
a''_5=\,&a_5\, .
\end{aligned}
\end{equation}
and 
\begin{equation}\label{eq:rulesctilde}
\begin{aligned}
{\delta {\tilde c}}_{1}=\,&\delta d_{1}-\frac{5}{3} \left(4 a_1-4 b_1-20 b_2-{\delta \tilde c}_0+\delta d_{0}\right)\,,\\
{\delta {\tilde c}}_{2}=\,& \delta d_2+\frac{1}{3} \left(-8 a_1+24 a_2+8 b_1+40 b_2+240 b_3+48 b_4-72 b_5+36 b_7+{\delta \tilde c}_0-\delta d_0\right)\,,\\
{\delta {\tilde c}}_{3}=\,&{\delta d}_{3}\,.
\end{aligned}
\end{equation}
Let us note that the coefficients in front of the Ricci scalar, controlled by ${\delta c}_{0}$ and ${\delta \tilde c}_{0}$, are arbitrary parameters that we can fix at will. This corresponds to adding a term $\sim \alpha g^2{\cal E}$ to the action through a perturbative field redefinition consisting of a constant rescaling of the metric.

We emphasize that this applies in a general context, since we have not yet imposed supersymmetry of the action. This is what we will do next.

%%%%%%%%%%%%%%%%%%%%%%%%%%%%%%%%%%%%%%%%%%%%%%%%%%%%%%%%%%%%%%%%%%%%%%%%%%%%%%%%%%%%%%%%%%%%%%%%%%%%%%%%%
\subsection{Off-shell supersymmetric invariants}
\label{sec:susyinvariants}

We consider the formulation of off-shell ${\cal N}=2$ Poincar\'e supergravity based on the standard Weyl multiplet~\cite{Bergshoeff:2004kh}. Out of the three independent supersymmetric invariants, only two are explicitly known in components in this formulation. These correspond to the supersymmetric completion of the Weyl squared term $C^2$ \cite{Hanaki:2006pj} and of the Ricci-scalar squared term $R^2$ \cite{Ozkan:2013nwa}.\footnote{The third one was given in superspace in~\cite{Butter:2014xxa} and corresponds to the supersymmetric completion of the Ricci tensor squared term.} (They can also be seen as the supersymmetrizations  of the $\epsilon^{\mu\nu\rho\sigma\lambda}R_{\mu\nu\alpha\beta}R_{\rho\sigma}{}^{\alpha\beta} A_{\lambda}$ and $\epsilon^{\mu\nu\rho\sigma\lambda}F_{\mu\nu}F_{\rho\sigma}A_{\lambda}$ Chern-Simons terms). We thus consider a linear combination of these two invariants (setting e.g. $\lambda_3=0$): 
\begin{equation}\label{eq:L4dWeyl2R2}
{\cal L}_{4\partial}=\lambda_1 \,{\mathcal L}_{C^2}+\lambda_2\, {\mathcal L}_{R^2}\, .
\end{equation}
As already mentioned, at the end of this section we will give an argument proving that in doing so there is in fact no loss of generality.

\subsubsection{Two-derivative off-shell theory}

Our starting point is the action for off-shell $\mathcal{N}=2$ Poincar\'e gauged supergravity in five dimensions as given in \cite{Ozkan:2013nwa},\footnote{Our normalization of the gauge coupling constant differs from the one used in \cite{Ozkan:2013nwa}. The relation between the two is  $g^{\rm{there}}=\sqrt{\frac{2}{3}}\, g^{\rm{here}}$.} 
\begin{equation}\label{eq:offshell2daction}
\begin{aligned}
S_{\rm{off\text{-}shell}}^{(0)}=&\frac{1}{16\pi G}\int \diff^5x \,e\left[\frac{1}{4}\left({\cal C}+3\right)R+\frac{1}{4}{\cal C}'' F^2+\frac{16}{3}\left(13{\cal C}-1\right)T_{\mu\nu}T^{\mu\nu}-8{\cal C}' F_{\mu\nu}T^{\mu\nu}\right.\\
&+8\left({\cal C}-1\right)D+2{\tilde V}_{\mu}{}^{ij}{\tilde V}^{\mu}{}_{ij}-2\sqrt{2}V_{\mu} P^{\mu}-2 P_{\mu}P^{\mu}-2N^2+\frac{1}{2}{\cal C}'' (\partial \rho)^2-{\cal C}'' Y^{ij}Y_{ij}\\
&\left.+\frac{1}{24}{\cal C}'''\epsilon^{\mu\nu\rho\sigma\lambda}F_{\mu\nu}F_{\rho\sigma} A_{\lambda}-2\sqrt{3}\,g Y_{ij}\delta^{ij}-2\sqrt{6}\,g A_{\mu}P^{\mu}-2\sqrt{6}\,g \rho N \right]\, ,
\end{aligned}
\end{equation}
where $D,\rho, T_{\mu\nu}=T_{[\mu\nu]},N,Y_{ij},P_\mu,V_\mu,{\tilde V}_{\mu}{}^{ij}$ are auxiliary fields that we are going to integrate out (here $ij$ denotes SU(2) triplets).  ${\tilde V}_{\mu}{}^{ij}$ and $V_{\mu}$ correspond, respectively, to the traceless and trace parts of the SU$(2)$ vector $V_{\mu}{}^{ij}$ that gauges the SU$(2)$ R-symmetry in the parent superconformal theory, namely 
\begin{equation}
V_{\mu}{}^{ij}={\tilde V}_{\mu}{}^{ij}+\frac{1}{2}\delta^{ij}V_{\mu}\, ,
\end{equation}
where $V_{\mu}=V_{\mu}{}^{ij}\delta_{ij}$. Finally, the cubic polynomial $\mathcal{C}$ is given by ${\cal C}(\rho)\propto \rho^3$. We refer to \cite{Ozkan:2013nwa} for additional details on this theory.

We now solve the auxiliary field equations of motion and then plug the solution back into the action. Let us first note that the scalar $\rho$ is fixed to an arbitrary constant by the equation of motion of $D$, which is satisfied by
\begin{equation}
{\cal C}=1\, , 
\end{equation}
and which further implies 
\begin{equation}
{\cal C}'=\frac{3}{\rho}\, , \hspace{1cm}{\cal C}''=\frac{6}{\rho^2}\, , \hspace{1cm} {\cal C}'''=\frac{6}{\rho^3}\,.
\end{equation}
The remaining equations of motion are solved by 
\begin{equation}\label{eq:auxfields}
\begin{aligned}
{\tilde V}_{\mu}^{ij}=\,&0\, , \hspace{1cm} V_{\mu}=\,-\sqrt{3}g A_{\mu}\, , \\
P_{\mu}=\,\,&0\, ,\\
N=\,&-\frac{\sqrt{3} g\rho}{\sqrt{2}}\, ,\\
Y_{ij}=&-\frac{g\rho^2}{2\sqrt{3}}\, \delta_{ij}\, ,\\
T_{\mu\nu}=\,&\frac{3}{16\rho}F_{\mu\nu}\, ,\\
D=\,&-\frac{1}{32}\left(R-\frac{1}{4\rho^2}F^2+\frac{20}{3}g^2 \rho^2\right)\, .
\end{aligned}
\end{equation}
\noindent
Inserting these expressions back into the action, we find 
\begin{equation}
S^{(0)}=\frac{1}{16\pi G}\int \diff^5x \, e \left[R-\frac{3}{4\rho^2}F^2+4g^2\rho^2 +\frac{1}{4\rho^3}\epsilon^{\mu\nu\rho\sigma\lambda}F_{\mu\nu}F_{\rho\sigma} A_{\lambda}\right]\, ,
\end{equation}
which is nothing but the bosonic action of minimal gauged supergravity. The normalization for the gauge field $A_{\mu}$ employed in \eqref{eq:2daction} is recovered once $\rho$ is fixed to the  value
\begin{equation}
\rho=-\sqrt{3}\, .
\end{equation}
Using this value in \eqref{eq:auxfields}, we get
\begin{equation}\label{eq:auxfields2}
\begin{aligned}
{\tilde V}_{\mu}^{ij}=\,&0\, , \hspace{1cm} V_{\mu}=\,-\sqrt{3}\,g A_{\mu}\, , \\
P_{\mu}=\,\,&0\, ,\\
N=\,&\frac{3\, g}{\sqrt{2}}\, ,\\
Y_{ij}=&-\frac{\sqrt{3}\, g}{2}\, \delta_{ij} \, ,\\
T_{\mu\nu}=\,&-\frac{\sqrt{3}}{16}F_{\mu\nu}\, ,\\
D=\,&-\frac{1}{32}\left(R-\frac{1}{12}F^2+20g^2 \right)=-\frac{1}{32}{\cal E}\, .
\end{aligned}
\end{equation}

\subsubsection{Four-derivative corrections}

Let us now consider the following action, 
\begin{equation}
S_{\rm{off\text{-}shell}}=S_{\rm{off\text{-}shell}}^{(0)}+\frac{\alpha}{16\pi G}\int \diff^5x\,e \left[\lambda_1\, {\cal L}^{\rm{off\text{-}shell}}_{C^2}+\lambda_{2}\,{\cal L}^{\rm{off\text{-}shell}}_{R^2}\right]\, ,
\end{equation}
where $S_{\rm{off\text{-}shell}}^{(0)}$ is again given by \eqref{eq:offshell2daction} and ${\cal L}^{\rm{off\text{-}shell}}_{C^2}$ and ${\cal L}^{\rm{off\text{-}shell}}_{R^2}$ are the off-shell four-derivative invariants constructed in \cite{Hanaki:2006pj, Ozkan:2013nwa}. In order to obtain the on-shell action, we repeat the procedure followed in the two-derivative case. It is important to note, however, that in order to obtain the on-shell action at first order in $\alpha$, we just need the solution to the two-derivative equations of motion of the auxiliary fields, eq.~\eqref{eq:auxfields2}. This is due to the fact that all the terms that involve the corrections to the auxiliary fields are proportional to the two-derivative equations of motion of the latter, hence they become effectively of order ${\cal O}(\alpha^2)$. Therefore, all we have to do in order to obtain the on-shell action is to substitute the values of the auxiliary fields in the four-derivative invariants. 

Let us first consider the Weyl squared invariant, which is given for instance in Eq.~(7.1) of \cite{Ozkan:2013nwa}.\footnote{We have absorbed an overall factor of $-\frac{\sqrt{3}}{8}$ in $\lambda_{1}$ for convenience.} Substituting the value of the auxiliary fields, integrating by parts and making use of the Bianchi identities, we obtain
\begin{equation}\label{eq:Weyl2}
\begin{aligned}
&{\cal L}^{\rm{off\text{-}shell}}_{C^2}|_{\eqref{eq:auxfields2}}=\,\,R_{\mu\nu\rho\sigma}R^{\mu\nu\rho\sigma}+\tfrac{1}{2}\,R_{\mu\nu\rho\sigma}F^{\mu\nu}F^{\rho\sigma}-\tfrac{5}{32} \,\left(F^2\right)^2+\tfrac{5}{8}\,F^4\\
&-\tfrac{1}{2\sqrt{3}}\, \epsilon^{\mu\nu\rho\sigma\lambda}R_{\mu\nu\alpha\beta}R_{\rho\sigma}{}^{\alpha\beta} A_{\lambda}
-\tfrac{4}{3}\,R_{\mu\nu}R^{\mu\nu}+\tfrac{1}{6}\,R^2+\tfrac{5}{12} \,RF^2-\tfrac{10}{3}\, R^{\mu\nu}F_{\mu\rho}F_{\nu}{}^{\rho}-\nabla_{\mu}F_{\nu\rho}\nabla^{\mu}F^{\nu\rho}\\
&+4\,\nabla_{\rho}F^{\rho\mu}\nabla_{\sigma}F^{\sigma}{}_{\mu}-\tfrac{\sqrt{3}}{4} \,\epsilon^{\mu\nu\rho\sigma\lambda}F_{\nu\rho}F_{\sigma\lambda}\nabla_{\delta}F^{\delta}{}_{\mu}-\tfrac{1}{2\sqrt{3}}\,\epsilon^{\mu\nu\rho\sigma\lambda}F_{\mu\nu}F_{\rho}{}^{\alpha}\nabla_{\sigma}F_{\lambda\alpha}-\tfrac{1}{6}{\cal E} F^2+\tfrac{1}{6}{\cal E}^2\\
&+\tfrac{g^2}{\sqrt{3}}\epsilon^{\mu\nu\rho\sigma\lambda}F_{\mu\nu}F_{\rho\sigma} A_{\lambda}\, .
\end{aligned}
\end{equation}
We observe that the elimination of the auxiliary fields in ${\cal L}^{\rm{off\text{-}shell}}_{C^2}$ gives rise not only to four-derivative corrections but also to ${\cal O}(\alpha g^2)$ (two-derivative) corrections. 

In turn, the evaluation of the $R^2$ term ---which is given in eq.~(7.4) of \cite{Ozkan:2013nwa}--- yields\footnote{Again, there is an overall factor of $\frac{9\sqrt{3}}{4}$ which we have absorbed in $\lambda_2$ for convenience.}
\begin{equation}\label{eq:R2}
{\cal L}^{\rm{off\text{-}shell}}_{R^2}|_{\eqref{eq:auxfields2}}=\,g^2\left(R-12g^2-\tfrac{3}{4} F^2-\tfrac{\sqrt{3}}{9}\epsilon^{\mu\nu\rho\sigma\lambda}F_{\mu\nu}F_{\rho\sigma}A_{\lambda}+\tfrac{8}{3}{\cal E}\right) -\tfrac{1}{9}{\cal E}^2\, .
\end{equation}
As we can see, the contribution of this invariant to the four-derivative terms is trivial, as all of the four-derivative terms appear inside ${\cal E}^{2}$, which could have been dropped since it does not contribute to first order in $\alpha$. This means that, as we are going to discuss further, the contribution of this invariant boils down to a change in the coefficients of the two-derivative terms. 

At this stage we are ready to extract the coefficients $a_{i}$ and $b_{i}$ in \eqref{eq:genexpL4d} from the four-derivative terms in \eqref{eq:Weyl2}. We obtain the following values\footnote{We ignore the contributions from ${\cal E}F^2$ and ${\cal E}^2$ since they can be directly eliminated with a field redefinition without changing the rest of the action.}
\begin{equation}
\begin{aligned}
a_1=&\,\lambda_1\,, \hspace{0.5cm}a_2=\frac{\lambda_1}{2}\, ,\hspace{0.5cm}a_3=-\frac{5\lambda_1}{32}\, ,\hspace{0.5cm}a_4=\frac{5\lambda_1}{8}\,, \hspace{0.5cm}a_5=-\frac{\lambda_1}{2\sqrt{3}}\, ,\\[1mm]
b_1=&-\frac{4\lambda_1}{3}\,, \hspace{0.5cm}b_2=\frac{1}{6}\lambda_1\, ,\hspace{0.5cm}b_3=\frac{5\lambda_1}{12}\, ,\hspace{0.5cm}b_4=-\frac{10\lambda_1}{3}\,, \hspace{0.5cm}b_5=-\lambda_1\, ,\\[1mm]
b_6=&\,4\lambda_1\,, \hspace{0.5cm}b_7=0\, ,\hspace{0.5cm}b_8=-\frac{\sqrt{3}\lambda_1}{4}\, ,\hspace{0.5cm}b_9=-\frac{\lambda_1}{2\sqrt{3}}\,.
\end{aligned}
\end{equation}
In turn, from the coefficients in front of the two-derivative terms in \eqref{eq:Weyl2} and \eqref{eq:R2}, we  can read the $\delta d_{i}$ coefficients, getting
\begin{equation}
\delta d_0=\lambda_{2}\,, \hspace{0.5cm} \delta d_1=-\lambda_{2}\,,\hspace{0.5cm}\delta d_3=3\lambda_{2}\,, \hspace{0.5cm}\delta d_4=-12\lambda_1+4\lambda_{2}\,.
\end{equation}
Therefore, we are ready to use \eqref{eq:rulesa'}, \eqref{eq:rulesc}, \eqref{eq:rulesa''} and \eqref{eq:rulesctilde} in order to obtain 
\begin{equation}
\begin{aligned}
a'_{1}=&\lambda_1\, , \hspace{0.5cm} a'_{2}=-\frac{\lambda_1}{2}\, , \hspace{0.5cm}a'_{3}=\frac{5\lambda_1}{36}\, , \hspace{0.5cm} a'_{4}=-\frac{13\lambda_1}{64}\, ,\hspace{0.5cm} a'_5=-\frac{\lambda_1}{2\sqrt{3}}\, ,\\[1mm]
a''_{1}=&\lambda_1\, , \hspace{0.5cm} a''_{2}=-\frac{\lambda_1}{2}\, , \hspace{0.5cm}a''_{3}=0\, , \hspace{0.5cm} a''_{4}=\frac{\lambda_1}{8}\, ,\hspace{0.5cm} a''_5=-\frac{\lambda_1}{2\sqrt{3}}\, ,
\end{aligned}
\end{equation}
and, choosing ${\delta c}_0={\delta \tilde c}_0=4\lambda_2$,
\begin{equation}
\begin{aligned}
\delta c_{1}=&-\frac{10\lambda_1}{3}+4\lambda_2 \,, \hspace{1cm} {\delta c}_{2}=\frac{32\lambda_1}{3}+4\lambda_2 \,, \hspace{1cm}{\delta c}_{3}=-12\lambda_1+4\lambda_2 \,,\\[1mm]
{\delta\tilde c}_{1}=&-10 \lambda_1+4\lambda_2 \,, \hspace{1cm} {\delta\tilde c}_{2}=4\lambda_1+4\lambda_2 \,, \hspace{1cm}{\delta\tilde c}_{3}={\delta c}_3 \,,
\end{aligned}
\end{equation}
as anticipated in \eqref{eq:4daction} and \eqref{eq:4daction2}.

So far we have checked that the corrected action can be written as we anticipated in \eqref{eq:4daction2} when the third supersymmetric invariant is not taken into account.\footnote{What we are going to say equally applies to the alternative form~\eqref{eq:4daction1} of the action.} However, we have not explained yet how we have come to the conclusion that this is in fact the most general action compatible with supersymmetry. 
This is not obvious given that \eqref{eq:4daction2} contains two independent parameters $\lambda_{1, 2}$ while the most general four-derivative Lagrangian compatible with supersymmetry \eqref{eq:L4dSUSY} contains three. Let us explain it. On general grounds, the third invariant will give us two types of corrections with four and two derivatives, as we have seen. The four-derivative corrections can always be reduced upon the use of field redefinitions to the same combination as in \eqref{eq:4daction2}, up to a shift in $\lambda_{1}$. As for the two-derivative terms, which are encoded in the ${\tilde c}_{i}$ coefficients, a naive expectation would be that supersymmetry only fixes one of these coefficients in terms of the other two, therefore the final action would depend on three parameters which can always be traced back to the original $\lambda_{1}$, $\lambda_{2}$ and $\lambda_{3}$ in \eqref{eq:L4dSUSY}. However, we have checked that actually supersymmetry fixes two parameters in terms of the third one, say $\tilde{c}_{3}$, hence the final action only depends on two paramaters, which one can always choose as in \eqref{eq:4daction2}. The way in which this has been checked is by imposing the vanishing of the Gibbs free energy for the black hole of~\cite{Gutowski:2004ez}.\footnote{When doing so we are assuming that the corrected non-extremal black hole admits a BPS limit in presence of the third invariant. } Further details are given at the end of Section~\ref{sec:BPSlimit}.

%One last subtlety that we would like to comment on is that we expect that field redefinitions, although very powerful and convenient for certain purposes, change the form of the supersymmetry transformations. This aspect certainly deserves further study.

%%%%%%%%%%%%%%%%%%%%%%%%%%%%%%%%%%%%%%%%%%%%%%%%%%%%%%%%%%%%%%%%%%%%%%%%%%%%%%%%%%%%%%
\section{Matching the supersymmetric black hole action with the dual index}\label{sec:onshellact}

In this section we compute the on-shell action for the black hole of \cite{Chong:2005hr} at order $\alpha$. Then we impose supersymmetry and match the dual index at the corresponding order in the large-$N$ expansion. 

\subsection{The two-derivative solution is enough for evaluating the action}

When setting up the computation of the on-shell action at linear order in $\alpha$, there are two crucial points that should be kept in mind. The first is that since we are working in the grand-canonical ensemble, the inverse temperature $\beta$ and the chemical potentials, $\Phi$, $\Omega_{1}$, $\Omega_{2}$, must be held fixed to their zeroth-order values given in Sec.~\ref{TwoDerReview}. Another way to say this is that we hold fixed the values of the supergravity fields at the boundary of the asymptotically AdS solution, so as to maintain the original Dirichlet variational problem. On the other hand, the action, the entropy and the conserved charges are allowed to receive corrections.

The second point is that the corrections to the bulk metric and gauge field are not needed in order to compute the $\Oa$ corrections to the thermodynamics, as recently argued in \cite{Reall:2019sah}. It is worth revisiting the argument of \cite{Reall:2019sah}, adapting it to the case at hands.

At zeroth-order in $\alpha$, the renormalized Euclidean action has three contributions:
\begin{equation}
I^{(0)}=I^{(0)}_{\text{bulk}}+I^{(0)}_{\text{GH}}+I^{(0)}_{\text{count}}\, .
\end{equation}
The first is the bulk contribution, which in our case is given by 
\begin{equation}
I^{(0)}_{\text{bulk}}=-\frac{1}{16\pi G}\int_{\cal M} \diff^5x \, e\, \left(R+12 g^2-\frac{1}{4}F^2-\frac{1}{12\sqrt{3}}\epsilon^{\mu\nu\rho\sigma\lambda}F_{\mu\nu}F_{\rho\sigma} A_{\lambda}\right)\, .
\end{equation}
The second is the Gibbons-Hawking boundary term that renders the Dirichlet variational problem for the metric well posed, 
\begin{equation}
I^{(0)}_{\text{GH}}=-\frac{1}{8\pi G}\int_{\partial {\cal M}}\diff^4x\sqrt{h}\, {K}\, ,
\end{equation}
where $h$ is the determinant of $h_{\mu\nu}=g_{\mu\nu}-n_{\mu}n_{\nu}$, the induced metric at $\partial {\cal M}$, and $K=h^{\mu\nu}K_{\mu\nu}$ is the trace of the extrinsic curvature $K_{\mu\nu}=\nabla_{(\mu}n_{\nu)}$, being $n_{\mu}$ is the unit normal to the boundary. Finally, $I^{(0)}_{\text{count}}$ denotes the boundary counterterms needed to remove the divergences due to the non-compactness of the space,
\begin{equation}
I^{(0)}_{\text{count}}=\frac{1}{8\pi G}\int_{\partial {\cal M}}\, \diff^4x \sqrt{h} \, \left(3g+\frac{1}{4g}\,{\cal R}\right)\, ,
\end{equation}
where ${\cal R}$ is the Ricci scalar of the induced metric $h_{ij}$. Taking now into account the four-derivative corrections, the full Euclidean action reads
\begin{equation}
I=I^{(0)}+\alpha \,I^{(1)}\,,
\end{equation}
where $I^{(1)}$ will be again the sum of bulk and boundary contributions, as we will discuss momentarily. Let us now assume we have a solution to the corrected equations of motion. This must be of the form,
\begin{equation}\label{eq:perturbation}
g_{\mu\nu}=g^{(0)}_{\mu\nu}+\alpha \,g^{(1)}_{\mu\nu}\, , \hspace{1cm} A_{\mu}=A^{(0)}_{\mu}+\alpha \,{A}^{(1)}_{\mu}\,, 
\end{equation}
being $\left\{g^{(0)}_{\mu\nu}, A^{(0)}_{\mu}\right\}$ a solution of the zeroth-order equations of motion. Evaluating the action on the corrected solution and expanding in $\alpha$ yields
\begin{equation}
I=I^{(0)}|_{\alpha=0}+\alpha\left(\partial_{\alpha} I^{(0)}+I^{(1)}\right)|_{\alpha=0}+{\cal O}\left(\alpha^2\right)\, .
\end{equation}
As we see, out of the three contributions that arise, the corrections to the metric and gauge field are only needed to compute $(\partial_{\alpha} I^{(0)})|_{\alpha=0}$. However, it is possible to show that this contribution actually vanishes if one fixes the boundary conditions appropriately \cite{Reall:2019sah}. Although the authors of \cite{Reall:2019sah} focus on the asymptotically-flat case, they argue (see footnote~6) that the same will hold in asymptotically locally AdS spacetimes. Let us demonstrate that this is in fact the case.\footnote{We have also made an explicit check of the claim of \cite{Reall:2019sah} by computing the $\Oa$ corrections to the Reissner-Nordstr\"om-AdS solution (which at $\mathcal{O}(\alpha^0)$ is obtained from the general solution in Sec.~\ref{TwoDerReview} by setting $a=b=0$). The corrections contain new free parameters that can be chosen so as to ensure that the boundary metric and gauge field (and therefore the inverse temperature $\beta$ and the electrostatic potential $\Phi$) do not receive corrections. An explicit evaluation of $I^{(0)}$ on the corrected solution then shows it is indeed independent of $\alpha$. See also~\cite{Cremonini:2019wdk}. } Integrating by parts, one finds that 
\begin{equation}
\begin{aligned}
\delta I^{(0)}=\,&\frac{1}{16\pi G}\int_{\cal M} \diff^5x\, e\, \left[\left({\cal E}_{\mu\nu}-\frac{1}{2}g_{\mu\nu}\,{\cal E}\right)\delta g^{\mu\nu}+{\cal E}^{\mu} \,\delta A_{\mu} \right]\\[1mm]
&+\frac12\int_{\partial {\cal M}} \diff^{4}x \,\sqrt{h}\, T_{ij}\, \delta h^{ij}+\int_{\partial {\cal M}} \diff^{4}x \,\sqrt{h}\, j^{i}\, \delta A_{i}\,,
\end{aligned}
\end{equation}
where we have introduced the (zeroth-order) Brown-York energy momentum tensor \cite{Brown:1992br,Balasubramanian:1999re} $T_{ij}$ and the electric current $j^i$, which are defined as follows
\begin{equation}
T_{ij}=\frac{2}{\sqrt{h}}\frac{\delta I^{(0)}}{\delta h^{ij}}\,, \hspace{1cm} j^i=\frac{1}{\sqrt{h}}\frac{\delta I^{(0)}}{\delta A_{i}}\,.
\end{equation}
Regarding now the corrected solution \eqref{eq:perturbation} as a perturbation over the leading-order solution, we immediately see that the bulk term vanishes at order $\alpha$ because of the zeroth-order equations of motion. This is perhaps less obvious for the boundary terms, though.  In the case at hands, the boundary is the hypersurface $r=r_{\text{bdry}}$, where $r_{\text{bdry}}$ is a regulator that eventually we will send to infinity, $r_{\text{bdry}}\to \infty$. The behavior of $T_{ij}$ and $j^{i}$ for large $r_{\text{bdry}}$ in asymptotically locally AdS$_{5}$ solutions is \cite{deHaro:2000vlm,Bianchi:2001kw}
\begin{equation}
\sqrt{h}\, T_{ij}\sim {\cal O}\left(r_{\text{bdry}}^{2}\right)\,, \hspace{1cm} \sqrt{h}\, j^{i}\sim {\cal O}\left(r_{\text{bdry}}^{0}\right)\, ,
\end{equation}
implying that the boundary terms also vanish if one imposes the following asymptotic behavior on the corrections,
\begin{equation}\label{eq:boundaryconditions}
{h}^{(1)}_{ij}={\cal O}\left(r_{\text{bdry}}^0\right)\,, \hspace{1cm}  A^{(1)}_{i}={\cal O}{\left(r_{\text{bdry}}^{-2}\right)}\, .
\end{equation}
The first can always be achieved by a suitable rescaling of the Euclidean time coordinate $\tau=it$ \cite{Cremonini:2009ih}, namely 
\begin{equation}
\tau\to \frac{\tau}{g\ell}\,, 
\end{equation}
where $\ell$ is the corrected AdS radius. As for the asymptotic condition on the gauge field, there is always a gauge in which it is satisfied. Let us show that this is also the gauge in which the $A_{\mu}$ is regular at the horizon. Calling $r_++\alpha\delta r_+$ the corrected position of the horizon, we must verify that
\be
V^{\mu}A_{\mu}|_{r=r_{+}+\alpha \delta r_{+}}=0\,,
\ee
where $V^\mu$ is the Killing vector generating the horizon.
Since we are working in the grand-canonical ensemble, we need to keep the electric potential $\Phi = V^{\mu}A_{\mu}|_{r=r_{+}+\alpha \delta r_{+}} - V^{\mu}A_{\mu}|_{r=\infty}$ fixed, that is
\begin{equation}
\delta \Phi=\alpha\, V^{\mu}\left( \partial_r A^{(0)}_\mu|_{r=r_+}\delta r_+ +  A^{(1)}_{\mu}|_{r=r_{+}}-A^{(1)}_{\mu}|_{r=\infty}\right)=0\, .
\end{equation}
 Together with \eqref{eq:boundaryconditions}, this leads to 
\begin{equation}
\partial_r A^{(0)}_\mu|_{r=r_+}\delta r_+ +  A^{(1)}_{\mu}|_{r=r_{+}}=V^{\mu}A^{(1)}_{\mu}|_{r=\infty}=0\, .
\end{equation}
It follows that
\begin{equation}
V^{\mu}A_{\mu}|_{r=r_{+}+\alpha \delta r_{+}}= V^{\mu}A^{(0)}_{\mu}|_{r=r_{+}}\,,
\end{equation}
implying that the regularity condition is satisfied as it is assumed such in the uncorrected solution.

We indicate the behavior~\eqref{eq:boundaryconditions} by saying that the corrections preserve the asymptotic boundary conditions on the fields.

\subsection{Higher-derivative boundary terms}

Having established that the action can be evaluated on the uncorrected solution, it only remains to specify the boundary terms that supplement the bulk contribution in the $\Oa$ action $I^{(1)}$.
 The boundary terms associated with a generic higher-derivative bulk action such as \eqref{eq:genexpL4d} are not known, although several effective prescriptions have been discussed in the literature before, see e.g.~\cite{Cremonini:2009ih, Bueno:2018xqc, Cano:2022ord} and references therein. In our simpler effective action \eqref{eq:4daction2}, the only four-derivative invariant for which the associated Gibbons-Hawking term $I^{(1)}_{\rm{GH}}$ is known is the Gauss-Bonnet one, \cite{Teitelboim:1987zz, Myers:1987yn}. However this is enough for our purposes here, given the asymptotic behavior of the field strength in the solutions under study. As a matter of fact, only the Gauss-Bonnet term contains divergences, as previously noted also in \cite{Cremonini:2019wdk} in the static case. Therefore for $I^{(1)}_{\rm{GH}}$ we take \cite{Teitelboim:1987zz, Myers:1987yn}
\begin{equation}\label{eq:GHGB}
\begin{aligned}
I^{(1)}_{\rm{GH}}=\,&\frac{\lambda_1}{8\pi G}\int_{\partial {\cal M}} \diff^{4}x\,\sqrt{h} \,\left[\frac{2}{3}K^3-2K K_{ij}K^{ij}+\frac{4}{3}K_{ij}K^{jk}K_{k}{}^{i}+4  {\cal G}_{ij}K^{ij}\right]\\[1mm]
&-\frac{\lambda_2g^2}{2\pi G}\int_{\partial {\cal M}}\diff^4x\sqrt{h}\, {K}\,,
\end{aligned}
\end{equation}
where ${\cal G}_{ij}={\cal R}_{ij}-\frac{1}{2}h_{ij}{\cal R}$ is the Einstein tensor of the induced metric. 
Regarding the boundary counterterms, we follow the prescription of \cite{Cremonini:2009ih}, which amounts to shift the coefficients in front of the boundary counterterms already present at zeroth order in $\alpha$, namely
\begin{equation}
I^{(1)}_{\text{count}}=\frac{1}{8\pi G}\int_{\partial {\cal M}}\, \diff^4x \sqrt{h} \, \left(3\mu_1g +\frac{\mu_2}{4 g}\,{\cal R}\right)\, ,
\end{equation}
where $\mu_1$ and $\mu_2$ are chosen so as to cancel the $r^4$ and $r^2$ divergences. For the effective action \eqref{eq:4daction2}, we find the following values 
\begin{equation}\label{eq:mus}
\mu_1=\frac{4g^2}{3}\left(-4\lambda_1+3\lambda_{2}\right)\,, \hspace{1cm}\mu_2=4g^2\left(2\lambda_1+\lambda_{2}\right)\, .
\end{equation}

Actually, one can check that the sum of the boundary terms $I^{(1)}_{\rm{GH}}+I^{(1)}_{\text{count}}$  is independent of whether we add the Gibbons-Hawking term associated with the Gauss-Bonnet or not.\footnote{Actually, the same occurs with the standard Gibbons-Hawking terms if we leave the coefficients in front of the counterterms free and fix them afterwards imposing the cancellation of the divergences.} Of course, this would change the values of $\mu_1$ and $\mu_2$ in \eqref{eq:mus}, but the final result is exactly the same. In other words, if we add a control parameter in front of the first term in \eqref{eq:GHGB}, the finite contribution to $I^{(1)}_{\rm{GH}}+I^{(1)}_{\text{count}}$ is independent of this parameter and, in particular, it can be set to zero without altering the final result.  While we do not expect this to hold for more general boundary metrics, it does so in the case at hands.   This allows us to evaluate the on-shell action for the effective action \eqref{eq:4daction1} with the Riemann squared term for which the associated Gibbons-Hawking term is not known. The results agree with the ones obtained from the action \eqref{eq:4daction2}, giving evidence of the claim made in Section~\ref{FourDerAction} regarding the invariance of the  black hole thermodynamics under the class of field redefinitions we have used to simplify the four-derivative action.

\subsection{Results}
Given the setup above, it is technically demanding but otherwise straightforward to evaluate the  action on the two-derivative solution \eqref{5met}, \eqref{gaugepot}. The next step is to impose supersymmetry. We do so by imposing the same condition as at the two-derivative level, that is~\eqref{susyCCLP}. One way to see that this is the correct condition even at order $\alpha$ is that the linear relation \eqref{linear_relation} between $\omega_1,\omega_2,\varphi$ must be satisfied, and none of these quantities depend on $\alpha$. 
  After imposing the supersymmetry condition, we find that remarkably the action only depends on the combinations of the parameters $a,b,r_+$ which enter in the supersymmetric chemical potentials $\omega_1,\omega_2$ and $\varphi=\frac{1}{2}(\omega_1+\omega_2 -2 \pi i)$ given in \eqref{susychemicalpotentials}. Specifically, we obtain
\begin{equation}\label{action_almost_done}
I=\frac{2\pi}{27 G  g^3} \left(1- 4 ( 3 \lambda_1 - \lambda _2)\alpha  g^2\right) \frac{\varphi^3 }{\omega_1\omega_2}+\frac{2\pi  \alpha  \lambda _1}{3  G g} \frac{\varphi \left(\omega_1^2+\omega_2^2-4 \pi ^2\right)}{\omega_1\omega_2}\,.
\end{equation}
The fact that $\beta$ drops out of this expression indicates the validity of the supersymmetric thermodynamics reviewed in Section~\ref{TwoDerReview} at linear order in the corrections.
We can now convert this result in field theory units. The dimensionless quantities $G g^3$, $\alpha g^2$ have a holographic counterpart in the dual SCFT central charges $\aa$, $\cc$, or equivalently in the R-symmetry anomaly coefficients ${\rm Tr}\mathcal{R}^3$, ${\rm Tr}\mathcal{R}$. In Appendix~\ref{sec:holographicdictionary} we show that for the bulk action~\eqref{eq:4daction2} we have been studying, the dictionary between the gravitational  and the field theory coefficients is 
\be
\begin{aligned}
{\rm Tr}\,\mathcal{R}^3 \,&=\, \frac{16}{9}(5\aa-3\cc)\,=\, \frac{4\pi}{9G g^3} \left( 1 -4(3\lambda_1-\lambda_2) \alpha g^2  \right)\,, \\[1mm]
{\rm Tr}\,\mathcal{R} \,&=\, 16(\aa-\cc)\,=\, -\frac{16\pi \alpha\lambda_1}{G g}   \,.
\end{aligned}
\ee
Plugging this in~\eqref{action_almost_done} and eliminating $\varphi$ via \eqref{linear_relation} yields
\begin{equation}\label{susy_action_om1om2}
I\,=\, {\rm Tr}\,\mathcal{R}^3\, \frac{(\omega_1+\omega_2 -2 \pi i )^3 }{48\,\omega_1\omega_2}- {\rm Tr}\,\mathcal{R} \,\frac{(\omega_1 +\omega_2 -2 \pi i) \left(\omega_1^2+\omega_2^2-4 \pi ^2\right)}{48\,\omega_1\omega_2}\,,
\end{equation}
which precisely reproduces the prediction from the index on the second sheet reported in Section~\ref{sec:intro} upon identifying $I = -\log\mathcal{I}$.

%Note we do not need to assume that $\omega_1,\omega_2$ are small here.

In the remainder of this subsection we illustrate the computation further by providing some details in the more manageable case where $a=b$. In this case the angular ${\rm U}(1)\times{\rm U}(1)$ symmetry of the solution is enhanced to ${\rm SU}(2)\times {\rm U}(1)$, hence there is only one independent angular momentum. Then the two-derivative solution reviewed in Section~\ref{TwoDerReview} reduces to the one first given in \cite{Cvetic:2004hs} (also discussed in~\cite{Kunduri:2005zg}). In the BPS limit, it gives the supersymmetric black hole of \cite{Gutowski:2004ez}.

For the non-supersymmetric on-shell action we find 
\begin{equation}\label{onshellactionCCLP}
\begin{aligned}
I=\,&\frac{\pi\beta\left(1+4 \lambda _2\alpha g^2 \right)}{4G\Xi_a^2}
\Big[m - g^2 (r_+^2 + a^2)^2-
\frac{q^2 r_+^2}{(r_+^2 + a^2)^2+a^2q}\Big]+\frac{\pi\lambda_1\alpha\beta}{G\Xi_a^2}\left\{-\frac{3 g^2 q^2}{r_+^2+a^2}\right.\\[1mm]
&+\frac{-9 m^2+a^2 g^2 q^2 \left(9+\sqrt{3} \xi
   \right)}{\left(r_+^2+a^2\right){}^2}+\frac{12 m q^2+4a^2\left[(2 m+q)^2- a^2 g^2 q^2 \right] \left(2+ \frac{\xi}{\sqrt{3}}\right)}{\left(r_+^2+a^2\right){}^3}\\[1mm]
&-\frac{3 q^4+8 a^4 (m+q)^2 \left(3+\sqrt{3} \xi \right)+a^2 q^2 \left[8 m \left(3+\sqrt{3} \xi \right)+q\left(15+4 \sqrt{3} \xi \right)\right]}{\left(r_+^2+a^2\right){}^4}\\[1mm]
&\left.+\frac{\sqrt{3} a^2 q^2 \left[3 q^2+8 a^2
   (m+q)\right] \xi }{\left(r_+^2+a^2\right){}^5}-\frac{a^4 q^4 \left(-3+\sqrt{3} \xi \right)}{\left(r_+^2+a^2\right){}^6}\right\}\,,
\end{aligned}
\end{equation}
where $m$ and $\xi$ are fixed in terms of $r_{+}$, $q$, $a$ as given in Footnote~\ref{foot:solm} and in Eq.~\eqref{regularxi}, respectively. In the above expression for the on-shell action, the contribution of the AdS vacuum has been subtracted, the latter being 
\begin{equation}
I_{\text{AdS}}=\frac{3\pi\beta}{32G g^2 }\left(1+4\lambda_2\alpha g^2\right)=\frac{3g\beta}{4} \aa \,,
\end{equation}
where in the second equality we have made use of \eqref{a_c_high_der} to observe {\it en passant} that it computes the corrected superconformal anomaly coefficient $\aa$.
 At this stage, everything is settled to impose supersymmetry, which amounts to fix the parameter $q$ as in \eqref{susyCCLP}, which we quote here again for the sake of convenience,
\begin{equation}\label{susy_again}
q= -(a- i r_+)^2 (1- i gr_+)\,.
\end{equation}
 Now, from~\eqref{susychemicalpotentials} one can find $r_{+}$ in terms of $\omega = \omega_1+\omega_2$ and $a$,
\begin{equation}
r_{+}=\frac{4\pi\left(-1+ag\right)-i\omega \left(1+ag\right)}{3g\omega}\,,
\end{equation}
and, substituting this expression in the Euclidean on-shell action, verify that all the dependence on the parameter $a$ (and therefore on $\beta$) disappears, yielding the simple expression
\begin{equation}
I=\frac{ \pi}{27 G g^3 } \left(1- 4 ( 3 \lambda_1 - \lambda _2)\alpha  g^2\right) \frac{(\omega -2i \pi )^3 }{\omega^2}+\frac{2 \pi  \alpha  \lambda _1}{3 G g} \frac{(\omega -2i \pi ) \left(\omega ^2-8 \pi ^2\right)}{\omega ^2}\,.
\end{equation}
which is just \eqref{action_almost_done} with $\omega_1=\omega_2\equiv\omega/2$ and $2\varphi=\omega - 2\pi i$.

%%%%%%%%%%%%%%%%%%%%%%%%%%%%%%%%%%%%%%%%%%%%%%%%%%%%%%%%%%%%%%%%%%%%%%%%%%%%%%%%%%%%%%%%%%%%%%%%%%%%%%%%%%%%%%%%%%%%%%%%%%
\section{Corrected BPS charges and entropy from the on-shell action}\label{sec:BPSlimit}

Assuming that the first law \eqref{firstlaw} and the quantum statistical relation \eqref{QSR_CCLP} remain valid at linear order in the four-derivative corrections, we can derive the non-supersymmetric charges and the entropy by varying the Euclidean on-shell action $I(\beta,\Omega_1,\Omega_2,\Phi)$ (or the Gibbs free energy $\mathcal{G}(T,\Omega_1,\Omega_2,\Phi)=I/\beta$) with respect to its arguments, as in \eqref{charges_from_I}. Concretely, this is done by varying with respect to the independent parameters $r_+,a,b,q$, and using the chain rule to express these in terms of variations of the thermodynamic potentials.
The resulting expressions are quite cumbersome; we provide them in Appendix~\ref{app:non_susy_charges} for the simpler case where $a=b$.

We have checked that after imposing supersymmetry through~\eqref{susyCCLP}, the corrected energy, charge and angular momenta, which now depend on $a$, $b$ and $r_+$, satisfy the supersymmetric relation \eqref{susyrel_charges}, while $\beta$ is still finite. Because of the choice~\eqref{susyCCLP}, all quantities have complex values in this case.

In the rest of this section we focus on the supersymmetric and extremal (BPS) limit of the conserved charges and entropy. The BPS solution is found by imposing the conditions \eqref{susyCCLP} and \eqref{r_BPS} on the parameters. The limiting procedure is however not unique and certain quantities such as the Euclidean on-shell action depend on it, as discussed in \cite{Cabo-Bizet:2018ehj}.  The prescription to take the limit that we follow here is the same as in~\cite{Silva:2006xv}, namely we fix $q$ to its BPS value, that is
$q = q^*=g^{-1}(a+b)(1+ag)(1+bg)$, and parametrize deviations with respect to the BPS locus by setting 
\begin{equation}
r_{+}=r_{*}+\epsilon\,, 
\end{equation}
where $r_{*}$ is given in \eqref{r_BPS} and $\epsilon$ is the expansion parameter. %\footnote{Since we are expressing $m$ in terms of $r_+$ as in Footnote~\ref{foot:solm}, we have that $m$ contains $\epsilon$ too.}
Recall that the expression for the temperature is the same as at zeroth order in $\alpha$, hence the extremality condition remains the same.

\subsection{Equal rotation parameters}\label{aeqb_sect}

In this subsection we  discuss the case where $a=b$, which gives one independent angular momentum $J\equiv J_1 = J_2$, with conjugate angular velocity $\Omega\equiv2\Omega_1=2\Omega_2$.
In the $\epsilon\to0$ limit, all the charges remain real and finite, and read
\begin{eqnarray}
Q^{*}&=&\frac{\sqrt{3} \pi a}{2Gg (1-ag)^2}\left[1+4  \lambda _2\alpha g^2+4\lambda_1\alpha g^2\frac{1+8 a g+36 a^2 g^2+44 a^3 g^3+19 a^4 g^4}{ag\left(-1+8 a
   g+11 a^2 g^2\right)}\right]\,, \label{QfromI}\\[2mm]
J^{*}&=&\frac{\pi a^2 (3+a g)}{2 Gg (1-ag)^3}\left[1+4
   \lambda _2 \alpha g^2+24 \lambda_1\alpha g^2\frac{1+9 a g+29 a^2 g^2+25 a^3 g^3+8 a^4 g^4}{ag (3+a g) \left(-1+8 a g+11 a^2 g^2\right)}\right],\qquad\\[2mm]
 E^* &=&  2g J^* + \sqrt{3}\, Q^* ,
\end{eqnarray}
with $E^*$ satisfying the supersymmetry relation \eqref{susyrel_charges},
while the entropy is
\be\label{SfromI}
{\cal S}^{*}\,=\, \frac{\pi ^2 a \sqrt{ag (a g+2)}}{g^{2} G (1-ag)^2}\left[1+4 \lambda _2\alpha  g^2 +48\lambda _1 \alpha  g^2 \frac{2 a^2 g^2+5 a g+2}{11 a^2 g^2+8 a g-1}\right]\,.
\ee
Let us note that, at the zeroth-order in $\alpha$, these agree with the BPS limit of the charges and the entropy given in Section~\ref{TwoDerReview}, as they should. 

We find that the BPS entropy as a function of the charges is given by 
\be\label{S_BPScorrected}
\mathcal{S}^* = \frac{\pi}{g} \sqrt{4{Q^*}^2 -\frac{2\pi}{Gg} J^*}\left[1-2\lambda_2 \alpha g^2 \,\frac{\frac{2\pi}{G g} J^*}{4 {Q^*}^2 -\frac{2\pi}{G g} J^*}\right]\, ,
\ee  
which, remarkably, agrees with the $\Oa$ expansion of
\be\label{eq:microcanonicalentropyCFT}
\mathcal{S}^* = \pi \sqrt{3(Q^*_{R})^2 - 16\,\aa \,J^*}\ ,
\ee
where we have also used the canonically normalized R-charge \eqref{QandRcharge} for the ease of comparison with field theory expressions.

In addition to this, we have been able to find the $\Oa$ corrections to the non-linear relation between the BPS charges given in \eqref{nonlinear_rel_BPScharges}. The corrected relation reads
\be
\begin{aligned}
&\left[\frac{2\sqrt{3}Q^*}{g}   + \frac{\pi}{2G g^3 }\left(1-8\lambda_1\alpha g^2+4\lambda_2\alpha g^2 \right) \right]\left[\frac{4{Q^*}^2}{g^2} - \frac{2\pi J^*}{G g^3 }\left(1+8\lambda_1\alpha g^2+4\lambda_2\alpha g^2 \right)\right]=\\[1mm]
&=\frac{8{Q^*}^3}{3\sqrt{3}g^3} + \frac{2\pi {J^*}^2}{G g^3 }\left(1+24\lambda_1\alpha g^2+4\lambda_2\alpha g^2 \right)\,.
\end{aligned}
\ee
We note that this is only valid at linearized level in $\alpha$, so one has to ignore both explicit and implicit ${\cal O}{\left(\alpha^2\right)}$ corrections. In field theory language, this relation translates into 
\be\label{corrected_nonlinear_rel_BPScharges}
\left[3 Q_R^*  + 4\left(2\aa-\cc\right) \right]\left( 3 Q_{R}^*{}^2 -  16\cc \,J^*\right) = Q_{R}^*{}^3 + 16 \left(3\cc-2\aa\right){J^*}^2\ ,
\ee
where again we have made use of \eqref{QandRcharge}. 

\subsection{Independent rotation parameters}\label{aneqb_sect}

We next give the more involved expressions of the corrected BPS quantities for the case with two independent rotation parameters, $a\neq b$.\footnote{These results have been added in the v2 of the present paper.}
Here we set $g=1$ for simplicity.

The electric charge reads
\begin{equation}
Q^* = \frac{\sqrt{3}\pi(a+b)}{4G(1-a)(1-b)}\Big[1+ 4\lambda_1\alpha\,\Delta Q^{*} + 4\lambda_2\alpha\Big]\,\,,
\label{eq:Q}\end{equation}
with
\[\small
\begin{aligned}
\Delta Q^{*} &= \frac{2}{3 \mathfrak{D}}\Big[ a^7 \left(5 b^2+4 b-1\right)-a^6 \left(7 b^3+34 b^2+57 b+26\right)-b (b+1)^2 \left(b^4+24 b^3+31 b^2+14 b+3\right)\\[1mm]
&-a^5 \left(55 b^4+244 b^3+427 b^2+322 b+80\right)-a^4 \left(55 b^5+400 b^4+985 b^3+1142 b^2+586 b+100\right)\\[1mm]
&-a^3 \left(7 b^6+244 b^5+985 b^4+1688 b^3+1410 b^2+516 b+62\right)\\[1mm]
&+a^2 \left(5 b^7-34 b^6-427 b^5-1142 b^4-1410 b^3-844 b^2-220 b-20\right)\\[1mm]
&+a \left(4 b^7-57 b^6-322 b^5-586 b^4-516 b^3-220 b^2-44 b-3\right)
\Big]\,,
\end{aligned}\]
and 
\begin{equation}
\begin{aligned}
\mathfrak{D} &= (1+a)(1+b)(a+b)^2 [a^4-2 a^3 (b+1)-3 a^2 \left(3 b^2+8 b+3\right)-2 a \left(b^3+12 b^2+12 b+1\right)\\[1mm]
&\quad +b^4-2 b^3-9 b^2-2 b+1]\,\,.
\end{aligned}
\label{eq:denominator}\end{equation}

The angular momentum $J_1$ is given by
\begin{equation}
J_1^* = \frac{\pi(a+b)\left(b+2a + ab\right)}{4G(1-a)^2(1-b)}\Big[ 1+ 4\lambda_1\alpha\,\Delta J_1^{*} + 4\lambda_2\alpha\Big]\,\,,
\label{eq:J1}\end{equation}
where
\[\small
\begin{aligned}
\Delta J_1^{*} =& \frac{-2}{ \mathfrak{D}(b+2a+ab)}\Big[a^8 \left(-3 b^2-2 b+1\right)+a^7 \left(4 b^3+38 b^2+68 b+30\right)\\[1mm]
&+2 a^6 \left(19 b^4+111 b^3+238 b^2+196 b+50\right) +2 a^5 \left(24 b^5+184 b^4+543 b^3+707 b^2+377 b+63\right)\\[1mm]
&+2 a^4 \left(6 b^6+109 b^5+509 b^4+1032 b^3+917 b^2+324 b+33\right)\\[1mm]
&+a^3 \left(-2 b^7+36 b^6+402 b^5+1398 b^4+2026 b^3+1234 b^2+294 b+24\right)\\[1mm]
&+a^2 \left(-b^8-2 b^7+58 b^6+428 b^5+1074 b^4+1076 b^3+442 b^2+76 b+5\right)\\[1mm]
&+2 a b \left(-b^7+23 b^5+118 b^4+204 b^3+131 b^2+32 b+3\right)-b^8+12 b^6+48 b^5+52 b^4+16 b^3+b^2
\Big]\,\,.
\end{aligned}\]
The angular momentum $J_2$ is obtained from $J_1$ by exchanging the rotational parameters, $a\leftrightarrow b$. 
The mass $E^*$ satisfies the supersymmetric relation \eqref{susyrel_charges},
while the entropy is
\begin{equation}
\mathcal S^* = \frac{\pi^2(a+b)\sqrt{a+b+ab}}{2G(1-a)(1-b)}\Big[1+ 4\lambda_1\alpha \,\Delta\mathcal S^{*} + 4\lambda_2 \alpha \Big]\,\,,
\label{eq:entropy0}\end{equation}
with
\[\small
\begin{aligned}
\Delta\mathcal S^{*} =& \frac{2}{\mathfrak{D}}\Big[ a^7 (b+1)-a^6 (b+1)-a^5 \left(12 b^3+48 b^2+63 b+23\right)-a^4 \left(26 b^4+156 b^3+312 b^2+245 b+59\right)\\[1mm]
-&a^3 \left(12 b^5+156 b^4+496 b^3+620 b^2+313 b+45\right)-a^2 \left(48 b^5+312 b^4+620 b^3+504 b^2+159 b+11\right)\\[1mm]
+&a b \left(b^6-b^5-63 b^4-245 b^3-313 b^2-159 b-26\right)+b^2 \left(b^5-b^4-23 b^3-59 b^2-45 b-11\right)
\Big]\,\,.
\end{aligned}\]
We find that the BPS entropy can be written in terms of the BPS charges as
\begin{equation}
\mathcal S^* = \pi\sqrt{4Q^*{}^2-\frac{\pi}{G}(1+4\lambda_2\alpha)(J_1^* + J_2^*)+\frac{3\pi^2}{2G^2}\lambda_1\alpha\,\frac{(J_1^* -J_2^*)^2}{Q^*{}^2-\frac{3\pi}{16G}(J_1^* + J_2^*)}
}\,\,.
\label{eq:microcanent}\end{equation}
As before, this expression is understood to be valid at $\mathcal{O}(\alpha)$.
Using the holographic dictionary~\eqref{a_c_high_der} together with \eqref{QandRcharge}, this can be re-expressed as
\begin{equation}
\mathcal S^* = \pi\sqrt{3Q_R^*{}^2 -8 \mathtt a (J_1^* + J_2^*) - 16\, \mathtt a (\mathtt{a}-\mathtt c) \frac{(J_1^*-J_2^*)^2}{Q_R^*{}^2-2\mathtt a(J_1^* + J_2^*)}}\,\,.
\label{eq:microcanent1}\end{equation}
Moreover, we find that the non-linear constraint on the BPS charges now becomes
\be
\begin{aligned}
& \left[3 Q_R^*  + 4\left(2\,\aa-\cc\right) \right]\left[ 3 Q_R^*{}^2 -  8\cc\, (J_1^*+J_2^*)\right]   \\
&= Q_R^*{}^3 + 16 \left(3\cc-2\aa\right)J_1^*J_2^* +\,64\aa\, (\aa-\cc)\frac{(Q_R^*+\aa)(J_1^*-J_2^*)^2}{Q_R^*{}^2-2\aa (J_1^*+J_2^*)}\,.
\end{aligned}
\ee

\subsection{Generality of effective action}

Finally, we close this section giving the details of the physical reasoning to argue that the effective action \eqref{eq:4daction2} is the most general one compatible with supersymmetry, already anticipated in Section~\ref{FourDerAction}. To this aim, we note that the linear supersymmetric relation on the charges \eqref{susyrel_charges} further implies that the Gibbs free energy $\mathcal{G}= I/\beta$ must identically vanish in the BPS limit, as indeed occurs for the action \eqref{eq:4daction2}. This a highly non-trivial test of the supersymmetric nature of the effective action \eqref{eq:4daction2}.  As a matter of fact, if we now consider an action such as \eqref{eq:4daction2} but with arbitrary coefficients ${\tilde c}_{i}=1+\alpha g^2 \delta {\tilde c}_{i}$, we obtain that the Gibbs free energy does not vanish unless they are related by
\begin{equation}
\delta {\tilde c}_{1}=-\frac{1}{3}\left(54\lambda_1-5\delta {\tilde c}_{0}+2\delta {\tilde c}_{3}\right)\,, \hspace{0.5cm} \delta {\tilde c}_{2}=\frac{1}{3}\left(36\lambda_1+\delta {\tilde c}_{0}+2\delta {\tilde c}_{3}\right)\, .
\end{equation}
Choosing $\delta {\tilde c}_{0}=12\lambda_1+\delta {\tilde c}_{3}$, we note that this is nothing but a reparametrization of the effective action \eqref{eq:4daction2} in terms of $\delta {\tilde c}_{3}$ instead of $\lambda_{2}$. This physical argument completes our proof that it is always possible to absorb $\lambda_3$ (the coupling in front of the third supersymmetric invariant) in a redefinition of $\lambda_1$ and $\lambda_2$. Then, \eqref{eq:4daction2} is the most general four-derivative action compatible with supersymmetry.
%%%%%%%%%%%%%%%%%%%%%%%%%%%%%%%%%%%%%%%%%%%%%%%%%%%%%%%%%%%%%%%%%%%%%%%%%%%%%%%%%%%%%%

\section{Wald entropy from the near-horizon geometry}\label{sec:near_horizon}

It is well known that in higher-derivative gravity the entropy is given by the Wald formula \cite{Wald:1993nt, Iyer:1994ys}. However, in order to be able to use this formula we need to first find the corrected black hole solution.\footnote{In particular, given our boundary conditions in the grand-canonical ensemble, we must allow the position of the horizon to fluctuate when including the higher-derivative corrections.} In the case at hands this is a very arduous task even in the $a=b$ case, and therefore we leave it for future investigations. This issue can be circumvented by studying corrections to the near-horizon geometry for the black hole of~\cite{Gutowski:2004ez}, which can be obtained from the general solution of \cite{Chong:2005hr} by taking $a=b$ and then the BPS limit. The near-horizon geometry of this BPS black hole with a single angular momentum involves a compact space with SU(2)$\times$U(1) isometry, compatible with the geometry of a three-sphere squashed by the rotation along an axis, fibred over an AdS$_2$ factor.

In what follows we find the corrected near-horizon solution of the equations of motion  from the action~\eqref{eq:4daction2} ---which we derive in Appendix~\ref{app:eoms}--- and then we use it to compute the black hole entropy, showing that the result obtained from this alternative method nicely agrees with the computation from the action presented in the previous section. 

\subsection{Near-horizon geometry}

Assuming the isometries of the solution are not spoiled by the corrections, the corrected solution must fit into the following ansatz 
\begin{eqnarray}
\diff s^2&=&v_1\left(-\varrho^2 \diff t^2+\frac{\diff\varrho^2}{\varrho^2}\right)+\frac{v_2}{4}\left[\sigma_1^2+\sigma_2^2+v_3\left(\sigma_3+w\, \varrho\, \diff t\right)^2\right]\,,\\[1mm]
A&=&e \,\varrho \, \diff t+ p\,\left(\sigma_3+w \,\varrho\, \diff t\right)\, ,
\end{eqnarray}
where the $\sigma$'s are the left-invariant Maurer-Cartan 1-forms of SU$(2)$,\footnote{The coordinates $\hat\theta,\hat\phi,\hat\psi$ are Euler angles on $S^3$, and are related in a simple way to those used in the rest of this work.}
\begin{equation}
\begin{aligned}
\sigma_1=\,&\cos\hat\psi \,\diff\hat\theta+\sin\hat\psi\sin\hat\theta \,\diff\hat\phi\,,\\[1mm]
\sigma_2=\,&-\sin\hat\psi \,\diff\hat\theta+\cos\hat\psi\sin\hat\theta \,\diff\hat\phi\,,\\[1mm]
\sigma_3=\,&\diff\hat\psi+\cos\hat\theta \,\diff\hat\phi\,,
\end{aligned}
\end{equation}
while
\begin{equation}
\begin{aligned}
&v_1=\,\frac{\chi ^2}{4g^2 \left(1+3 \chi ^2\right)}+\alpha \,\delta v_1\, , \hspace{0.5cm} v_2=\frac{\chi ^2}{g^2}+\alpha \,\delta v_2\, , \hspace{0.5cm}v_3=1+\frac{3 \chi ^2}{4}+\alpha \,\delta v_3\, , \\[1mm]
&p=\frac{\sqrt{3} \chi ^2}{4g}+\alpha \,\delta p\, , \hspace{0.25cm} w=\, \frac{3 \chi }{\left(1+3 \chi ^2\right)\sqrt{4+3 \chi ^2}}+\alpha \,\delta w\, , \hspace{0.25cm}
e=\frac{\sqrt{3} \chi}{g \left(1+3 \chi ^2\right)\sqrt{4+3 \chi ^2}}+\alpha\, \delta e\, ,
\end{aligned}
\end{equation}
contain the deviations from the $\alpha=0$ solution. Here, the (dimensionless) parameter  $\chi$ is related to the only parameter $R_0$ of the solution in \cite{Gutowski:2004ez} by $\chi=g\,R_0$. The relation between $\chi$ and the parameter $a$ $(=b)$ of the BPS solution of \cite{Chong:2005hr} is 
\begin{equation}\label{eq:GRtoCCLP}
\chi^2=\frac{2ag}{1-ag}\, .
\end{equation}

Working perturbatively in $\alpha$, the problem of solving the equations of motion reduces to a linear system of algebraic equations,
\begin{equation}
{\cal M}{\cal X}={\cal N}\, ,
\end{equation}
where ${\cal M}={\cal M}(\chi; g)$ is a degenerate $6\times 6$ matrix, ${\cal X}=(\delta v_1, \delta v_2, \delta v_3, \delta p, \delta e, \delta w)^T$ and ${\mathcal N}={\mathcal N}(\chi; g)$ is a vector which encodes the contribution to the equations of motion of the corrections. The general solution to this equation is
\begin{equation}
{\cal X}={\cal X}^{H}+{\cal X}^{P}\,,
\end{equation}
where ${\cal X}^{H}$ is the homogeneous solution, ${\mathcal M}{\cal X}^{H}=0$, and ${\cal X}^{P}$ is a particular solution. The latter carries information about the new physics while the homogeneous solution (which is non-trivial since the matrix ${\cal M}$ is not invertible) parametrizes the freedom that we still have to fix the boundary conditions.\footnote{Here we have already fixed the temperature to zero.} This will be used later on to fix the electric potential and angular velocity at the horizon but, before doing so, let us write down the general solution. Solving the homogeneous system in terms of $\delta e$ and $\delta w$ yields the following solution
\begin{equation}
\begin{aligned}
{\delta v}^{H}_1=\,&\frac{\chi  \left(9 \chi ^6-12 \chi ^4-68 \chi ^2-48\right)\delta e}{2 g \sqrt{9 \chi ^2+12} \left(27 \chi
   ^8+162 \chi ^6+144 \chi ^4+4 \chi ^2-12\right)}\\[1mm]
&-\frac{\chi ^3 \left(9 \chi ^6+78 \chi ^4+154 \chi ^2+88\right)\delta w}{4 g^2 \sqrt{3 \chi ^2+4} \left(27 \chi
   ^8+162 \chi ^6+144 \chi ^4+4 \chi ^2-12\right)}\,,\\[1mm]
{\delta v}^{H}_2=\,&4\left(1+3\chi^2\right)^2\delta v_1^{H}\,,\\[1mm]
{\delta v}^{H}_3=\,&-\frac{\sqrt{3}  g \chi ^3 \left(3 \chi ^2+4\right)^{3/2} \left(45 \chi ^4+66 \chi ^2+17\right)\delta e}{2
   \left(3 \chi ^4+16 \chi ^2+6\right) \left(9 \chi ^4+6 \chi ^2-2\right)}\\[1mm]
&-\frac{3 \chi  \left(3 \chi ^2+1\right) \left(3 \chi ^2+4\right)^{3/2} \left(3 \chi ^6+7 \chi
   ^4+2\right)\delta w}{4 \left(3 \chi ^4+16 \chi ^2+6\right) \left(9 \chi ^4+6 \chi ^2-2\right)}\,,\\[1mm]
\delta p^{H}=\,&\frac{\chi  \left(3 \chi ^2+1\right) \sqrt{3 \chi ^2+4} \left(9 \chi ^6-48 \chi ^4-62 \chi
   ^2-6\right)\delta e}{2 \left(3 \chi ^4+16 \chi ^2+6\right) \left(9 \chi ^4+6 \chi ^2-2\right)}\\[1mm]
&-\frac{\chi  \left(3 \chi ^2+1\right) \sqrt{9 \chi ^2+12} \left(9 \chi ^8+69 \chi ^6+70 \chi ^4+10
   \chi ^2+4\right)\delta w}{4 g \left(3 \chi ^4+16 \chi ^2+6\right) \left(9 \chi ^4+6 \chi ^2-2\right)}\,,\\[1mm]
{\delta e}^{H}=\;\,&\delta e\, ,\\[1mm]
{\delta w}^{H}=\,\,&\delta w\, ,
\end{aligned}
\end{equation}
while a particular solution is given by 
\begin{equation}\label{eq:particularsol}
\begin{aligned}
{\delta v}^{P}_1=\,&\frac{\lambda _1 \left(18 \chi ^6+21 \chi ^4+14 \chi
   ^2+2\right)}{27 \chi ^6+27 \chi ^4-2}\,,\\[1mm]
{\delta v}^{P}_2=\,&\frac{4 \lambda _1 \left(36 \chi ^6+51 \chi ^4+10 \chi ^2+6\right)}{9 \chi ^4+6 \chi
   ^2-2}\,,\\[1mm]
{\delta v}^{P}_3=\,&-\frac{\lambda _1 g^2  \left(3 \chi ^2+4\right) \left(63 \chi ^4+78 \chi ^2-38\right)}{2 \left(9 \chi ^4+6 \chi
   ^2-2\right)}\,,\\[1mm]
\delta p^{P}=\,&\frac{\sqrt{3} \lambda _1 g  \left(3 \chi ^2+4\right) \left(27 \chi ^4-6 \chi ^2+10\right)}{4 \left(9 \chi ^4+6 \chi
   ^2-2\right)}\, ,\\[1mm]
\delta e^{P}=\,&0\, , \\[1mm]
\delta w^{P}=\,&0\, .
\end{aligned}
\end{equation}
As discussed in \cite{Morales:2006gm, Dias:2007dj} building on the formalism of~\cite{Sen:2005wa,Sen:2008vm}, the variables $e$ and $w$ are identified with the thermodynamical variables conjugated to the electric charge and angular momentum. Thus, if we want to obtain the corrections in the grand-canonical ensemble, we must impose the following choice of boundary conditions,
\begin{equation}
\delta e=0\, , \hspace{1cm} \delta w=0\, ,
\end{equation}
in which case the homogeneous part of the solution vanishes and the solution is simply equal to the particular one in \eqref{eq:particularsol}.

%%%%%%%%%%%%%%%%%%%%%%%%%%%%%%%%%%%%%%%%%%%%%%%%%%%%%%%%%%%%%%%%%%%%%%%%%%%%%%%%%%%%%%%%%%%%%%%%%%%%%%%%%%%%%%%%%%%%%%%%%%%%%%

\subsection{Black hole entropy}

Having found the corrected near-horizon geometry, we can now compute the entropy of the black hole. As anticipated, the black-hole entropy ${\cal S}$ in higher-derivative gravity can be computed by means of the Wald formula \cite{Wald:1993nt, Iyer:1994ys}, 
\begin{equation}\label{eq:WaldIyerformula}
{\cal S}=-2\pi \int_{\Sigma} \diff^3x \sqrt{\gamma} \, {\cal E}^{\mu\nu\rho\sigma}_{R}\, \epsilon_{\mu\nu}\epsilon_{\rho\sigma}\, ,
\end{equation}
where $\gamma$ is the determinant of the induced metric on an arbitrary space-like cross-section of the horizon $\Sigma$, $\epsilon_{\mu\nu}$ is the binormal normalized so that $\epsilon_{\mu\nu}\epsilon^{\mu\nu}=-2$ and ${\cal E}^{\mu\nu\rho\sigma}_{R}$ is obtained varying the action with respect to the Riemann tensor as if it were an independent field. Using  results derived in Appendix~\ref{app:eoms}, we get
\begin{equation}
{\cal E}^{\mu\nu\rho\sigma}_{R}\equiv\frac{\delta S}{e\,\delta R_{\mu\nu\rho\sigma}}=\frac{1}{16\pi G}\left(P^{\mu\nu\rho\sigma}+{\Pi}^{\mu\nu\rho\sigma}\right)\, ,
\end{equation}
where 
\begin{equation}
\begin{aligned}
{P}^{\mu\nu\rho\sigma}=\,&\left(1+4\lambda_2\alpha g^2\right)g^{\mu[\rho}g^{\sigma]\nu}+\alpha\,\lambda_1\left[2R^{\mu\nu\rho\sigma}-4\left(R^{\mu[\rho}g^{\sigma]\nu}-R^{\nu[\rho}g^{\sigma]\mu}\right)+2 g^{\mu[\rho}g^{\sigma]\nu} R\right.\\[1mm]
&\left. -\frac{1}{2}F^{\mu\nu}F^{\rho\sigma}-\frac{1}{12}g^{\mu[\rho}g^{\sigma]\nu} F^2+\frac{1}{3}\left(F^{\mu\alpha}F^{[\rho}{}_{\alpha}g^{\sigma]\nu}-F^{\nu\alpha}F^{[\rho}{}_{\alpha}g^{\sigma]\mu}\right)\right]\,,
\end{aligned}
\end{equation}
and 
\begin{equation}
\Pi^{\mu\nu\rho\sigma}=-\frac{\alpha\,\lambda_1}{\sqrt{3}}\epsilon^{\mu\nu\alpha\beta\gamma}R_{\alpha\beta}{}^{\rho\sigma}A_{\gamma}\, .
\end{equation}
The tensor $\Pi^{\mu\nu\rho\sigma}$ is the contribution from the mixed Chern-Simons term in the action, and as such it depends explicitly on the gauge field $A$. 
 This means that the entropy obtained using Wald's formula would not be gauge invariant in general.\footnote{This issue has been recently addressed in \cite{Elgood:2020xwu, Elgood:2020nls} in the context of the heterotic superstring theory, where a gauge-invariant generalization of the Wald formula has been obtained. One way to preserve gauge invariance, at the expense of breaking diffeomorphisms, would be to start from a five-dimensional action where the mixed gauge-gravitational Chern-Simons term has been integrated by parts, and use the prescription of \cite{Tachikawa:2006sz} for the entropy. We do not do this, as it would correspond to a scheme where the mixed gauge-gravitational anomaly manifests itself in the non-conservation of the holographic energy-momentum tensor, rather than of the R-current. In any case, we have checked that the prescription of \cite{Tachikawa:2006sz} yields the same result as ours, since the total derivative does not contribute in the present case.
%It was also previously addressed in \cite{Tachikawa:2006sz}, though, as far as we understand, the prescription given in that reference follows from integrating by parts the mixed Chern-Simons term and applying Wald's formula. Although this solves the issue of the gauge invariance, the resulting formula is not manifestly invariant under diffeomorphisms. In any case, we have checked that the prescription of \cite{Tachikawa:2006sz} yields the same result as ours, since the total derivative does not contribute in the present case.
} In the case at hands, there is only one gauge transformation which affects the entropy, which is $A\to A+\diff \hat \psi$. However, this gauge transformation breaks the SU$(2)\times$U$(1)$ symmetry of the near-horizon solution and in fact is not globally well defined. Hence, it is reasonable to expect that the direct application of the Wald formula in the gauge we are using should produce the right result. This is exactly the case, as we are going to show next.

As it is well known, Wald's formula encodes two types of corrections. On the one hand, we have the corrections to the area law, which are captured by the $\Oa$ contributions to the tensor $P^{\mu\nu\rho\sigma}$ and by $\Pi^{\mu\nu\rho\sigma}$, which is already of $\Oa$. Clearly, this set of corrections can be evaluated without knowing the corrected solution. However, there is a second set for which the previous statement is not true. They come from the first term in $P^{\mu\nu\rho\sigma}$, which gives rise to the Bekenstein-Hawking term,  and from the fact that the area of the horizon receives $\Oa$ corrections, namely:
\begin{equation}
\begin{aligned}
A_{\Sigma}=\frac{\pi ^2\chi ^3\sqrt{4+3 \chi ^2}}{g^3}\left[1+ \lambda _1\alpha g^2\frac{
    153 \chi ^6+228 \chi ^4+98 \chi ^2+36}{\chi^2 \left(9 \chi ^4+6 \chi ^2-2\right)}\right]\,.
\end{aligned}
\end{equation}
In order to evaluate the corrections to the Bekenstein-Hawking term, we choose $\Sigma$ to be a $t=const$ slice of the horizon, as it is well known that the entropy is independent of this choice, \cite{Jacobson:1993vj}. Thus, the non-vanishing components of the binormal are

\begin{equation}
\epsilon_{t\varrho}=v_1\,.
\end{equation}
The final expression for the Wald entropy is,
\begin{equation}
{\cal S}=\frac{\pi ^2\chi ^3\sqrt{4+3 \chi ^2}}{4 Gg^3}\left[1+4 \lambda _2\alpha  g^2 +24 \lambda _1\alpha  g^2 \frac{9 \chi ^4+18 \chi ^2+8}{9 \chi ^4+6 \chi ^2-2}\right]\, ,
\end{equation}
which in terms of the parameter $a$ ---see eq.~\eqref{eq:GRtoCCLP}--- reads

\begin{equation}
{\cal S}=\frac{\pi ^2 ag \sqrt{ag (a g+2)}}{Gg^{3}\left(1-ag\right)^2}\left[1+4 \lambda _2\alpha  g^2 +48\lambda _1 \alpha  g^2 \frac{2 a^2 g^2+5 a g+2}{11 a^2 g^2+8 a g-1}\right]\,,
\end{equation}
 nicely matching the expression obtained from the on-shell action, \eqref{SfromI}. This is a very robust check of the validity of our results. 

Before closing this section, let us remark the fact that the correction to the entropy is positive (which is what one would naively expect) if the couplings $\alpha\lambda_1$ and $\alpha\lambda_2$ are positive.\footnote{Strictly speaking, it is positive for $a>a_{\rm{crit}}=\frac{1}{11}\left(-4+3\sqrt{3}\right)\approx 0.1087$, which is something we have assumed when deriving the corrected solution. Note that this is always satisfied for large AdS black holes.} This kind of observations have been previously made in the literature in the context of the weak gravity conjecture \cite{Cremonini:2019wdk}, pointing out an intriguing connection with violations of the Kovtun-Son-Starinets (KSS) bound \cite{Kovtun:2003wp} studied before in \cite{Buchel:2008vz}. There it was shown that the latter occur whenever $\cc-\aa$ is positive.  Using the holographic dictionary, we find that  
\begin{equation}
\frac{\cc-\aa}{\cc}=8\lambda_1\alpha g^2\,,
\end{equation}
hence in our effective theory violations of the KSS bound would occur whenever $\lambda_{1}\alpha >0$. It would be interesting to see if one can constrain the sign of the second parameter $\lambda_2\alpha$ to be positive, so as to be able to say something more rigorous about the positiveness of the corrections to the entropy.

%%%%%%%%%%%%%%%%%%%%%%%%%%%%%%%%%%%%%%%%%%%%%%%%%%%%%%%%%%%%%%%%%%%%%%%%%%%%%%%%%%%%%%
\section{The constrained Legendre transform}\label{constrained_transform}

In this section we match our expression for the entropy by directly evaluating the Legendre transform of the grand-canonical function~\eqref{susy_action_om1om2}, at linear order in ${\rm Tr}\,\mathcal{R}$.\footnote{This section has been added in v2.} As we have discussed, this function can be seen as either the supersymmetric black hole on-shell action at linear order in the four-derivative corrections, or as the terms in the log of the superconformal index on the second sheet that have power-law dependence on $\omega_1,\omega_2$ when these are taken small, cf.~Eq.~\eqref{cardyresult}.

We will show that, interestingly, the procedure of evaluating a constrained Legendre transform, described in Appendix B of~\cite{Cabo-Bizet:2018ehj} for the leading-order term, extends to the corrected expression. 
The advantage of this procedure compared to the one followed in Section~\ref{sec:BPSlimit} is that one can directly reach the final expression for the BPS entropy, with no need to explicitly solve for the relation between the charges and the potentials.

We start by observing that expression~\eqref{susy_action_om1om2} can be rephrased as a homogeneous function of degree 1 in terms of the variables $\omega_1,\omega_2,\varphi$. This is done using the constraint 
\be\label{constraint}
\omega_1 + \omega_2 - 2\varphi = 2\pi i\,,
\ee
to eliminate the factors of $2\pi i$ (including the term $-4\pi^2 = (2\pi i)^2$),
so that
\be
\begin{aligned}\label{expressI}
I 
\,&=\, ({\rm Tr}\mathcal{R}^3-{\rm Tr}\mathcal{R})\, \frac{  \varphi^3}{6\,\omega_1\omega_2}  -  {\rm Tr}\mathcal{R}\,\frac{  \varphi  \left[  - 2\varphi(\omega_1+\omega_2)+ \omega_1^2 + \omega_2^2 +\omega_1\omega_2\right]}{12\,\omega_1\omega_2}\,.  
\end{aligned}
\ee

The entropy is given by the following extremized function:
\be
\mathcal{S} = {\rm ext}_{\{\omega_1,\omega_2,\varphi,\Lambda\}} \left[ -I -\omega_1J_1-\omega_2J_2  -\varphi\, Q_R -\Lambda(\omega_1+\omega_2-2\varphi-2\pi i)\right]\,,
\ee
where the Lagrange multiplier $\Lambda$ implements the linear constraint \eqref{constraint}.\footnote{There is a slightly different way to impose supersymmetry that leads to the replacement $i\to -i$ in the expressions above. The analysis of this other case is completely analogous to the one we are presenting, one just has to pick the opposite sign from the pair of purely imaginary solutions for $\Lambda$ in the final step.} The extremization equations are:
\be\label{derivativesI}
-\frac{\partial{I}}{\partial\omega_1} = J_1 + \Lambda  \,,\qquad
-\frac{\partial{I}}{\partial\omega_2} = J_2 + \Lambda  \,,\qquad
-\frac{\partial{I}}{\partial\varphi} = Q_R -2 \Lambda  \,,
\ee
together with \eqref{constraint}.
It follows that
\be
\begin{aligned}\label{SfromLegTransf}
\mathcal{S} \,&=\, {\rm ext}\left[ -I +\omega_1 \frac{\partial{I}}{\partial\omega_1}+\omega_2 \frac{\partial{I}}{\partial\omega_2}  +\varphi \frac{\partial{I}}{\partial\varphi} + 2\pi i\Lambda\right]\\
\,&=\, {\rm ext} \left[ 2\pi i\Lambda\right]\,,
\end{aligned}
\ee
where to reach the second line it is sufficient to recall Euler's theorem for homogeneous functions. A real entropy is only obtained if the solution for $\Lambda$ is purely imaginary. As we are going to discuss, this means that the equation for $\Lambda$ has to factorize as $(\Lambda^2+ X)({\rm rest})=0$ for some positive $X$. The factorization condition turns out to be equivalent to the non-linear constraint among the charges.

\subsection{Review of the case ${\rm Tr}\mathcal{R}=0$}
It is useful to recall the procedure presented in~\cite[App.~B]{Cabo-Bizet:2018ehj}, valid when one sets ${\rm Tr}\mathcal{R}=0$ in~\eqref{expressI}. For some theories, such as $\mathcal{N}=4$ SYM, ${\rm Tr}\mathcal{R}=0$ is an exact relation, otherwise more generally this is a consequence of considering a holographic SCFT at leading-order in the large-$N$ expansion (or, equivalently, of working at the two derivative level in the gravitational theory). In this case, the equation for $\Lambda$ is obtained by evaluating the derivatives of $I$ explicitly and noting that
\be
\frac{9}{2}\,{\rm Tr }\mathcal{R}^3 \, \frac{\partial{I}}{\partial\omega_1}\frac{\partial{I}}{\partial\omega_2} - \left(\frac{\partial{I}}{\partial\varphi}\right)^3 = 0\,,
\ee
which using \eqref{derivativesI} is equivalent to the cubic
\be
p_0 + p_1 \Lambda + p_2 \Lambda^2 + \Lambda^3 = 0\,,
\ee
with coefficients
\be
\begin{aligned}
p_0 &=  -\frac{1}{8} \left(Q_R^3 + \frac{9}{2} \,{\rm Tr}\mathcal{R}^3  J_1J_2 \right)\,,\\
p_1 &= \frac{1}{4}\left(3Q_R^2 - \frac{9}{4}\,{\rm Tr}\mathcal{R}^3  \, (J_1+J_2)\right)\,,\\
p_2 &= -\frac{3}{2}Q_R - \frac{9}{16}\,{\rm Tr}\mathcal{R}^3  \,.\\
\end{aligned}
\ee
Now one uses the fact that, assuming the microcanonical charges are chosen real, the expression for $\mathcal{S}$ obtained from \eqref{SfromLegTransf} is real only if one imposes the factorization condition 
\be\label{nonl_constraint}
p_0=p_1p_2\,,
\ee 
so that the cubic becomes
\be
( p_1 +\Lambda^2)( p_2 +\Lambda) = 0
\ee
and admits purely imaginary roots $\Lambda = \pm i \sqrt{p_1}$ for $p_1>0$.  Choosing the appropriate sign so that the resulting entropy is positive, one obtains the expression
\be
\begin{aligned}
\mathcal{S} &=  2\pi \sqrt{p_1}\\[2mm]
&=\pi \sqrt{3Q_R^2 - \frac{9}{4}\,{\rm Tr}\mathcal{R}^3  \, (J_1+J_2)}\,.
\end{aligned}
\ee
 One also has that~\eqref{nonl_constraint} is the non-linear constraint between the charges, again at leading order in the expansion.

\subsection{Back to ${\rm Tr}\mathcal{R}\neq 0$}

We now include the first subleading terms in the large-$N$ expansion, namely we work at linear order in ${\rm Tr}\mathcal{R}\neq 0$. Equivalently, we can say we work at linear order in the four-derivative corrections in the gravitational solution.
 We will see that the same method works, however with non-trivial corrections. 
We find that $I$ in \eqref{expressI} satisfies
\begin{equation}
\begin{aligned}\label{eq:constraint1}
&\frac{9}{2}\left(\text{Tr}\mathcal R^3-\text{Tr}\mathcal R \right) \frac{\partial I}{\partial\omega_1}\frac{\partial I}{\partial\omega_2}-\Big(\frac{\partial I}{\partial\varphi}\Big)^3-\frac{3}{2}\,\text{Tr}\mathcal R \left(\frac{\partial I}{\partial\omega_1} + \frac{\partial I}{\partial\omega_2} \right)\frac{\partial I}{\partial\varphi}-\frac{3}{4}\,\text{Tr}\mathcal R\,\Big( \frac{\partial I}{\partial \varphi}\Big)^2   \\[2mm]
&\,\simeq\,\frac{9}{8}\,\text{Tr}\mathcal R^3\,\text{Tr}\mathcal R\, \frac{\left(\frac{\partial I}{\partial\omega_1} - \frac{\partial I}{\partial \omega_2}\right)^2 }{\frac{\partial I}{\partial\varphi}} \,,
\end{aligned}
\end{equation}
where by the symbol $\simeq$ we indicate that the equality holds up to terms of order $({\rm Tr}\mathcal{R})^2$. The same notation is used in the rest of this section.

Using~\eqref{derivativesI}, the equation above can be written as an equation for $\Lambda$,
\begin{equation}
  \frac{p_{-1}}{\Lambda-\frac{1}{2}Q_R} + p_0 + p_1 \Lambda  + p_2 \Lambda^2 + \Lambda^3  \simeq 0\,\,,
\label{eq:lambda1}\end{equation}
where the coefficients are given by
\be
\begin{aligned}
p_{-1} \,&=\, 4\aa \,(\aa-\cc)(J_1-J_2)^2\\[1mm]
p_0 &=  -\frac{1}{8} \left[Q_R^3  + 12 (\cc- \aa) Q_R (Q_R+2(J_1+J_2))+ 16 (3 \cc - 2 \aa) J_1J_2 \right]\,,\\[1mm]
p_1 &= \frac{3}{4}Q_R^2-2\aa\, (J_1+J_2)\,,\\[1mm]
p_2 &= -\frac{3}{2}Q_R-2\aa\,.
\end{aligned}
\ee
 Here, we used the dictionary \eqref{relacTrR} to express the R-symmetry anomaly coefficients $\text{Tr}\mathcal R^3,\,\text{Tr}\mathcal R$ in terms of the conformal anomalies $\aa,\,\cc$, since the resulting expressions are slightly more compact.

\subsubsection{The case $J_1=J_2$}

We first consider the case where the angular momenta are chosen equal, $J_1=J_2\equiv J$.
Then the coefficient $p_{-1}$ appearing in \eqref{eq:lambda1} vanishes and the equation for $\Lambda$ is third order, as in the two-derivative case. Hence we can proceed precisely as in that case, the only difference being in the corrected  expressions for $p_{0,1,2}$. 
The factorization condition  $p_0=p_1p_2$ now can be expressed as the non-linear constraint between the charges
\be
 \left[3 Q_R  + 4\left(2\aa-\cc\right) \right]\left( 3 Q_R^2 -  16\cc \,J\right) \,\simeq\, Q_R^3 + 16 \left(3\cc-2\aa\right){J}^2\ ,
\ee
while the entropy is given by
\be
\begin{aligned}
\mathcal{S} \,&\simeq\,  2\pi \sqrt{p_1} \\[2mm]
&\,=\, \pi \sqrt{3Q_R^2-16\aa\, J}\,,
\end{aligned}
\ee
These match the expressions we found in Subsection~\ref{aeqb_sect}.

\subsubsection{The case $J_1\neq J_2$}

We finally come to the case of general angular momenta, where $p_{-1}$ does not vanish and (\ref{eq:lambda1}) is a quartic equation for $\Lambda$. 
In analogy with the discussion above, we choose the microcanonical charges real and require that there exist two purely imaginary roots of opposite sign, so that a real and positive entropy is obtained. This means that the equation has to factorize as
\begin{equation}
(\Lambda^2 + X) (\Lambda^2 + Y\Lambda + Z) \,\simeq\, 0\,. 
\label{eq:lambda2}\end{equation}
Comparing with (\ref{eq:lambda1}), we can read the coefficients
\begin{equation}
X \,=\, \frac{p_0- \tfrac12 Q_R\, p_1}{p_2- \tfrac12 Q_R}\,\,,\qquad
Y \,=\, p_2 -\tfrac{1}{2}Q_R \,\,,\qquad
Z \,=\, -\tfrac{1}{2}Q_R\,p_2 +\frac{p_1 p_2-p_0}{p_2-\frac{1}{2}Q_R}\,\,,
\label{eq:coefficients}\end{equation}
and find the factorization condition,
\begin{equation}
p_{-1} \left(p_2 - \tfrac12 Q_R \right) - (p_1 p_2-p_0)\left( p_1+ \tfrac14 Q_R^2 \right) + \frac{ (p_1p_2- p_0)^2}{p_2 - \tfrac12 Q_R}\,\simeq\, 0\,.
\label{eq:factorization}\end{equation}
Since $p_{-1} = \mathcal{O}(\aa-\cc)\sim \mathcal{O}({\rm Tr}\,\mathcal{R}) $, we see that the equation is solved demanding that $p_0-p_1 p_2 = \mathcal{O}(\aa-\cc)$ too. Then the last term is higher-order and should be dropped, so at linear order the equation is solved by taking
\begin{equation}\label{nonl_constraint_general}
p_0 \,\simeq\, p_1 p_2 - \frac{p_{-1} \left(p_2 - \tfrac12 Q_R \right)}{p_1+ \tfrac14 Q_R^2} \,.
\end{equation}
Substituting the expressions for the $p$'s, this condition can be written as the constraint
\be
\begin{aligned}
& \left[3 Q_R  + 4\left(2\,\aa-\cc\right) \right]\left[ 3 Q_R^2 -  8\cc\, (J_1+J_2)\right]   \\
&\simeq Q_R^3 + 16 \left(3\cc-2\aa\right)J_1J_2 +\,64\aa\, (\aa-\cc)\frac{(Q_R+\aa)(J_1-J_2)^2}{Q_R^2-2\aa (J_1+J_2)}\,.
\end{aligned}
\ee
The entropy is then given by
\begin{equation}
\begin{aligned}
\mathcal S &= 2\pi\sqrt{X} \,\simeq\, 2\pi\,\sqrt{p_1 -\frac{p_{-1}}{p_1+\frac14 Q_R^2}}\\[2mm]
&\simeq\pi\sqrt{3Q_R^2 -8 \mathtt a \left(J_1 + J_2\right) - 16\, \mathtt a \left(\mathtt{a}-\mathtt c\right) \frac{(J_1-J_2)^2}{Q_R^2-2\,\mathtt a\left(J_1 + J_2\right)}}\,\,,
\end{aligned}
\label{eq:entropy}\end{equation}
where in the second step we have used \eqref{nonl_constraint_general}.
Again we find perfect agreement with the expressions found by first varying the non-supersymmetric on-shell action and subsequently taking the BPS limit, cf. Subsection~\ref{aneqb_sect}.

%%%%%%%%%%%%%%%%%%%%%%%%%%%%%%%%%%%%%%%%%%%%%%%%%%%%%%%%%%%%%%%%%%%%%%%%%%%%%%%%%%%%%%
\section{Discussion}\label{sec:discussion}

In this paper, we have studied the four-derivative corrections to the thermodynamics of asymptotically AdS black hole solutions to five-dimensional minimal gauged supergravity. We have computed the on-shell action at linear order in the corrections. Then, after restricting to supersymmetric configurations using the approach of~\cite{Cabo-Bizet:2018ehj}, we have showed that the action can be expressed in terms of the supersymmetric chemical potentials $\omega_1,\omega_2$ and matches the prediction from the Cardy-like limit of the dual superconformal index on the second sheet, reported in \eqref{cardyresult}. The corrections to the (non-supersymmetric) thermodynamics, including the corrected entropy and conserved charges, have been derived using the identification of the on-shell action with the logarithm of the grand-canonical partition function and taking its variations with respect to the potentials. We also gave the expression for the microcanonical BPS entropy, that is the supersymmetric and extremal entropy as a function of the charges,  at linear order in the corrections. When $J_1=J_2$, it takes the same functional form as at the two derivative level, with the corrections being entirely incorporated in the anomaly coefficients of the dual superconformal field theory. However for $J_1\neq J_2$ the correction involves a new term depending on the charges. 
 
 We provided a very direct derivation of the BPS entropy in the general case by evaluating the Legendre transform of the supersymmetric on-shell action. Finally, in the $J_1=J_2$ case we have confirmed our result for the BPS entropy by computing the corrected near-horizon geometry of the BPS black hole of \cite{Gutowski:2004ez} and evaluating Wald's formula. 

It is interesting to discuss how our effective field theory can arise from top-down constructions accounting for different stringy and/or quantum effects. In particular, one can focus on the Chern-Simons terms, which directly capture the central charges $\aa,\cc$ of the dual field theory. The higher-dimensional origin of the $\epsilon^{\mu\nu\rho\sigma\lambda} {\rm} R_{\mu\nu}{}^{\alpha\beta} R_{\rho\sigma\alpha\beta}A_\lambda$ term for type IIB compactifications on Sasaki-Einstein manifolds was studied e.g.\ in~\cite{Liu:2010gz,ArabiArdehali:2013jiu,ArabiArdehali:2013vyp}. Even when the coefficient $\lambda_1\sim (\aa-\cc)$ of this term vanishes exactly---as for type IIB on $S^5$---there are still corrections to $\epsilon^{\mu\nu\rho\sigma\lambda} F_{\mu\nu}  F_{\rho\sigma}A_\lambda$ controlled by $\lambda_2$,  the other parameter  in our effective action. These arise from quantum effects in the Kaluza-Klein towers. In the case of type IIB on $S^5$, they yield a shift $N^2\to N^2-1$ in the expression for $\aa=\cc$\ \cite{Bilal:1999ph}.

A different type of corrections, arising from $\alpha'{}^3$ eight-derivative terms in type IIB on $S^5$, has been analyzed in~\cite{Melo:2020amq}. These do not generate the terms in our effective action. They encode corrections in the 't Hooft coupling and thus cannot contribute to the holographic description of the SCFT index, which is independent of continuous parameters. In fact, the authors of~\cite{Melo:2020amq} evaluated these corrections on the black hole solution of~\cite{Chong:2005hr}, finding that they yield a vanishing contribution to the on-shell action when supersymmetry is imposed.

It will be interesting to extend our study to the case where vector multiplets of five-dimensional supergravity are included, so as to incorporate multiple electric charges in the discussion. 

It would also be intriguing to study how the four-derivative corrections affect the near-BPS thermodynamics that has recently been studied, from different viewpoints, in \cite{Boruch:2022tno,Castro:2018ffi,Larsen:2019oll}. In particular, one could extend the Schwarzian theory of \cite{Boruch:2022tno} and determine the contributions of our $\lambda_1,\lambda_2$ terms to the gap found there.

%It is interesting to observe that the supersymmetry invariants contributing to the on-shell action are supersymmetry completions of Chern-Simons invariants, as in the 3d effective approach on the SCFT side. Maybe there are no other terms contributing, even when more than four derivatives are considered. 

We use the rest of the present section to discuss two subtle open issues of the analysis.

In order to obtain our effective action we have used the standard Weyl multiplet formulation of ${\cal N}=2$ off-shell supergravity. Working with this formulation has the limitation that only two out of three four-derivative supersymmetric invariants are explicitly known in components. 
 Nevertheless, we have given an argument indicating that our two-parameter effective action is in fact the most general one compatible with supersymmetry. This is based on imposing the vanishing of the Gibbs free energy ${\cal G}=I/\beta$ for the BPS black hole of~\cite{Gutowski:2004ez}, which gives enough conditions to conclude that the third invariant would not give  contributions of a new form.
One way to test this conclusion would have been to use the alternative dilaton Weyl multiplet formulation of  supergravity \cite{Fujita:2001kv,Bergshoeff:2001hc} (see also \cite{Coomans:2012cf}), as recently done in \cite{Liu:2022sew}. Indeed in this formulation the three supersymmetric invariants were given in components in \cite{Ozkan:2013nwa}. There are however a number of technical issues why we have decided not to do so in this paper. The first and more important is that the off-shell four-derivative invariants in \cite{Ozkan:2013nwa} were obtained for the ungauged theory, $g=0$. Then, it is reasonable to expect there are ${\cal O}(\alpha g^2)$ corrections missing, which  as we have seen play a key r\^ole in the story. In spite of this issue, what one can check is whether the combination of four-derivative terms (which does not depend on $g$) that one obtains in the dilaton Weyl multiplet is the same as in \eqref{eq:4daction2}. This question was answered in \cite{Liu:2022sew} in the affirmative and we have also made an independent check of their findings.   A second issue is that, when the auxiliary fields are integrated out, one does not arrive directly to the minimal supergravity theory. This remains coupled to an additional vector multiplet, and a further truncation is needed. While the corrected values of the fields in the additional vector multiplet are not needed, we do not see an immediate way to argue that the truncation is still consistent at $\Oa$, although the results of \cite{Liu:2022sew} would indicate that this is the case.\footnote{It should however be noted that the Ricci scalar squared invariant was actually treated in the standard Weyl formalism there.}

Another issue that should be clarified regards the BPS limit of the corrected charges. The BPS limit we have taken in Section~\ref{sec:BPSlimit} is the same as the one discussed in \cite{Silva:2006xv}, while it differs from the one that reaches the BPS point along a supersymmetric trajectory in parameter space. As discussed in \cite{Cabo-Bizet:2018ehj}, the different limiting procedures lead to different expressions for the BPS on-shell action and chemical potentials $\omega_1,\omega_2$. However, the two limits should lead to the same BPS values of the charges and the entropy. While this can be verified at the two-derivative level, we have not been able to prove so for the corrected charges. Perhaps revisiting the entropy function formalism \cite{Sen:2005wa,Sen:2008vm} in the present context will shed light on this question. We leave the clarification of this point to future work.

%%%%%%%%%%%%%%%%%%%%%%%%%%%%%%%%%%%%%%%%%%%%%%%%%%%%%%%%%%%%%%%%%%%%%%%%%%%%%%%%%%%%%%
\section*{Acknowledgments}
DC would like to thank the organizers and participants in the workshop ``Crossing Horizons'' (King's College London, 23 May - 1 June 2022) for many interesting discussions on related topics.
AR is supported by a postdoctoral fellowship associated to the MIUR-PRIN contract 2017CC72MK003.

%%%%%%%%%%%%%%%%%%%%%%%%%%%%%%%%%%%%%%%%%%%%%%%%%%%%%%%%%%%%%%%%%%%%%%%%%%%%%%%%%%%%%%
%%%%%%%%%%%%%%%%%%%%%%%%%%%%%%%%%%%%%%%%%%%%%%%%%%%%%%%%%%%%%%%%%%%%%%%%%%%%%%%%%%%%%%%%%%%%
\appendix

%%%%%%%%%%%%%%%%%%%%%%%%%%%%%%%%%%%%%%%%%%%%%%%%%%%%%%%%%%%%%%%%%%%%%%%%%%%%%%%%%%%%%%%%%%%%
\section{Dictionary with superconformal anomalies}\label{sec:holographicdictionary}

In this appendix, we derive the dictionary between the dimensionless gravitational quantities $g^3 G$ and $\alpha g^2$ and the dual superconformal anomaly coefficients.
We stress that this dictionary is universal, namely it is valid for any holographic $\mathcal{N}=1$ SCFT; the details of the SCFT only affect the explicit expression of the anomaly coefficients in terms of the field theory data.

$\mathcal{N}=1$ SCFT's have a superconformal anomaly controlled by coefficients $\aa,\cc$. This shows up in the trace of the energy-momentum tensor $T_{ij}$ and in the divergence of the R-current $J^i$ as \cite{Anselmi:1997am,Cassani:2013dba}
\begin{align}\label{correcttrace}
 T_i{}^i  \, & =\, \,- \,
\frac{\aa}{16\pi^2}  \hat E + \frac{\cc}{16\pi^2}  \hat{C}^2    \, - \,  \frac{\cc}{6\pi^2}\,   \hat{F}^2  ~, \\[2mm]
\nabla_i  J^i \, & =\, \frac{\cc-\aa}{24\pi^2}\, \frac{1}{2}\,\epsilon^{ijkl}\hat{R}_{ijab} \hat{R}_{kl}{}^{ab}\, + \,\frac{5\aa-3\cc}{27\pi^2} \,\frac{1}{2}\,\epsilon^{ijkl}\hat{F}_{ij} \hat{F}_{kl}\,,
 \label{correctchiral}
 \end{align}
where $\hat C^2$ and $\hat E = \hat R_{ijkl}\hat R^{ijkl}-4 \hat R_{ij}\hat R^{ij}+\hat R^2$ denote the Weyl$^2$ and Euler invariants of the four-dimensional background geometry, respectively, while $\hat F_{ij}$ is the field strength of the background U(1) gauge field that canonically couples to the R-current.

Comparing with the general expression for the anomaly of a U(1) current with cubic and linear coefficients denoted by ${\rm Tr}\mathcal{R}^3$ and ${\rm Tr}\mathcal{R}$, respectively, 
\be
\nabla_i  J^i  \, =\,-\frac{{\rm Tr}\mathcal{R}}{384\pi^2}  \, \frac{1}{2}\,\epsilon^{ijkl}\hat{R}_{ijab} \hat{R}_{kl}{}^{ab}  + \frac{ {\rm Tr}\mathcal{R}^3 }{48\pi^2}\, \,\frac{1}{2}\,\epsilon^{ijkl}\,\hat{F}_{ij} \hat{F}_{kl}\,,
\ee
one finds the relations
\be\label{relacTrR}
\aa = \frac{3}{32}(3\,{\rm Tr}\mathcal{R}^3 - {\rm Tr}\mathcal{R})\,, \qquad \cc = \frac{1}{32}(9\,{\rm Tr}\mathcal{R}^3 - 5\, {\rm Tr}\mathcal{R})\,.
\ee

In the large-$N$ expansion, these anomaly coefficients are related to the dimensionless gravitational quantities $g^3 G$ and $\alpha g^2$. Let us determine the precise dictionary in our setup. We proceed as in~\cite{Cremonini:2008tw} and use results of \cite{Fukuma:2001uf} (see also~\cite{Baggio:2014hua,Bobev:2021qxx}). 

The Weyl anomaly can be read off from the logarithmically divergent term in the on-shell action~\cite{Henningson:1998gx}. 
It is sufficient to switch off the gauge field and only consider the gravitational part of the bulk action, which in general may take the form
\be
e^{-1} \mathcal{L} \,=\, \frac{1}{16\pi G_{\rm{eff}}} \left(R+12g_{\rm eff}^2 +\alpha_1 R^2 + \alpha_2 R_{\mu\nu}R^{\mu\nu} + \alpha_3 R_{\mu\nu\rho\sigma}R^{\mu\nu\rho\sigma} \right)\,,
\ee
where, importantly, $G_{\rm eff}$ and $g_{\rm eff}$ may also contain terms of order $\alpha$.
It is easy to check that the radius $\ell$ of the AdS$_5$ solution is given by
%\be
%g_{\rm eff}\, =\, \frac{1}{\ell}\left[1-\frac{1}{3\ell^2}(10\alpha_1+2\alpha_2+\alpha_3) \right]\,,
%\ee
\be
\ell \,=\, \frac{1}{g_{\rm eff}} \left[1- \frac{g_{\rm eff}^2}{3} (10\alpha_1+2\alpha_2+\alpha_3) \right]\,.
\ee
This equality, as well as the following ones, is meant to hold at linear order in the $\alpha$ corrections.
Identifying the logarithmically divergent piece of the action evaluated on a general solution with the conformal anomaly of the dual SCFT, one obtains the corrected holographic formulae for the anomaly coefficients,
\begin{equation}
\begin{aligned}
\aa \,&=\, \frac{\pi\ell^3}{8G_{\rm{eff}}} \left[ 1-\frac{4}{\ell^2} (10\alpha_1+2\alpha_2+\alpha_3) \right] \,,\\[1mm]
\cc \,&=\, \frac{\pi\ell^3}{8G_{\rm{eff}}} \left[ 1-\frac{4}{\ell^2} (10\alpha_1+2\alpha_2-\alpha_3) \right] \,.
\end{aligned}
\end{equation}

In the main text, we focus on the supersymmetric action~\eqref{eq:4daction2}, where the only purely gravitational four-derivative term is the Gauss-Bonnet one. This fixes 
\be
\alpha_1 =\alpha\lambda_1\,,\quad \alpha_2 =-4\alpha\lambda_1\,,\quad \alpha_3 =\alpha\lambda_1\,,
\ee
while from the Ricci scalar and cosmological constant term we read 
\be
\begin{aligned}
G_{\rm eff}=\,&G\left(1-4\lambda_2 \alpha g^2\right)\,,\\[1mm]
g_{\rm eff} =\,& g\left(1-5\lambda_1\alpha g^2\right)\,.
\end{aligned}
\ee

Plugging these values in the above expressions, we obtain the AdS radius
%\be
%g_{\rm eff}\, =\, \frac{1}{\ell}\left(1-\frac{\alpha}{\ell^2} \right)\,,
%\ee
\be
\ell \,=\, \frac{1}{g} \left(1+ 4\lambda_1\alpha g^2 \right)\,,
\ee
and the anomaly coefficients
\begin{equation}\label{a_c_high_der}
\begin{aligned}
\aa \,&=\, \frac{\pi}{8Gg^3} \left( 1 +4\lambda_2 \alpha g^2  \right)\,,\\[1mm]
\cc \,&=\, \frac{\pi}{8Gg^3} \left( 1 +4(2\lambda_1+\lambda_2)\alpha g^2 \right) \,.
\end{aligned}
\end{equation}
Note that at leading order this matches the familiar two-derivative result $\aa = \cc = \frac{\pi}{8G g^3 }$.

Inverting \eqref{relacTrR}, we can also give the holographic expression of the R-symmetry anomaly coefficients:
\be
\begin{aligned}\label{TrRintermsofmu}
{\rm Tr}\,\mathcal{R}^3 \,&=\, \frac{16}{9}(5\aa-3\cc)\,=\, \frac{4\pi}{9G g^3} \left( 1 -4(3\lambda_1-\lambda_2) \alpha g^2  \right)\,, \\[1mm]
{\rm Tr}\,\mathcal{R} \,&=\, 16(\aa-\cc)\,=\, -\frac{16\pi \alpha\lambda_1}{G g}   \,.
\end{aligned}
\ee

We can check the dictionary above by matching the non-invariance of the bulk action~\eqref{eq:4daction2} under a U(1)$_R$ gauge transformation with the R-current anomaly~\cite{Witten:1998qj}. One can see that the boundary value $A_i$, $i=0,\ldots,3$, of the bulk gauge field is to be identified with the background gauge field $\hat A_i$ coupling canonically to the R-current of the dual SCFT as $A_i = \frac{2}{\sqrt{3}\,g}\,\hat A_i$. Making the gauge transformation $\delta A_i = \frac{2}{\sqrt{3}g}\,\partial_i \lambda$, the bulk action~\eqref{eq:4daction2} transforms as
\begin{equation}
\begin{aligned}
\delta_\lambda S
\,=&\,\frac{1}{24 \pi g^3 G}\int_{\partial {\cal M}} \diff^4x \,\hat{e}\, \lambda \left( -\frac{\tilde{c}_3}{9}\,\hat{\epsilon}^{\,ijkl}\hat{F}_{ij}\hat{F}_{kl} \,-\, \frac{\alpha\lambda_1 g^2}{2}\,\hat{\epsilon}^{\,ijkl}\hat{R}_{ijab}\hat{R}_{kl}{}^{ab}\right)\,,%\\[1mm]
%\,=&\,\frac{1}{24 \pi g^3 G}\int_{\partial {\cal M}}  \lambda \left( \frac{4}{9}\tilde{c}_3\,\hat{F}\wedge\hat{F} \,+\, 2\alpha g^2\,\hat{R}_{ab}\wedge\hat{R}^{ab}\right)\, ,
\end{aligned}
\end{equation}
where the hat symbol denotes that the quantities are evaluated at the boundary.

Through the identification $\rme^{i S}=Z_{\rm grav} = Z_{\rm CFT}$, this should match the anomalous variation of the field theory partition function $Z$ under the background gauge transformation $\delta_\lambda\hat A_i =\partial_i\lambda\,$,% (we assume there is no contribution from the SCFT dynamical fields to the anomaly), 
\begin{equation}
\begin{aligned}
\delta_\lambda \log Z_{\rm CFT} \,&=\, -i  \int_{\partial {\cal M}} \diff^4 x\, \hat{e} \, \lambda\, \nabla_i J^i \\[1mm]
\,&=\, -i \int_{\partial {\cal M}} \diff^4 x\, \hat{e} \, \lambda \left( \frac{5\aa-3\cc}{27\pi^2} \,\frac{1}{2}\,\epsilon^{ijkl}\hat{F}_{ij} \hat{F}_{kl} \, + \,  \frac{\cc-\aa}{24\pi^2}\, \frac{1}{2}\,\epsilon^{ijkl}\hat{R}_{ijab} \hat{R}_{kl}{}^{ab}\right)\,.%\\
%\,&=\, -i  \int_{\rm bdy} \diff^4 x\, \hat{e} \, \lambda\, \left( -\frac{{\rm Tr}\mathcal{R}}{384\pi^2}  \, \frac{1}{2}\,\epsilon^{ijkl}\hat{R}_{ijab} \hat{R}_{kl}{}^{ab}  + \frac{ {\rm Tr}\mathcal{R}^3 }{48\pi^2}\, \,\frac{1}{2}\,\epsilon^{ijkl}\,\hat{F}_{ij} \hat{F}_{kl}\right)\,,
\end{aligned}
\end{equation}
Imposing the holographic matching condition $\delta S= -i\, \delta\log Z_{\rm CFT}$ and recalling the value of $\tilde c_3$ given in \eqref{ctildes}, we obtain the following $\mathcal{O}(\alpha)$ dictionary between the coefficients:
\be
5\aa-3\cc \,=\, \frac{\pi \tilde{c}_3}{4 Gg^3} = \frac{\pi}{4 Gg^3}\left( 1-4(3\lambda_1-\lambda_2)\alpha g^2\right)\,,\qquad \aa-\cc \,=\, -\frac{\pi \alpha \lambda_1}{ Gg}\,,
\ee
which perfectly agrees with the dictionary found by considering the Weyl anomaly. This can be seen as a consistency check for the supersymmetry of our action~\eqref{eq:4daction2}.

\section{Corrected entropy, electric charge and angular momentum for $a=b$}\label{app:non_susy_charges}

We give here the expressions for the corrected non-supersymmetric entropy, electric charge and angular momentum that we determined from the thermodynamics starting from the on-shell action with equal rotational parameters, given in~\eqref{onshellactionCCLP}. The energy can be deduced from the quantum statistical relation.  Here we set $g=1$ to make the expressions a little simpler.

For the entropy we find:
\begin{equation}
{\cal S}=\,\frac{\pi ^2 \left[r_+^4+a^4+a^2 \left(q+2 r_+^2\right)\right] }{2Gr_+\left(1-a^2\right)^2}\left(1+4 \lambda _2\alpha\right)+\lambda_1\alpha \Delta {\cal S}\,,
\end{equation}
with
{\small\begin{equation}
\begin{aligned}
\Delta {\cal S}\!&=\,F_{S}\left\{a^4 \left(a^2+q\right)^6 \left(a^2+2 q\right)+a^4 \left(a^2+q\right)^4 \left(2 a^6-5 a^2 q+3 a^4 q-3q^2\right) r_+^2\right.\\[1mm]
&+a^2\left(a^2+q\right)^3 \left[-7 a^8+a^{10}-38 a^2 q^2-4 q^3-2 a^4 q (37+8 q)-2 a^6 (14+13q)\right] r_+^4\\[1mm]
&\hspace{-3mm}-a^2 \left(a^2+q\right)^2 \left[14 a^{10}+13 q^3+a^2 q^2 (104+5 q)+3 a^8 (32+7 q)+16 a^6 (7+13q)+a^4 q (211+108 q)\right] r_+^6\\[1mm]
&-a^2 \left(a^2+q\right) \left[108 a^{10}+7 a^{12}+q^3 (23+2 q)+5 a^2 q^2 (31+4
   q)+4 a^8 (77+38 q)+6 a^4 q (55+53 q)\right.\\[1mm]
&\left.+a^6 \left(210+593q+34 q^2\right)\right] r_+^8-a^2 \left[40 a^{12}+2 (11-4 q) q^3+2a^{10} (140+9 q)+a^2 q^2 (155+48 q)\right.\\[1mm]
&\left.+a^8 (476+451 q)+a^4 q (347-2 (-260+q) q)+a^6 (224+q (920+153 q))\right]
   r_+^{10}-a^2 \left[84 a^{10}\right.\\[1mm]
&\left.+a^8 (322-15 q)+(14-23 q) q^2+14 a^6 (27+20 q)+a^4 (140+(409-55 q) q)\right.\\[1mm]
&\left.+a^2 q (110+(65-17
   q) q)\right] r_+^{12}+\left[-56 a^{10}+3 q^3+6 a^4 (-2+3 q) (4+7 q)+4 a^8 (-21+25 q)\right.\\[1mm]
&\left.+a^6 (-112+73 q)+a^2 q (-13+q(28+3 q))\right] r_+^{14}+\left[70 a^8+3 q^2+44 a^4 (1+4 q)+a^6 (196+135 q)\right.\\[1mm]
&\left.+a^2 \left(-7+45 q+48 q^2\right)\right]
   r_+^{16}+\left[168 a^6+3 q (3+q)+a^2 (42+75 q)+a^4 (232+78 q)\right] r_+^{18}\\[1mm]
&\left.+\left[140 a^4+9 (1+q)+a^2 (105+17
   q)\right] r_+^{20}+2 \left(9+28 a^2\right) r_+^{22}+9 r_+^{24}\right\}\,,
\end{aligned}
\end{equation}}
where 
\begin{equation}
{F}_{\cal S}=-\frac{2\pi^2}{Gr_+^3\left(1-a^2\right)^2\left(a^2+r_+^2\right){}^2{\cal D}}\,,
\end{equation}
and
\begin{equation}
\begin{aligned}
{\cal D}=\,&a^2 \left(a^2+q\right)^3\left(a^2+2 q\right)+a^2 \left(a^2+q\right)^2 \left(5 a^2+a^4+6 q\right) r_+^2\\[1mm]
&\hspace{-8mm}+\left(a^2+q\right) \left[3 a^6+8 a^2q-q^2+a^4 (10+q)\right] r_+^4-\left[-10 a^4-2 a^6+2 a^8+a^2 (-8+q) q+q^2\right] r_+^6\\[1mm]
& -\left[2 a^4+8 a^6+(-1+q)
   q+a^2 (-5+4 q)\right] r_+^8-\left(-1+3 a^2+12 a^4+2 q\right) r_+^{10}\\[1mm]
&-\left(1+8 a^2\right) r_+^{12}-2
   r_+^{14}\,.
\end{aligned}
\end{equation}

The electric charge is given by:
\begin{equation}
Q =\,\frac{\sqrt{3} \pi q}{4 G\left(1-a^2\right)^2 }\left(1+4  \lambda _2\alpha\right)+\lambda_1\alpha  \Delta Q\,,
\end{equation}
with 
{\small\begin{equation}
\begin{aligned}
\Delta Q\!&= F_{Q}\left\{a^4 q^8+2 a^2 \left(a^2-2 r_+^2\right) \left(1+r_+^2\right){}^3 \left(a^2+r_+^2\right){}^7 \left(a^2+r_+^2-2
   r_+^4\right)+a^2 q^7 \left(9 a^4+2 a^2 r_+^2+r_+^4\right)\right.\\[1mm]
&+q^4 \left(a^2+r_+^2\right){}^2 \left[3 r_+^{10}+5 a^8
   \left(21+14 r_+^2+r_+^4\right)-a^4 r_+^4 \left(87+193 r_+^2+3 r_+^4\right)\right.\\[1mm]
&\left.+a^2 r_+^6 \left(-22-3 r_+^2+10r_+^4\right)-a^6 r_+^2 \left(40+213 r_+^2+24 r_+^4\right)\right]+q^6 \left[-3 r_+^8+a^6 r_+^2 \left(19-22r_+^2\right)\right.\\[1mm]
&\hspace{-5mm}\left.-a^2 r_+^6 \left(9+2 r_+^2\right)+a^8 \left(35+4 r_+^2\right)-6 a^4 \left(r_+^4+2r_+^6\right)\right]+q^2 \left(a^2+r_+^2\right){}^4 \left[a^8 \left(1+r_+^2\right) \left(49+31 r_+^2+2r_+^4\right)\right.\\[1mm]
&\left.+3 r_+^8 \left(1+3 r_+^2+8 r_+^4\right)-a^4 r_+^4 \left(70+273 r_+^2+151 r_+^4+2 r_+^6\right)-a^6 r_+^2
   \left(29+125 r_+^2+84 r_+^4+6 r_+^6\right)\right.\\[1mm]
&\left.+a^2 r_+^6 \left(11-27 r_+^2+54 r_+^4+38 r_+^6\right)\right]+q^5
   \left(a^2+r_+^2\right) \left[-107 a^6 r_+^4+a^2 r_+^6 \left(-32+3 r_+^2\right)\right.\\[1mm]
&\left.-21 a^4 r_+^4 \left(2+3r_+^2\right)+a^8 \left(77+26 r_+^2\right)-3 \left(r_+^8+r_+^{10}\right)\right]\\[1mm]
&+q \left(1+r_+^2\right)
   \left(a^2+r_+^2\right){}^6 \left[a^6 \left(1+r_+^2\right) \left(15+4 r_+^2\right)+3 a^2 r_+^4 \left(-3-9 r_+^2+2r_+^4\right)\right.\\[1mm]
&\left.-a^4 r_+^2 \left(21+49 r_+^2+22 r_+^4\right)+3 \left(r_+^6+3 r_+^8+8 r_+^{10}\right)\right]+q^3\left(a^2+r_+^2\right){}^3 \left[a^8 \left(91+100 r_+^2+21 r_+^4\right)\right.\\[1mm]
&\left.+a^2 r_+^6 \left(13-15 r_+^2+44
   r_+^4\right)-a^4 r_+^4 \left(99+323 r_+^2+68 r_+^4\right)-a^6 r_+^2 \left(53+221 r_+^2+72 r_+^4\right)\right.\\[1mm]
&\left.\left.+3
   \left(r_+^{10}+r_+^{12}\right)\right]\right\}\, .
\end{aligned}
\end{equation}}
where
\begin{equation}
F_{Q}=-\frac{\pi }{\sqrt{3}G(1-a^2)^2\,r_{+}^4\left(a^2+r_{+}^2\right)^3{\cal D}}\, .
\end{equation}

The angular momentum reads:
\begin{equation}
J=\frac{a \pi \left(1+4 \lambda_2\alpha \right)}{4 \left(1-a^2\right)^3 G r_+^2}\left[\left(a^2+q\right)^2+\left(a^4+q+a^2 (2+q)\right) r_+^2+\left(1+2a^2\right) r_+^4+r_+^6\right] +\lambda_1\alpha \Delta J\,,
\end{equation}
with
{\small\begin{equation}
\begin{aligned}
\Delta J\!&=F_{J}\left\{-2 a^2 q^8+q^6 \left[-49 a^6+a^4 \left(33+7 a^2\right) r_+^2+51 a^2 \left(1+a^2\right) r_+^4+\left(9+7 a^2\right) r_+^6+3r_+^8\right]\right.\\[1mm]
&+q^7 \left[4 r_+^4+a^4 \left(-15+r_+^2\right)+a^2 r_+^2 \left(5+r_+^2\right)\right]-\left(1+r_+^2\right){}^2\left(a^2+r_+^2\right){}^7 \left[a^4 \left(1-3 r_+^2\right)\right.\\
&\hspace{-5mm}\left.-2 a^2 r_+^2 \left(4+19 r_+^2+7 r_+^4\right)+r_+^4\left(7+13 r_+^2+18 r_+^4\right)\right]+q^5 \left[-2 r_+^8 \left(-5+r_+^2\right)+a^4 r_+^4 \left(233+281 r_+^2-4 r_+^4\right)\right.\\
&\left.+a^2 r_+^6 \left(93+47 r_+^2-2
   r_+^4\right)+a^8 \left(-91+21 r_+^2+8 r_+^4\right)+a^6 r_+^2 \left(91+317 r_+^2+38 r_+^4\right)\right]\\
&+q^4 \left[r_+^{10} \left(-1+r_+^2\right)+a^2 r_+^8 \left(71+103 r_+^2-66 r_+^4\right)+2 a^4 r_+^6 \left(181+361 r_+^2-14r_+^4\right)\right.\\[1mm]
&\left.+a^{10} \left(-105+35 r_+^2+34 r_+^4\right)+5 a^8 r_+^2 \left(27+169 r_+^2+60 r_+^4\right)+2 a^6 r_+^4\left(265+715 r_+^2+152 r_+^4\right)\right]\\[1mm]
&+q \left(1+r_+^2\right) \left(a^2+r_+^2\right){}^5 \left[a^6 \left(-9+16 r_+^2+9 r_+^4\right)-2 r_+^6 \left(3+17 r_+^2+26r_+^4\right)\right.\\[1mm]
&\left.+2 a^4 r_+^2 \left(29+140 r_+^2+90 r_+^4+7 r_+^6\right)-a^2 r_+^4 \left(19-22 r_+^2+57 r_+^4+34r_+^6\right)\right]\\[1mm]
&+q^2 \left(a^2+r_+^2\right){}^3 \left[8 a^4 r_+^4 \left(13+82 r_+^2+58 r_+^4-2 r_+^6\right)+a^8 \left(-35+21 r_+^2+53r_+^4+9 r_+^6\right)\right.\\[1mm]
&\hspace{-10mm}\left.-r_+^8 \left(1+25 r_+^2+57 r_+^4+21 r_+^6\right)-4 a^2 r_+^6 \left(7+4 r_+^2+46 r_+^4+37r_+^6\right)+4 a^6 r_+^2 \left(40+215 r_+^2+209 r_+^4+46 r_+^6\right)\right]\\[1mm]
&+q^3 \left(a^2+r_+^2\right){}^2 \left[-2 r_+^8 \left(4+3 r_+^2+23 r_+^4\right)+a^4 r_+^4 \left(209+791 r_+^2+231 r_+^4-39r_+^6\right)\right.\\[1mm]
&\left.+5 a^2 r_+^6 \left(3+17 r_+^2-31 r_+^4-5 r_+^6\right)+a^8 \left(-77+35 r_+^2+59 r_+^4+3 r_+^6\right)\right.\\[1mm]
&\left.\left.+a^6r_+^2 \left(269+1119 r_+^2+687 r_+^4+53 r_+^6\right)\right]\right\}\,,
\end{aligned}
\end{equation}}
where
\begin{equation}
{F}_{J}=\frac{a\pi}{2G(1-a^2)^3(a^2+r_+^2)^2r_+^4 {\cal D}}\,.
\end{equation}

\section{Equations of motion}\label{app:eoms}

In this appendix we derive the equations of motion from the higher-derivative action. We start with a general discussion, that may also be useful in other contexts, for a higher-derivative action constructed out of the metric and a gauge field, also allowing for Chern-Simons terms. Then we specialize to the action \eqref{eq:4daction2} studied in the main text.

Let us consider the following general action (where we set $16\pi G=1$ for simplicity), 
\begin{equation}
S=\int \diff^5x \,e\, {\mathcal L}'\left(R_{\mu\nu\rho\sigma}, F_{\mu\nu}\right)+S_{\rm{CS}}\, ,
\end{equation}
where ${\mathcal L}'\left(R_{\mu\nu\rho\sigma}, F_{\mu\nu}\right)$ denotes a Lagrangian constructed out of arbitrary contractions (via the inverse metric $g^{\mu\nu}$) of the Riemann tensor $R_{\mu\nu\rho\sigma}$ and the field strength $F_{\mu\nu}$, while
\begin{equation}
S_{\rm{CS}}=\int \diff^5x \, e\, \left[-\frac{\tilde c_3}{12\sqrt{3}}\epsilon^{\mu\nu\rho\sigma\lambda}F_{\mu\nu}F_{\rho\sigma}A_{\lambda}-\frac{\lambda_1\,\alpha}{2\sqrt{3}}\epsilon^{\mu\nu\rho\sigma\lambda}R_{\mu\nu\alpha\beta}R_{\rho\sigma}{}^{\alpha\beta}A_{\lambda}\right]\, ,
\end{equation}
are the Chern-Simons terms, which will be treated separately for convenience. Here the choice of the coefficients $\tilde{c}_3,\lambda_1$ reflects the one in the main text. The variation of the Lagrangian with respect to the inverse metric $g^{\mu\nu}$ and the gauge field $A_{\mu}$~is
\begin{equation}
\begin{aligned}
\delta S=\,&\int \diff^5x\, e\, \left [\left(\frac{\partial {\cal L}'}{\partial g^{\mu\nu}}-\frac{1}{2}g_{\mu\nu}{\cal L}'\right)\delta g^{\mu\nu}+P^{\mu\nu\rho\sigma}\delta R_{\mu\nu\rho\sigma}-2\nabla_{\mu}\left(\frac{\partial {\cal L}'}{\partial F_{\mu\nu}}\right)\delta A_{\nu}\right. \\[1mm]
&+\nabla_\mu \Xi^\mu\bigg]+\delta S_{\rm{CS}}\, ,
\end{aligned}
\end{equation}
where $\Xi^\mu$ is a boundary term
\begin{equation}
\Xi^\mu=2\frac{\partial {\cal L}'}{\partial F_{\mu\nu}}{\delta A}_{\nu}\, ,
\end{equation}
and the tensor $P^{\mu\nu\rho\sigma}$ is defined as
\begin{equation}
P^{\mu\nu\rho\sigma}=\frac{\partial {\cal L}'}{\partial R_{\mu\nu\rho\sigma}}\, ,
\end{equation}
assuming it inherits the following symmetries of the Riemann tensor
\begin{equation}
P_{\mu\nu\rho\sigma}=-P_{\nu\mu\rho\sigma}\, ,\hspace{0.75cm} P_{\mu\nu\rho\sigma}=-P_{\mu\nu\sigma\rho}\, ,\hspace{0.75cm}P_{\mu\nu\rho\sigma}=P_{\rho\sigma\mu\nu}\, .
\end{equation}
Recalling the well-known variation of the Riemann tensor
\be
\delta R^\mu{}_{\nu \rho\sigma} = 2 \nabla_{[\rho} \,\delta \Gamma^\mu_{\sigma]\nu}\,,
\ee
\be
\delta \Gamma^\mu_{\sigma\nu} = \frac{1}{2}g^{\mu\kappa}\left(\nabla_\sigma \delta g_{\kappa\nu} + \nabla_\nu \delta g_{\kappa\sigma} - \nabla_{\kappa}\delta g_{\sigma\nu} \right)\,,
\ee
and performing integrations by parts, we obtain that 
\begin{align}
\int \!\diff^5 x\,e\, P^{\mu\nu\rho\sigma}\, \delta R_{\mu\nu\rho\sigma}\,& =\, \int \!\diff^5 x\,e \left[\left(-P^{\alpha\beta\gamma}{}_{\mu}R_{\alpha\beta\gamma \nu}-2\nabla^{\alpha}\nabla^{\beta} P_{\beta \mu\nu\alpha}\right) \delta g^{\mu\nu} + \nabla_{\nu} \Theta^\nu \right]\,,
\end{align}
where the boundary term reads
\begin{align}\label{bdry_term_metric_var}
\Theta^\nu \,&=\, 2 P_\mu{}^{\lambda\nu\sigma}\,  \delta \Gamma^\mu_{\lambda\sigma} -  2\nabla_{\lambda} P^{\rho\nu\lambda\sigma} \delta g_{\rho\sigma}\nn\\[1mm]
&=\,  2P^{\rho\lambda\nu\sigma}\,   \nabla_\lambda \delta g_{\rho\sigma}    -  2\nabla_{\lambda} P^{\rho\nu\lambda\sigma} \delta g_{\rho\sigma}\,.
\end{align}
Then, 
\begin{equation}\label{eq:deltaS}
\begin{aligned}
\delta S=\,&\int \diff^5x\, e\, \left [\left(\frac{\partial {\cal L}'}{\partial g^{\mu\nu}}-\frac{1}{2}g_{\mu\nu}{\cal L}'-P^{\alpha\beta\gamma}{}_{\mu}R_{\alpha\beta\gamma \nu}-2\nabla^{\alpha}\nabla^{\beta} P_{\beta \mu\nu\alpha}\right)\delta g^{\mu\nu}-2\nabla_{\mu}\left(\frac{\partial {\cal L}'}{\partial F_{\mu\nu}}\right)\delta A_{\nu}\right. \\[1mm]
&\left.+\nabla_\mu (\Theta^\mu+\Xi^\mu)\right]+\delta S_{\rm{CS}}\, ,
\end{aligned}
\end{equation}
This first line can be expressed exclusively in terms of $\frac{\partial {\cal L}'}{\partial F^{\mu\nu}}$ and $P_{\mu\nu\rho\sigma}$, once $\frac{\partial {\cal L}'}{\partial g^{\mu\nu}}$ is expressed in terms of the latter \cite{Padmanabhan:2011ex}. To this aim, let us write the Lie derivative of the Lagrangian in two different ways. First, we can write it as 
\begin{equation}\label{eq:LieL1}
\begin{aligned}
{\mathsterling}_{\xi}{\mathcal L}'=\,\xi^{\alpha}\partial_{\alpha}{\mathcal L}'=\,&\xi^{\alpha}\left(\frac{\partial {\cal L}'}{\partial R_{\mu\nu\rho\sigma}}\nabla_{\alpha}R_{\mu\nu\rho\sigma}+\frac{\partial {\cal L}'}{\partial g^{\mu\nu}}\nabla_{\alpha}g^{\mu\nu}+\frac{\partial {\cal L}'}{\partial F_{\mu\nu}}\nabla_{\alpha}F_{\mu\nu}\right)\\[1mm]
=\,&\xi^{\alpha}\left(P^{\mu\nu\rho\sigma}\nabla_{\alpha}R_{\mu\nu\rho\sigma}+\frac{\partial {\cal L}'}{\partial F_{\mu\nu}}\nabla_{\alpha}F_{\mu\nu}\right)\,.
\end{aligned}
\end{equation}
However, another possibility is 
\begin{equation}\label{eq:LieL2}
{\mathsterling}_{\xi}{\mathcal L}'=\frac{\partial {\cal L}'}{\partial R_{\mu\nu\rho\sigma}}\mathsterling_{\xi} R_{\mu\nu\rho\sigma}+\frac{\partial {\cal L}'}{\partial g^{\mu\nu}}\mathsterling_{\xi} g^{\mu\nu}+\frac{\partial {\cal L}'}{\partial F_{\mu\nu}}\mathsterling_{\xi} F_{\mu\nu}\, .
\end{equation}
After a bit of algebra, we can rewrite each of the terms appearing in this equation as follows,
\begin{eqnarray}
\frac{\partial {\cal L}'}{\partial R_{\mu\nu\rho\sigma}}\mathsterling_{\xi} R_{\mu\nu\rho\sigma}&=&\xi^{\alpha}P^{\mu\nu\rho\sigma}\nabla_{\alpha}R_{\mu\nu\rho\sigma}+4 P_{\mu}{}^{\alpha\beta\gamma}R_{\nu\alpha\beta\gamma}\nabla^{\mu}\xi^{\nu}\, ,\\[1mm]
\frac{\partial {\cal L}'}{\partial g^{\mu\nu}}\mathsterling_{\xi} g^{\mu\nu}&=&-2\frac{\partial {\cal L}'}{\partial g^{\mu\nu}}\nabla^{(\mu}\xi^{\nu)}\, ,\\[1mm]
\frac{\partial {\cal L}'}{\partial F_{\mu\nu}}\mathsterling_{\xi} F_{\mu\nu}&=&\xi^{\alpha}\frac{\partial {\cal L}'}{\partial F_{\mu\nu}}\nabla_{\alpha}F_{\mu\nu}+2\frac{\partial {\cal L}'}{\partial F^{\mu\rho}} F_{\nu}{}^{\rho}\nabla^{\mu}\xi^{\nu}\, .
\end{eqnarray}
Substituting in \eqref{eq:LieL2} and making use of \eqref{eq:LieL1}, we get the following identity
\begin{equation}
\nabla^{(\mu}\xi^{\nu)}\left(2P_{\mu}{}^{\alpha\beta\gamma}R_{\nu\alpha\beta\gamma}-\frac{\partial {\cal L}'}{\partial g^{\mu\nu}}+\frac{\partial {\cal L}'}{\partial F^{\mu\rho}} F_{\nu}{}^{\rho}\right)+\nabla^{[\mu}\xi^{\nu]}\left(2P_{\mu}{}^{\alpha\beta\gamma}R_{\nu\alpha\beta\gamma}+\frac{\partial {\cal L}'}{\partial F^{\mu\rho}} F_{\nu}{}^{\rho}\right)=0\,.
\end{equation}
Since this equality must be true for an arbitrary vector $\xi^{\mu}$, we conclude that the terms in brackets must vanish, which leads to 
\begin{eqnarray}
\label{eq:dL/dg}
\frac{\partial {\cal L}'}{\partial g^{\mu\nu}}&=&2P_{(\mu}{}^{\alpha\beta\gamma}R_{\nu)\alpha\beta\gamma}+\frac{\partial {\cal L}'}{\partial F^{(\mu|\rho}} F_{|\nu)}{}^{\rho}\, ,\\[1mm]
\frac{\partial {\cal L}'}{\partial F^{[\mu|\rho}} F_{|\nu]}{}^{\rho}&=&-2P_{[\mu}{}^{\alpha\beta\gamma}R_{\nu]\alpha\beta\gamma}\,.
\end{eqnarray}
Then, we can eliminate $\frac{\partial {\cal L}'}{\partial g^{\mu\nu}}$ in \eqref{eq:deltaS} by using \eqref{eq:dL/dg}, which yields 
\begin{equation}
\begin{aligned}
\delta S=\,&\int \diff^5x\, e\, \left [\left(-\frac{1}{2}g_{\mu\nu}{\cal L}'+P^{\alpha\beta\gamma}{}_{\mu}R_{\alpha\beta\gamma \nu}-2\nabla^{\alpha}\nabla^{\beta} P_{\beta \mu\nu\alpha}+\frac{\partial {\cal L}'}{\partial F^{\mu\rho}}F_{\nu}{}^{\rho}\right)\delta g^{\mu\nu}\right. \\[1mm]
&\left.-2\nabla_{\mu}\left(\frac{\partial {\cal L}'}{\partial F_{\mu\nu}}\right)\delta A_{\nu}+\nabla_\mu (\Theta^\mu+\Xi^\mu)\right]+\delta S_{\rm{CS}}\, .
\end{aligned}
\end{equation}
Now let us perform the variation of the Chern-Simons terms explicitly. They yield,
\begin{equation}
\begin{aligned}
\delta S_{\rm{CS}}=\,&\int \diff^5x\, e \left[\left(-\frac{\tilde c_3}{4\sqrt{3}}\epsilon^{\nu\alpha\beta\gamma\delta}F_{\alpha\beta}F_{\gamma\delta}-\frac{\lambda_1\,\alpha}{2\sqrt{3}}\epsilon^{\nu\alpha\beta\gamma\delta}R_{\alpha\beta\rho\sigma}R_{\gamma\delta}{}^{\rho\sigma}\right){\delta A}_{\nu}-2\nabla^{\alpha}\nabla^{\beta}\Pi_{\beta \mu\nu \alpha}{\delta g}^{\mu\nu}\right.\\[1mm]
&\left.+\nabla_{\mu}\left(\Theta^{\mu}_{\rm{CS}}+\Xi^{\mu}_{\rm{CS}}\right)\right]\,,
\end{aligned}
\end{equation}
where the boundary terms are given by 
\begin{eqnarray}
\Theta^{\mu}_{\rm{CS}}&=&2\Pi^{\mu\sigma\rho\lambda}\,   \nabla_\lambda \delta g_{\rho\sigma}-2\nabla_{\lambda} \Pi^{\lambda\sigma\rho\mu} \delta g_{\rho\sigma}\, ,\\
\Xi^{\mu}_{\rm{CS}}&=&\frac{\tilde c_3}{3\sqrt 3}\epsilon^{\nu\mu\rho\sigma\lambda}F_{\rho\sigma}A_\lambda{\delta A}_{\nu} \, .
\end{eqnarray}
and where we have defined
\begin{equation}
\Pi^{\mu\nu\rho\sigma}=-\frac{\lambda_1\,\alpha}{\sqrt{3}}\epsilon^{\mu\nu\alpha\beta\gamma}R_{\alpha\beta}{}^{\rho\sigma}A_{\gamma}\, .
\end{equation}
Arranging all the terms together, we arrive to the final form for the variation of the action,
\begin{equation}
\begin{aligned}
\delta S=\,&\int \diff^5x\, e\, \left \{\left[-\frac{1}{2}g_{\mu\nu}{\cal L}'+P^{\alpha\beta\gamma}{}_{\mu}R_{\alpha\beta\gamma \nu}-2\nabla^{\alpha}\nabla^{\beta} (P_{\beta \mu\nu\alpha}+\Pi_{\beta \mu\nu\alpha})+\frac{\partial {\cal L}'}{\partial F^{\mu\rho}}F_{\nu}{}^{\rho}\right]\delta g^{\mu\nu}\right. \\[1mm]
&\left.\left[-2\nabla_{\mu}\left(\frac{\partial {\cal L}'}{\partial F_{\mu\nu}}\right)-\frac{\tilde c_3}{4\sqrt{3}}\epsilon^{\nu\alpha\beta\gamma\delta}F_{\alpha\beta}F_{\gamma\delta}-\frac{\lambda_1\,\alpha}{2\sqrt{3}}\epsilon^{\nu\alpha\beta\gamma\delta}R_{\alpha\beta\rho\sigma}R_{\gamma\delta}{}^{\rho\sigma}\right]\delta A_{\nu}+\nabla_\mu v^{\mu}\right\}\,,
\end{aligned}
\end{equation} 
where 
\begin{equation}
v^\mu=\Theta^{\mu}+\Theta^{\mu}_{\rm{CS}}+\Xi^{\mu}+\Xi^{\mu}_{\rm{CS}}\, .
\end{equation}
Therefore, the equations of motion of  the metric and gauge field are given by
\begin{eqnarray}
-\frac{1}{2}g_{\mu\nu}{\cal L}'+P^{\alpha\beta\gamma}{}_{(\mu|}R_{\alpha\beta\gamma |\nu)}-2\nabla^{\alpha}\nabla^{\beta} (P_{\beta (\mu\nu)\alpha}+\Pi_{\beta (\mu\nu)\alpha})+\frac{\partial {\cal L}'}{\partial F^{(\mu|\rho}}F_{|\nu)}{}^{\rho}&=&0\, \label{EinsteinEqCorrected},\\[2mm]
2\nabla_{\mu}\left(\frac{\partial {\cal L}}{\partial F_{\mu\nu}}\right)+\frac{\tilde c_3}{4\sqrt{3}}\epsilon^{\nu\alpha\beta\gamma\delta}F_{\alpha\beta}F_{\gamma\delta}+\frac{\lambda_1\,\alpha}{2\sqrt{3}}\epsilon^{\nu\alpha\beta\gamma\delta}R_{\alpha\beta\rho\sigma}R_{\gamma\delta}{}^{\rho\sigma}&=&0\, .
\end{eqnarray}

We now specialize to the action \eqref{eq:4daction2} that we study in the main text.
The concrete expressions for $P_{\mu\nu\rho\sigma}$ and $\frac{\partial {\cal L}}{\partial F_{\mu\nu}}$ are the following,
\begin{equation}\label{Ptensor_explicit}
\begin{aligned}
P_{\mu\nu\rho\sigma}=&{\tilde c}_{0}g_{\mu[\rho}g_{\sigma]\nu}+\lambda_1\,\alpha\left[2R_{\mu\nu\rho\sigma}-4\left(R_{\mu[\rho}g_{\sigma]\nu}-R_{\nu[\rho}g_{\sigma]\mu}\right)+2 g_{\mu[\rho}g_{\sigma]\nu} R \right.\\[1mm]
&\left.-\frac{1}{2}F_{\mu\nu}F_{\rho\sigma}-\frac{1}{12}g_{\mu[\rho}g_{\sigma]\nu} F^2+\frac{1}{3}\left(F_{\mu\alpha}F_{[\rho}{}^{\alpha}g_{\sigma]\nu}-F_{\nu\alpha}F_{[\rho}{}^{\alpha}g_{\sigma]\mu}\right)\right]\, ,\\[1mm]
\frac{\partial {\cal L}'}{\partial F_{\mu\nu}}=&-\frac{\tilde c_2}{2}F^{\mu\nu}+\lambda_1\,\alpha \left(-C^{\mu\nu\rho\sigma}F_{\rho\sigma}+\frac{1}{2}F^{\mu\rho}F_{\rho\sigma}F^{\nu\sigma}\right)\, .
\end{aligned}
\end{equation}
%Let us note that the equations of motion of the theory \eqref{eq:4daction2} have reduced order: they are of third order, while on general grounds one expects fourth-order differential equations when Einstein-Hilbert action is supplemented with quadratic curvature terms. However, in the case at hands the quadratic curvature term is the Gauss-Bonnet invariant, which leads to second-order equations of motion as it is well known. In any case, since our approach here is perturbative in $\alpha$, the r\^ole played by the higher-derivative terms in the equations of motion for the metric and gauge field will be that of  an effective energy-momentum tensor and of an effective electromagnetic current.

\bibliographystyle{JHEP}
\bibliography{HighDer5d}

\providecommand{\href}[2]{#2}\begingroup\raggedright\begin{thebibliography}{10}

\bibitem{Adams:2006sv}
A.~Adams, N.~Arkani-Hamed, S.~Dubovsky, A.~Nicolis and R.~Rattazzi,
  \emph{{Causality, analyticity and an IR obstruction to UV completion}},
  \href{https://doi.org/10.1088/1126-6708/2006/10/014}{\emph{JHEP} {\bfseries
  10} (2006) 014} [\href{https://arxiv.org/abs/hep-th/0602178}{{\ttfamily
  hep-th/0602178}}].

\bibitem{Arkani-Hamed:2006emk}
N.~Arkani-Hamed, L.~Motl, A.~Nicolis and C.~Vafa, \emph{{The String landscape,
  black holes and gravity as the weakest force}},
  \href{https://doi.org/10.1088/1126-6708/2007/06/060}{\emph{JHEP} {\bfseries
  06} (2007) 060} [\href{https://arxiv.org/abs/hep-th/0601001}{{\ttfamily
  hep-th/0601001}}].

\bibitem{Baggio:2014hua}
M.~Baggio, N.~Halmagyi, D.R.~Mayerson, D.~Robbins and B.~Wecht, \emph{{Higher
  Derivative Corrections and Central Charges from Wrapped M5-branes}},
  \href{https://doi.org/10.1007/JHEP12(2014)042}{\emph{JHEP} {\bfseries 12}
  (2014) 042} [\href{https://arxiv.org/abs/1408.2538}{{\ttfamily 1408.2538}}].

\bibitem{Bobev:2021qxx}
N.~Bobev, K.~Hristov and V.~Reys, \emph{{AdS$_{5}$ holography and
  higher-derivative supergravity}},
  \href{https://doi.org/10.1007/JHEP04(2022)088}{\emph{JHEP} {\bfseries 04}
  (2022) 088} [\href{https://arxiv.org/abs/2112.06961}{{\ttfamily
  2112.06961}}].

\bibitem{Liu:2022sew}
J.T.~Liu and R.J.~Saskowski, \emph{{Four-derivative corrections to minimal
  gauged supergravity in five dimensions}},
  \href{https://doi.org/10.1007/JHEP05(2022)171}{\emph{JHEP} {\bfseries 05}
  (2022) 171} [\href{https://arxiv.org/abs/2201.04690}{{\ttfamily
  2201.04690}}].

\bibitem{Melo:2020amq}
J.a.F.~Melo and J.E.~Santos, \emph{{Stringy corrections to the entropy of
  electrically charged supersymmetric black holes with $\mathrm{AdS}_5\times
  S^5$ asymptotics}},
  \href{https://doi.org/10.1103/PhysRevD.103.066008}{\emph{Phys. Rev. D}
  {\bfseries 103} (2021) 066008}
  [\href{https://arxiv.org/abs/2007.06582}{{\ttfamily 2007.06582}}].

\bibitem{Bobev:2020egg}
N.~Bobev, A.M.~Charles, K.~Hristov and V.~Reys, \emph{{The Unreasonable
  Effectiveness of Higher-Derivative Supergravity in AdS$_4$ Holography}},
  \href{https://doi.org/10.1103/PhysRevLett.125.131601}{\emph{Phys. Rev. Lett.}
  {\bfseries 125} (2020) 131601}
  [\href{https://arxiv.org/abs/2006.09390}{{\ttfamily 2006.09390}}].

\bibitem{Bobev:2021oku}
N.~Bobev, A.M.~Charles, K.~Hristov and V.~Reys, \emph{{Higher-derivative
  supergravity, AdS$_{4}$ holography, and black holes}},
  \href{https://doi.org/10.1007/JHEP08(2021)173}{\emph{JHEP} {\bfseries 08}
  (2021) 173} [\href{https://arxiv.org/abs/2106.04581}{{\ttfamily
  2106.04581}}].

\bibitem{Genolini:2021urf}
P.B.~Genolini and P.~Richmond, \emph{{Supersymmetry of higher-derivative
  supergravity in AdS4 holography}},
  \href{https://doi.org/10.1103/PhysRevD.104.L061902}{\emph{Phys. Rev. D}
  {\bfseries 104} (2021) L061902}
  [\href{https://arxiv.org/abs/2107.04590}{{\ttfamily 2107.04590}}].

\bibitem{Gutowski:2004ez}
J.B.~Gutowski and H.S.~Reall, \emph{{Supersymmetric AdS(5) black holes}},
  \href{https://doi.org/10.1088/1126-6708/2004/02/006}{\emph{JHEP} {\bfseries
  02} (2004) 006} [\href{https://arxiv.org/abs/hep-th/0401042}{{\ttfamily
  hep-th/0401042}}].

\bibitem{Chong:2005hr}
Z.W.~Chong, M.~Cvetic, H.~Lu and C.N.~Pope, \emph{{General non-extremal
  rotating black holes in minimal five-dimensional gauged supergravity}},
  \href{https://doi.org/10.1103/PhysRevLett.95.161301}{\emph{Phys. Rev. Lett.}
  {\bfseries 95} (2005) 161301}
  [\href{https://arxiv.org/abs/hep-th/0506029}{{\ttfamily hep-th/0506029}}].

\bibitem{Zaffaroni:2019dhb}
A.~Zaffaroni, \emph{{AdS black holes, holography and localization}},
  \href{https://doi.org/10.1007/s41114-020-00027-8}{\emph{Living Rev. Rel.}
  {\bfseries 23} (2020) 2} [\href{https://arxiv.org/abs/1902.07176}{{\ttfamily
  1902.07176}}].

\bibitem{Cabo-Bizet:2018ehj}
A.~Cabo-Bizet, D.~Cassani, D.~Martelli and S.~Murthy, \emph{{Microscopic origin
  of the Bekenstein-Hawking entropy of supersymmetric AdS$_{5}$ black holes}},
  \href{https://doi.org/10.1007/JHEP10(2019)062}{\emph{JHEP} {\bfseries 10}
  (2019) 062} [\href{https://arxiv.org/abs/1810.11442}{{\ttfamily
  1810.11442}}].

\bibitem{Choi:2018hmj}
S.~Choi, J.~Kim, S.~Kim and J.~Nahmgoong, \emph{{Large AdS black holes from
  QFT}},  \href{https://arxiv.org/abs/1810.12067}{{\ttfamily 1810.12067}}.

\bibitem{Benini:2018ywd}
F.~Benini and P.~Milan, \emph{{Black Holes in 4D $\mathcal{N}$=4
  Super-Yang-Mills Field Theory}},
  \href{https://doi.org/10.1103/PhysRevX.10.021037}{\emph{Phys. Rev. X}
  {\bfseries 10} (2020) 021037}
  [\href{https://arxiv.org/abs/1812.09613}{{\ttfamily 1812.09613}}].

\bibitem{Romelsberger:2005eg}
C.~Romelsberger, \emph{{Counting chiral primaries in N = 1, d=4 superconformal
  field theories}},
  \href{https://doi.org/10.1016/j.nuclphysb.2006.03.037}{\emph{Nucl. Phys. B}
  {\bfseries 747} (2006) 329}
  [\href{https://arxiv.org/abs/hep-th/0510060}{{\ttfamily hep-th/0510060}}].

\bibitem{Kinney:2005ej}
J.~Kinney, J.M.~Maldacena, S.~Minwalla and S.~Raju, \emph{{An Index for 4
  dimensional super conformal theories}},
  \href{https://doi.org/10.1007/s00220-007-0258-7}{\emph{Commun. Math. Phys.}
  {\bfseries 275} (2007) 209}
  [\href{https://arxiv.org/abs/hep-th/0510251}{{\ttfamily hep-th/0510251}}].

\bibitem{Closset:2013vra}
C.~Closset, T.T.~Dumitrescu, G.~Festuccia and Z.~Komargodski, \emph{{The
  Geometry of Supersymmetric Partition Functions}},
  \href{https://doi.org/10.1007/JHEP01(2014)124}{\emph{JHEP} {\bfseries 01}
  (2014) 124} [\href{https://arxiv.org/abs/1309.5876}{{\ttfamily 1309.5876}}].

\bibitem{Assel:2014paa}
B.~Assel, D.~Cassani and D.~Martelli, \emph{{Localization on Hopf surfaces}},
  \href{https://doi.org/10.1007/JHEP08(2014)123}{\emph{JHEP} {\bfseries 08}
  (2014) 123} [\href{https://arxiv.org/abs/1405.5144}{{\ttfamily 1405.5144}}].

\bibitem{Cabo-Bizet:2019eaf}
A.~Cabo-Bizet and S.~Murthy, \emph{{Supersymmetric phases of 4d $ \mathcal{N} $
  = 4 SYM at large $N$}},
  \href{https://doi.org/10.1007/JHEP09(2020)184}{\emph{JHEP} {\bfseries 09}
  (2020) 184} [\href{https://arxiv.org/abs/1909.09597}{{\ttfamily
  1909.09597}}].

\bibitem{ArabiArdehali:2019orz}
A.~Arabi~Ardehali, J.~Hong and J.T.~Liu, \emph{{Asymptotic growth of the 4d $
  \mathcal{N} $ = 4 index and partially deconfined phases}},
  \href{https://doi.org/10.1007/JHEP07(2020)073}{\emph{JHEP} {\bfseries 07}
  (2020) 073} [\href{https://arxiv.org/abs/1912.04169}{{\ttfamily
  1912.04169}}].

\bibitem{Cabo-Bizet:2020nkr}
A.~Cabo-Bizet, D.~Cassani, D.~Martelli and S.~Murthy, \emph{{The large-$N$
  limit of the 4d $ \mathcal{N} $ = 1 superconformal index}},
  \href{https://doi.org/10.1007/JHEP11(2020)150}{\emph{JHEP} {\bfseries 11}
  (2020) 150} [\href{https://arxiv.org/abs/2005.10654}{{\ttfamily
  2005.10654}}].

\bibitem{Aharony:2021zkr}
O.~Aharony, F.~Benini, O.~Mamroud and P.~Milan, \emph{{A gravity interpretation
  for the Bethe Ansatz expansion of the $\mathcal{N}=4$ SYM index}},
  \href{https://doi.org/10.1103/PhysRevD.104.086026}{\emph{Phys. Rev. D}
  {\bfseries 104} (2021) 086026}
  [\href{https://arxiv.org/abs/2104.13932}{{\ttfamily 2104.13932}}].

\bibitem{Kim:2019yrz}
J.~Kim, S.~Kim and J.~Song, \emph{{A 4d $ \mathcal{N} $ = 1 Cardy Formula}},
  \href{https://doi.org/10.1007/JHEP01(2021)025}{\emph{JHEP} {\bfseries 01}
  (2021) 025} [\href{https://arxiv.org/abs/1904.03455}{{\ttfamily
  1904.03455}}].

\bibitem{Cabo-Bizet:2019osg}
A.~Cabo-Bizet, D.~Cassani, D.~Martelli and S.~Murthy, \emph{{The asymptotic
  growth of states of the 4d $ \mathcal{N}=1 $ superconformal index}},
  \href{https://doi.org/10.1007/JHEP08(2019)120}{\emph{JHEP} {\bfseries 08}
  (2019) 120} [\href{https://arxiv.org/abs/1904.05865}{{\ttfamily
  1904.05865}}].

\bibitem{Cassani:2021fyv}
D.~Cassani and Z.~Komargodski, \emph{{EFT and the SUSY Index on the 2nd
  Sheet}}, \href{https://doi.org/10.21468/SciPostPhys.11.1.004}{\emph{SciPost
  Phys.} {\bfseries 11} (2021) 004}
  [\href{https://arxiv.org/abs/2104.01464}{{\ttfamily 2104.01464}}].

\bibitem{ArabiArdehali:2021nsx}
A.~Arabi~Ardehali and S.~Murthy, \emph{{The 4d superconformal index near roots
  of unity and 3d Chern-Simons theory}},
  \href{https://doi.org/10.1007/JHEP10(2021)207}{\emph{JHEP} {\bfseries 10}
  (2021) 207} [\href{https://arxiv.org/abs/2104.02051}{{\ttfamily
  2104.02051}}].

\bibitem{GonzalezLezcano:2020yeb}
A.~Gonz\'alez~Lezcano, J.~Hong, J.T.~Liu and L.A.~Pando~Zayas,
  \emph{{Sub-leading Structures in Superconformal Indices: Subdominant Saddles
  and Logarithmic Contributions}},
  \href{https://doi.org/10.1007/JHEP01(2021)001}{\emph{JHEP} {\bfseries 01}
  (2021) 001} [\href{https://arxiv.org/abs/2007.12604}{{\ttfamily
  2007.12604}}].

\bibitem{Amariti:2021ubd}
A.~Amariti, M.~Fazzi and A.~Segati, \emph{{Expanding on the Cardy-like limit of
  the SCI of 4d $ \mathcal{N} $ = 1 ABCD SCFTs}},
  \href{https://doi.org/10.1007/JHEP07(2021)141}{\emph{JHEP} {\bfseries 07}
  (2021) 141} [\href{https://arxiv.org/abs/2103.15853}{{\ttfamily
  2103.15853}}].

\bibitem{Ohmori:2021sqg}
K.~Ohmori and L.~Tizzano, \emph{{Anomaly Matching Across Dimensions and
  Supersymmetric Cardy Formulae}},
  \href{https://arxiv.org/abs/2112.13445}{{\ttfamily 2112.13445}}.

\bibitem{Hosseini:2017mds}
S.M.~Hosseini, K.~Hristov and A.~Zaffaroni, \emph{{An extremization principle
  for the entropy of rotating BPS black holes in AdS$_{5}$}},
  \href{https://doi.org/10.1007/JHEP07(2017)106}{\emph{JHEP} {\bfseries 07}
  (2017) 106} [\href{https://arxiv.org/abs/1705.05383}{{\ttfamily
  1705.05383}}].

\bibitem{Reall:2019sah}
H.S.~Reall and J.E.~Santos, \emph{{Higher derivative corrections to Kerr black
  hole thermodynamics}},
  \href{https://doi.org/10.1007/JHEP04(2019)021}{\emph{JHEP} {\bfseries 04}
  (2019) 021} [\href{https://arxiv.org/abs/1901.11535}{{\ttfamily
  1901.11535}}].

\bibitem{Cremonini:2009ih}
S.~Cremonini, J.T.~Liu and P.~Szepietowski, \emph{{Higher Derivative
  Corrections to R-charged Black Holes: Boundary Counterterms and the
  Mass-Charge Relation}},
  \href{https://doi.org/10.1007/JHEP03(2010)042}{\emph{JHEP} {\bfseries 03}
  (2010) 042} [\href{https://arxiv.org/abs/0910.5159}{{\ttfamily 0910.5159}}].

\bibitem{Gibbons:1976ue}
G.W.~Gibbons and S.W.~Hawking, \emph{{Action Integrals and Partition Functions
  in Quantum Gravity}},
  \href{https://doi.org/10.1103/PhysRevD.15.2752}{\emph{Phys. Rev. D}
  {\bfseries 15} (1977) 2752}.

\bibitem{Cassani:2019mms}
D.~Cassani and L.~Papini, \emph{{The BPS limit of rotating AdS black hole
  thermodynamics}}, \href{https://doi.org/10.1007/JHEP09(2019)079}{\emph{JHEP}
  {\bfseries 09} (2019) 079}
  [\href{https://arxiv.org/abs/1906.10148}{{\ttfamily 1906.10148}}].

\bibitem{Bobev:2022bjm}
N.~Bobev, V.~Dimitrov, V.~Reys and A.~Vekemans, \emph{{Higher-Derivative
  Corrections and AdS$_5$ Black Holes}},
  \href{https://arxiv.org/abs/2207.10671}{{\ttfamily 2207.10671}}.

\bibitem{Gauntlett:2007ma}
J.P.~Gauntlett and O.~Varela, \emph{{Consistent Kaluza-Klein reductions for
  general supersymmetric AdS solutions}},
  \href{https://doi.org/10.1103/PhysRevD.76.126007}{\emph{Phys. Rev. D}
  {\bfseries 76} (2007) 126007}
  [\href{https://arxiv.org/abs/0707.2315}{{\ttfamily 0707.2315}}].

\bibitem{Cassani:2019vcl}
D.~Cassani, G.~Josse, M.~Petrini and D.~Waldram, \emph{{Systematics of
  consistent truncations from generalised geometry}},
  \href{https://doi.org/10.1007/JHEP11(2019)017}{\emph{JHEP} {\bfseries 11}
  (2019) 017} [\href{https://arxiv.org/abs/1907.06730}{{\ttfamily
  1907.06730}}].

\bibitem{Chen:2005zj}
W.~Chen, H.~Lu and C.N.~Pope, \emph{{Mass of rotating black holes in gauged
  supergravities}},
  \href{https://doi.org/10.1103/PhysRevD.73.104036}{\emph{Phys. Rev. D}
  {\bfseries 73} (2006) 104036}
  [\href{https://arxiv.org/abs/hep-th/0510081}{{\ttfamily hep-th/0510081}}].

\bibitem{deHaro:2000vlm}
S.~de~Haro, S.N.~Solodukhin and K.~Skenderis, \emph{{Holographic reconstruction
  of space-time and renormalization in the AdS / CFT correspondence}},
  \href{https://doi.org/10.1007/s002200100381}{\emph{Commun. Math. Phys.}
  {\bfseries 217} (2001) 595}
  [\href{https://arxiv.org/abs/hep-th/0002230}{{\ttfamily hep-th/0002230}}].

\bibitem{Bianchi:2001kw}
M.~Bianchi, D.Z.~Freedman and K.~Skenderis, \emph{{Holographic
  renormalization}},
  \href{https://doi.org/10.1016/S0550-3213(02)00179-7}{\emph{Nucl. Phys. B}
  {\bfseries 631} (2002) 159}
  [\href{https://arxiv.org/abs/hep-th/0112119}{{\ttfamily hep-th/0112119}}].

\bibitem{Papadimitriou:2005ii}
I.~Papadimitriou and K.~Skenderis, \emph{{Thermodynamics of asymptotically
  locally AdS spacetimes}},
  \href{https://doi.org/10.1088/1126-6708/2005/08/004}{\emph{JHEP} {\bfseries
  08} (2005) 004} [\href{https://arxiv.org/abs/hep-th/0505190}{{\ttfamily
  hep-th/0505190}}].

\bibitem{Kim:2006he}
S.~Kim and K.-M.~Lee, \emph{{1/16-BPS Black Holes and Giant Gravitons in the
  AdS(5) X S**5 Space}},
  \href{https://doi.org/10.1088/1126-6708/2006/12/077}{\emph{JHEP} {\bfseries
  12} (2006) 077} [\href{https://arxiv.org/abs/hep-th/0607085}{{\ttfamily
  hep-th/0607085}}].

\bibitem{Ozkan:2013nwa}
M.~Ozkan and Y.~Pang, \emph{{All off-shell $R^{2}$ invariants in five
  dimensional $\mathcal{N} =$ 2 supergravity}},
  \href{https://doi.org/10.1007/JHEP08(2013)042}{\emph{JHEP} {\bfseries 08}
  (2013) 042} [\href{https://arxiv.org/abs/1306.1540}{{\ttfamily 1306.1540}}].

\bibitem{Jacobson:1993vj}
T.~Jacobson, G.~Kang and R.C.~Myers, \emph{{On black hole entropy}},
  \href{https://doi.org/10.1103/PhysRevD.49.6587}{\emph{Phys. Rev. D}
  {\bfseries 49} (1994) 6587}
  [\href{https://arxiv.org/abs/gr-qc/9312023}{{\ttfamily gr-qc/9312023}}].

\bibitem{Cremonini:2008tw}
S.~Cremonini, K.~Hanaki, J.T.~Liu and P.~Szepietowski, \emph{{Black holes in
  five-dimensional gauged supergravity with higher derivatives}},
  \href{https://doi.org/10.1088/1126-6708/2009/12/045}{\emph{JHEP} {\bfseries
  12} (2009) 045} [\href{https://arxiv.org/abs/0812.3572}{{\ttfamily
  0812.3572}}].

\bibitem{Bergshoeff:2004kh}
E.~Bergshoeff, S.~Cucu, T.~de~Wit, J.~Gheerardyn, S.~Vandoren and
  A.~Van~Proeyen, \emph{{N = 2 supergravity in five-dimensions revisited}},
  \href{https://doi.org/10.1088/0264-9381/23/23/C01}{\emph{Class. Quant. Grav.}
  {\bfseries 21} (2004) 3015}
  [\href{https://arxiv.org/abs/hep-th/0403045}{{\ttfamily hep-th/0403045}}].

\bibitem{Hanaki:2006pj}
K.~Hanaki, K.~Ohashi and Y.~Tachikawa, \emph{{Supersymmetric Completion of an
  R**2 term in Five-dimensional Supergravity}},
  \href{https://doi.org/10.1143/PTP.117.533}{\emph{Prog. Theor. Phys.}
  {\bfseries 117} (2007) 533}
  [\href{https://arxiv.org/abs/hep-th/0611329}{{\ttfamily hep-th/0611329}}].

\bibitem{Butter:2014xxa}
D.~Butter, S.M.~Kuzenko, J.~Novak and G.~Tartaglino-Mazzucchelli,
  \emph{{Conformal supergravity in five dimensions: New approach and
  applications}}, \href{https://doi.org/10.1007/JHEP02(2015)111}{\emph{JHEP}
  {\bfseries 02} (2015) 111} [\href{https://arxiv.org/abs/1410.8682}{{\ttfamily
  1410.8682}}].

\bibitem{Cremonini:2019wdk}
S.~Cremonini, C.R.T.~Jones, J.T.~Liu and B.~McPeak, \emph{{Higher-Derivative
  Corrections to Entropy and the Weak Gravity Conjecture in Anti-de Sitter
  Space}}, \href{https://doi.org/10.1007/JHEP09(2020)003}{\emph{JHEP}
  {\bfseries 09} (2020) 003}
  [\href{https://arxiv.org/abs/1912.11161}{{\ttfamily 1912.11161}}].

\bibitem{Brown:1992br}
J.D.~Brown and J.W.~York, Jr., \emph{{Quasilocal energy and conserved charges
  derived from the gravitational action}},
  \href{https://doi.org/10.1103/PhysRevD.47.1407}{\emph{Phys. Rev. D}
  {\bfseries 47} (1993) 1407}
  [\href{https://arxiv.org/abs/gr-qc/9209012}{{\ttfamily gr-qc/9209012}}].

\bibitem{Balasubramanian:1999re}
V.~Balasubramanian and P.~Kraus, \emph{{A Stress tensor for Anti-de Sitter
  gravity}}, \href{https://doi.org/10.1007/s002200050764}{\emph{Commun. Math.
  Phys.} {\bfseries 208} (1999) 413}
  [\href{https://arxiv.org/abs/hep-th/9902121}{{\ttfamily hep-th/9902121}}].

\bibitem{Bueno:2018xqc}
P.~Bueno, P.A.~Cano and A.~Ruip\'erez, \emph{{Holographic studies of
  Einsteinian cubic gravity}},
  \href{https://doi.org/10.1007/JHEP03(2018)150}{\emph{JHEP} {\bfseries 03}
  (2018) 150} [\href{https://arxiv.org/abs/1802.00018}{{\ttfamily
  1802.00018}}].

\bibitem{Cano:2022ord}
P.A.~Cano, A.J.~Murcia, A.~Rivadulla~S\'anchez and X.~Zhang,
  \emph{{Higher-derivative holography with a chemical potential}},
  \href{https://doi.org/10.1007/JHEP07(2022)010}{\emph{JHEP} {\bfseries 07}
  (2022) 010} [\href{https://arxiv.org/abs/2202.10473}{{\ttfamily
  2202.10473}}].

\bibitem{Teitelboim:1987zz}
C.~Teitelboim and J.~Zanelli, \emph{{Dimensionally continued topological
  gravitation theory in Hamiltonian form}},
  \href{https://doi.org/10.1088/0264-9381/4/4/010}{\emph{Class. Quant. Grav.}
  {\bfseries 4} (1987) L125}.

\bibitem{Myers:1987yn}
R.C.~Myers, \emph{{Higher Derivative Gravity, Surface Terms and String
  Theory}}, \href{https://doi.org/10.1103/PhysRevD.36.392}{\emph{Phys. Rev. D}
  {\bfseries 36} (1987) 392}.

\bibitem{Cvetic:2004hs}
M.~Cvetic, H.~Lu and C.N.~Pope, \emph{{Charged Kerr-de Sitter black holes in
  five dimensions}},
  \href{https://doi.org/10.1016/j.physletb.2004.08.011}{\emph{Phys. Lett. B}
  {\bfseries 598} (2004) 273}
  [\href{https://arxiv.org/abs/hep-th/0406196}{{\ttfamily hep-th/0406196}}].

\bibitem{Kunduri:2005zg}
H.K.~Kunduri and J.~Lucietti, \emph{{Notes on non-extremal, charged, rotating
  black holes in minimal D=5 gauged supergravity}},
  \href{https://doi.org/10.1016/j.nuclphysb.2005.07.017}{\emph{Nucl. Phys. B}
  {\bfseries 724} (2005) 343}
  [\href{https://arxiv.org/abs/hep-th/0504158}{{\ttfamily hep-th/0504158}}].

\bibitem{Silva:2006xv}
P.J.~Silva, \emph{{Thermodynamics at the BPS bound for Black Holes in AdS}},
  \href{https://doi.org/10.1088/1126-6708/2006/10/022}{\emph{JHEP} {\bfseries
  10} (2006) 022} [\href{https://arxiv.org/abs/hep-th/0607056}{{\ttfamily
  hep-th/0607056}}].

\bibitem{Wald:1993nt}
R.M.~Wald, \emph{{Black hole entropy is the Noether charge}},
  \href{https://doi.org/10.1103/PhysRevD.48.R3427}{\emph{Phys. Rev. D}
  {\bfseries 48} (1993) R3427}
  [\href{https://arxiv.org/abs/gr-qc/9307038}{{\ttfamily gr-qc/9307038}}].

\bibitem{Iyer:1994ys}
V.~Iyer and R.M.~Wald, \emph{{Some properties of Noether charge and a proposal
  for dynamical black hole entropy}},
  \href{https://doi.org/10.1103/PhysRevD.50.846}{\emph{Phys. Rev. D} {\bfseries
  50} (1994) 846} [\href{https://arxiv.org/abs/gr-qc/9403028}{{\ttfamily
  gr-qc/9403028}}].

\bibitem{Morales:2006gm}
J.F.~Morales and H.~Samtleben, \emph{{Entropy function and attractors for AdS
  black holes}},
  \href{https://doi.org/10.1088/1126-6708/2006/10/074}{\emph{JHEP} {\bfseries
  10} (2006) 074} [\href{https://arxiv.org/abs/hep-th/0608044}{{\ttfamily
  hep-th/0608044}}].

\bibitem{Dias:2007dj}
O.J.C.~Dias and P.J.~Silva, \emph{{Euclidean analysis of the entropy functional
  formalism}}, \href{https://doi.org/10.1103/PhysRevD.77.084011}{\emph{Phys.
  Rev. D} {\bfseries 77} (2008) 084011}
  [\href{https://arxiv.org/abs/0704.1405}{{\ttfamily 0704.1405}}].

\bibitem{Sen:2005wa}
A.~Sen, \emph{{Black hole entropy function and the attractor mechanism in
  higher derivative gravity}},
  \href{https://doi.org/10.1088/1126-6708/2005/09/038}{\emph{JHEP} {\bfseries
  09} (2005) 038} [\href{https://arxiv.org/abs/hep-th/0506177}{{\ttfamily
  hep-th/0506177}}].

\bibitem{Sen:2008vm}
A.~Sen, \emph{{Quantum Entropy Function from AdS(2)/CFT(1) Correspondence}},
  \href{https://doi.org/10.1142/S0217751X09045893}{\emph{Int. J. Mod. Phys. A}
  {\bfseries 24} (2009) 4225}
  [\href{https://arxiv.org/abs/0809.3304}{{\ttfamily 0809.3304}}].

\bibitem{Elgood:2020xwu}
Z.~Elgood and T.~Ortin, \emph{{T duality and Wald entropy formula in the
  Heterotic Superstring effective action at first-order in
  \ensuremath{\alpha}'}},
  \href{https://doi.org/10.1007/JHEP10(2020)097}{\emph{JHEP} {\bfseries 10}
  (2020) 097} [\href{https://arxiv.org/abs/2005.11272}{{\ttfamily
  2005.11272}}].

\bibitem{Elgood:2020nls}
Z.~Elgood, T.~Ort\'\i{}n and D.~Pere\~n\'\i{}guez, \emph{{The first law and
  Wald entropy formula of heterotic stringy black holes at first order in
  $\alpha'$}}, \href{https://doi.org/10.1007/JHEP05(2021)110}{\emph{JHEP}
  {\bfseries 05} (2021) 110}
  [\href{https://arxiv.org/abs/2012.14892}{{\ttfamily 2012.14892}}].

\bibitem{Tachikawa:2006sz}
Y.~Tachikawa, \emph{{Black hole entropy in the presence of Chern-Simons
  terms}}, \href{https://doi.org/10.1088/0264-9381/24/3/014}{\emph{Class.
  Quant. Grav.} {\bfseries 24} (2007) 737}
  [\href{https://arxiv.org/abs/hep-th/0611141}{{\ttfamily hep-th/0611141}}].

\bibitem{Kovtun:2003wp}
P.~Kovtun, D.T.~Son and A.O.~Starinets, \emph{{Holography and hydrodynamics:
  Diffusion on stretched horizons}},
  \href{https://doi.org/10.1088/1126-6708/2003/10/064}{\emph{JHEP} {\bfseries
  10} (2003) 064} [\href{https://arxiv.org/abs/hep-th/0309213}{{\ttfamily
  hep-th/0309213}}].

\bibitem{Buchel:2008vz}
A.~Buchel, R.C.~Myers and A.~Sinha, \emph{{Beyond eta/s = 1/4 pi}},
  \href{https://doi.org/10.1088/1126-6708/2009/03/084}{\emph{JHEP} {\bfseries
  03} (2009) 084} [\href{https://arxiv.org/abs/0812.2521}{{\ttfamily
  0812.2521}}].

\bibitem{Liu:2010gz}
J.T.~Liu and R.~Minasian, \emph{{Computing 1/$N^{2}$ corrections in AdS/CFT}},
  \href{https://arxiv.org/abs/1010.6074}{{\ttfamily 1010.6074}}.

\bibitem{ArabiArdehali:2013jiu}
A.~Arabi~Ardehali, J.T.~Liu and P.~Szepietowski, \emph{{$1/N^2$ corrections to
  the holographic Weyl anomaly}},
  \href{https://doi.org/10.1007/JHEP01(2014)002}{\emph{JHEP} {\bfseries 01}
  (2014) 002} [\href{https://arxiv.org/abs/1310.2611}{{\ttfamily 1310.2611}}].

\bibitem{ArabiArdehali:2013vyp}
A.~Arabi~Ardehali, J.T.~Liu and P.~Szepietowski, \emph{{The shortened KK
  spectrum of IIB supergravity on $Y^{p,q}$}},
  \href{https://doi.org/10.1007/JHEP02(2014)064}{\emph{JHEP} {\bfseries 02}
  (2014) 064} [\href{https://arxiv.org/abs/1311.4550}{{\ttfamily 1311.4550}}].

\bibitem{Bilal:1999ph}
A.~Bilal and C.-S.~Chu, \emph{{A Note on the chiral anomaly in the AdS / CFT
  correspondence and 1 / N**2 correction}},
  \href{https://doi.org/10.1016/S0550-3213(99)00553-2}{\emph{Nucl. Phys. B}
  {\bfseries 562} (1999) 181}
  [\href{https://arxiv.org/abs/hep-th/9907106}{{\ttfamily hep-th/9907106}}].

\bibitem{Boruch:2022tno}
J.~Boruch, M.T.~Heydeman, L.V.~Iliesiu and G.J.~Turiaci, \emph{{BPS and
  near-BPS black holes in $AdS_5$ and their spectrum in $\mathcal{N}=4$ SYM}},
  \href{https://arxiv.org/abs/2203.01331}{{\ttfamily 2203.01331}}.

\bibitem{Castro:2018ffi}
A.~Castro, F.~Larsen and I.~Papadimitriou, \emph{{5D rotating black holes and
  the nAdS$_{2}$/nCFT$_{1}$ correspondence}},
  \href{https://doi.org/10.1007/JHEP10(2018)042}{\emph{JHEP} {\bfseries 10}
  (2018) 042} [\href{https://arxiv.org/abs/1807.06988}{{\ttfamily
  1807.06988}}].

\bibitem{Larsen:2019oll}
F.~Larsen, J.~Nian and Y.~Zeng, \emph{{AdS$_{5}$ black hole entropy near the
  BPS limit}}, \href{https://doi.org/10.1007/JHEP06(2020)001}{\emph{JHEP}
  {\bfseries 06} (2020) 001}
  [\href{https://arxiv.org/abs/1907.02505}{{\ttfamily 1907.02505}}].

\bibitem{Fujita:2001kv}
T.~Fujita and K.~Ohashi, \emph{{Superconformal tensor calculus in
  five-dimensions}}, \href{https://doi.org/10.1143/PTP.106.221}{\emph{Prog.
  Theor. Phys.} {\bfseries 106} (2001) 221}
  [\href{https://arxiv.org/abs/hep-th/0104130}{{\ttfamily hep-th/0104130}}].

\bibitem{Bergshoeff:2001hc}
E.~Bergshoeff, T.~de~Wit, R.~Halbersma, S.~Cucu, M.~Derix and A.~Van~Proeyen,
  \emph{{Weyl multiplets of N=2 conformal supergravity in five-dimensions}},
  \href{https://doi.org/10.1088/1126-6708/2001/06/051}{\emph{JHEP} {\bfseries
  06} (2001) 051} [\href{https://arxiv.org/abs/hep-th/0104113}{{\ttfamily
  hep-th/0104113}}].

\bibitem{Coomans:2012cf}
F.~Coomans and M.~Ozkan, \emph{{An off-shell formulation for internally gauged
  D=5, N=2 supergravity from superconformal methods}},
  \href{https://doi.org/10.1007/JHEP01(2013)099}{\emph{JHEP} {\bfseries 01}
  (2013) 099} [\href{https://arxiv.org/abs/1210.4704}{{\ttfamily 1210.4704}}].

\bibitem{Anselmi:1997am}
D.~Anselmi, D.Z.~Freedman, M.T.~Grisaru and A.A.~Johansen,
  \emph{{Nonperturbative formulas for central functions of supersymmetric gauge
  theories}}, \href{https://doi.org/10.1016/S0550-3213(98)00278-8}{\emph{Nucl.
  Phys. B} {\bfseries 526} (1998) 543}
  [\href{https://arxiv.org/abs/hep-th/9708042}{{\ttfamily hep-th/9708042}}].

\bibitem{Cassani:2013dba}
D.~Cassani and D.~Martelli, \emph{{Supersymmetry on curved spaces and
  superconformal anomalies}},
  \href{https://doi.org/10.1007/JHEP10(2013)025}{\emph{JHEP} {\bfseries 10}
  (2013) 025} [\href{https://arxiv.org/abs/1307.6567}{{\ttfamily 1307.6567}}].

\bibitem{Fukuma:2001uf}
M.~Fukuma, S.~Matsuura and T.~Sakai, \emph{{Higher derivative gravity and the
  AdS / CFT correspondence}},
  \href{https://doi.org/10.1143/PTP.105.1017}{\emph{Prog. Theor. Phys.}
  {\bfseries 105} (2001) 1017}
  [\href{https://arxiv.org/abs/hep-th/0103187}{{\ttfamily hep-th/0103187}}].

\bibitem{Henningson:1998gx}
M.~Henningson and K.~Skenderis, \emph{{The Holographic Weyl anomaly}},
  \href{https://doi.org/10.1088/1126-6708/1998/07/023}{\emph{JHEP} {\bfseries
  07} (1998) 023} [\href{https://arxiv.org/abs/hep-th/9806087}{{\ttfamily
  hep-th/9806087}}].

\bibitem{Witten:1998qj}
E.~Witten, \emph{{Anti-de Sitter space and holography}},
  \href{https://doi.org/10.4310/ATMP.1998.v2.n2.a2}{\emph{Adv. Theor. Math.
  Phys.} {\bfseries 2} (1998) 253}
  [\href{https://arxiv.org/abs/hep-th/9802150}{{\ttfamily hep-th/9802150}}].

\bibitem{Padmanabhan:2011ex}
T.~Padmanabhan, \emph{{Some aspects of field equations in generalised theories
  of gravity}}, \href{https://doi.org/10.1103/PhysRevD.84.124041}{\emph{Phys.
  Rev. D} {\bfseries 84} (2011) 124041}
  [\href{https://arxiv.org/abs/1109.3846}{{\ttfamily 1109.3846}}].

\end{thebibliography}\endgroup

\end{document}